\documentclass[longauth]{aa}  

\usepackage{graphicx}
\usepackage{txfonts}
\usepackage{longtable}
\usepackage{hyperref}
\usepackage{natbib}
\usepackage[capitalize]{cleveref}

\hypersetup{
        colorlinks=true,    
        linkcolor=blue,  
        citecolor=blue,    
        filecolor=cyan,     
        urlcolor=black,      
} 

\begin{document}

\title{NOEMA formIng Cluster survEy (NICE): Characterizing eight massive galaxy groups at $1.5< z<4$ in the COSMOS field\thanks{Tables C1 to C8 are only available in electronic form at the CDS via anonymous ftp to \url{cdsarc.u-strasbg.fr} (130.79.128.5) or via \url{http://cdsweb.u-strasbg.fr/cgi-bin/qcat?J/A+A/.}}}

\titlerunning{Massive galaxy groups at $1.5<z<4$}
\authorrunning{Sillassen et al.}

\author{
Nikolaj B. Sillassen\inst{1,2},
Shuowen Jin\inst{1,2,\thanks{Marie Curie Fellow}},
Georgios E. Magdis\inst{1,2,3},
Emanuele Daddi\inst{4},
Tao Wang\inst{5,6},
Shiying Lu\inst{4,5,6},
Hanwen Sun\inst{5,6},
Vinod Arumugam\inst{7},
Daizhong Liu\inst{8},
Malte Brinch\inst{1,2},
Chiara D'Eugenio\inst{9,10},
Raphael Gobat\inst{11},
Carlos Gómez-Guijarro\inst{4},
Michael Rich\inst{12},
Eva Schinnerer\inst{13},
Veronica Strazzullo\inst{14,15},
Qinghua Tan\inst{8},
Francesco Valentino\inst{1,16},
Yijun Wang\inst{5,6},
Mengyuan Xiao\inst{17},
Luwenjia Zhou\inst{5,6},
David Blánquez-Sesé\inst{1,2},
Zheng Cai\inst{18},
Yanmei Chen\inst{5,6},
Laure Ciesla\inst{4},
Yu Dai\inst{19},
Ivan Delvecchio\inst{20},
David Elbaz\inst{4},
Alexis Finoguenov\inst{21},
Fangyou Gao\inst{5,6},
Qiusheng Gu\inst{5,6},
Catherine Hale\inst{22},
Qiaoyang Hao\inst{5,6},
Jiasheng Huang\inst{19},
Matt Jarvis\inst{23,24},
Boris Kalita\inst{25},
Xu Ke\inst{5,6},
Aurelien Le Bail\inst{4},
Benjamin Magnelli\inst{4},
Yong Shi\inst{5,6},
Mattia Vaccari\inst{26},
Imogen Whittam\inst{23,24},
Tiancheng Yang\inst{5,6},
Zhiyu Zhang\inst{5,6}
}


   \institute{Cosmic Dawn Center (DAWN), Denmark\\
      \email{nbsi@space.dtu.dk}
    \and
            DTU Space, Technical University of Denmark, Elektrovej 327, DK-2800 Kgs. Lyngby, Denmark\\
       \email{shuji@dtu.dk}
    \and
            Niels Bohr Institute, University of Copenhagen, Jagtvej 128, DK-2200 Copenhagen, Denmark
    \and
            AIM, CEA, CNRS, Université Paris-Saclay, Université Paris Diderot, Sorbonne Paris Cité, F-91191 Gif-sur-Yvette, France
    \and
            School of Astronomy and Space Science, Nanjing University, Nanjing 210093, China\\
        \email{taowang@nju.edu.cn}
    \and
            Key Laboratory of Modern Astronomy and Astrophysics (Nanjing University), Ministry of Education, Nanjing 210093, China
    \and
            IRAM,300 rue de la piscine, F-38406 Saint-Martin d’Hères, France
    \and
            Purple Mountain Observatory, Chinese Academy of Sciences, 10 Yuanhua Road, Nanjing 210023, China
    \and
            Instituto de Astrofísica de Canarias, C. Vía Láctea, s/n, 38205 La Laguna, Tenerife, Spain
    \and 
            Universidad de La Laguna, Dpto. Astrofísica, 38206 La Laguna, Tenerife, Spain
    \and
            Instituto de Física, Pontificia Universidad Católica de Valparaíso, Casilla 4059, Valparaíso, Chile
    \and
            Department of Physics \& Astronomy, University of California Los Angeles, 430 Portola Plaza, Los Angeles, CA 90095, USA
    \and
            Max-Planck-Institut für Extraterrestrische Physik (MPE), Giessenbachstrasse 1, 85748 Garching, Germany
    \and
            INAF-Osservatorio Astronomico di Trieste, Via Tiepolo 11, 34131, Trieste, Italy
    \and    
            IFPU-Institute for Fundamental Physics of the Universe, Via Beirut 2, 34014, Trieste
    \and
            European Southern Observatory, Karl-Schwarzschild-Str. 2, D85748 Garching bei Munchen, Germany
    \and
            Department of Astronomy, University of Geneva, Chemin Pegasi 51, 1290 Versoix, Switzerland
    \and
            Department of Astronomy, Tsinghua University, Beijing 100084, China
    \and
            Chinese Academy of Sciences South America Center for Astronomy (CASSACA), National Astronomical Observatories of China (NAOC), 20A Datun Road
    \and
            INAF- Osservatorio Astronomico di Brera, via Brera 28, I-20121, Milano, Italy \& via Bianchi 46, I-23807, Merate, Italy
    \and
            Department of Physics, University of Helsinki, Gustaf Hällströmin katu 2, FI-00014 Helsinki, Finland
    \and
            School of Physics and Astronomy, Institute for Astronomy, University of Edinburgh, Royal Observatory, Blackford Hill, EH9 3HJ Edinburgh, UK 
    \and
            Sub-Department of Astrophysics, University of Oxford, Keble Road, Oxford OX1 3RH, UK 
    \and
            Department of Physics and Astronomy, University of the Western Cape, Robert Sobukwe Road, 7535 Bellville, Cape Town, South Africa 
    \and    
            Kavli IPMU (WPI), UTIAS, The University of Tokyo, Kashiwa, Chiba 277-8583, Japan
    \and
            Inter-university Institute for Data Intensive Astronomy, Department of Physics and Astronomy, University of the Western Cape, 7535 Bellville, Cape Town, South Africa
    }
   \date{Received XX / Accepted XX}

 \abstract
{The NOrthern Extended Millimeter Array (NOEMA) formIng Cluster survEy (NICE) is a NOEMA large programme targeting 69 massive galaxy group candidates at $z>2$ over six deep fields with a total area of 46 deg$^2$. Here we report the spectroscopic confirmation of eight massive galaxy groups at redshifts $1.65\leq z\leq3.61$ in the Cosmic Evolution Survey (COSMOS) field. Homogeneously selected as significant overdensities of red IRAC sources that have red {\it Herschel} colours, four groups in this sample are confirmed by CO and [CI] line detections of multiple sources with NOEMA 3mm observations, three are confirmed with Atacama Large Millimeter Array (ALMA) observations, and one is confirmed by H$\alpha$ emission from Subaru/FMOS spectroscopy. 
Using rich ancillary data in the far-infrared and sub-millimetre, we constructed the integrated  far-infrared spectral energy distributions for the eight groups, obtaining a total infrared star formation rate  (SFR) of 260-1300~${\rm M_\odot}$~yr$^{-1}$.
We adopted six methods for estimating the dark matter masses of the eight groups, including stellar mass to halo mass relations, overdensity with galaxy bias, and NFW profile fitting to radial stellar mass densities. We find that the radial stellar mass densities of the eight groups are consistent with a NFW profile, supporting the idea that they are collapsed structures hosted by a single dark matter halo. The best halo mass estimates are $\log(M_{\rm h}/{\rm M_\odot})=12.8-13.7$ with a general uncertainty of 0.3 dex.
Based on the halo mass estimates, we derived baryonic accretion rates (BARs) of $(1-8)\times10^{3}\,{\rm M_{\odot}/yr}$ for this sample. Together with massive groups in the literature, we find a quasi-linear correlation between the integrated SFR/BAR ratio and the theoretical halo mass limit for cold streams, $M_{\rm stream}/M_{\rm h}$, with ${\rm SFR/BAR}=10^{-0.46\pm0.22}\left({M_{\rm stream}/M_{\rm h}}\right)^{0.71\pm0.16}$ with a scatter of $0.40\,{\rm dex}$.
Furthermore, we compared the halo masses and the stellar masses with simulations, and find that the halo masses of all structures are  consistent with those of progenitors of $M_{\rm h}(z=0)>10^{14}\,{\rm M_{\odot}}$ galaxy clusters, and that the most massive central galaxies have stellar masses consistent with those of the brightest cluster galaxy progenitors in the TNG300 simulation. Above all, the results strongly {suggest} that these massive structures are in the process of forming massive galaxy clusters via baryonic and dark matter accretion.
}


\maketitle
 

\section{Introduction}

Galaxy clusters are the largest virialized structures in the local Universe, and their progenitors \citep[see][for a review]{Overzier2016} are suspected to be massive galaxy groups and protoclusters in the early Universe \citep{Muldrew2015_extended_protoclusters}. At high redshifts, massive groups and protoclusters represent the earliest massive collapsed structures hosted by massive dark matter halos \citep{Wang_T2016cluster,Willis2020cluster,DiMascolo2023_SZ_spiderweb}. Characterizing their dark matter halos provides key constraints on cosmological parameters that can be used to test structure formation theories \citep{Overzier2016}.
On the other hand, the evolutionary track between the early structures and local clusters is a major topic in modern astrophysics but remains poorly understood \citep{Chiang2013cluster,Shimakawa2018SW}. 
These early ($z>2$) massive structures host a large number of star-forming galaxies rich in dust and gas (e.g. \citealt{Dannerbauer2014LABOCA,Oteo2018cluster,Gobat2019_CLJ1449}). The most massive members of these structures are often found to have vigorous starbursts \citep[e.g.][]{Oteo2018cluster,Miller2018cluster_z4,Zhou2023_LHSBC3_NICE}, while some members are already quiescent at $z>2$ \citep[e.g.][]{Kubo2021quiescent,Kalita2021quiescent,McConachie2022,Ito2023,Jin2023_Cosmic_Vine}. The diverse populations of member galaxies and their rapid evolution make them an ideal laboratory for studying the formation of clusters and brightest cluster galaxies \citep[BCGs;][]{Shi2023_bcg_formation,Jin2023_Cosmic_Vine}.
Therefore, studying a large sample of galaxy groups and (proto-)clusters before cosmic noon \citep[i.e. the peak of star formation at $z\sim2$;][]{Madau2014a}, where groups and (proto-)clusters are expected to contribute significantly to the cosmic star formation rate (SFR) density, is essential to unveiling the evolution and formation of massive galaxies and clusters \citep{Chiang2017cluster}.

In the last decade, massive groups and (proto-)clusters have been discovered at cosmic noon and out to the epoch of re-ionization ($z\gtrsim6-8$; e.g. \citealt{Hu_WD2021NatAs,Brinch2023_allpczs,Jones2023,Arribas2023,Morishita2023}) using various techniques. 
To date, most structures have been selected by mapping overdensity of galaxies, including the overdensities of optical/near-infrared (NIR) sources \citep[e.g.][]{Gobat2011,Wang_T2016cluster}, dusty star-forming galaxies \citep[e.g.][]{Dannerbauer2014LABOCA}, and star-forming galaxies detected in the radio \citep[e.g.][]{Daddi2017,Daddi2021Lya}. However, a major limitation of the overdensity mapping is the line-of-sight projection \citep[e.g.][]{Chen2023_overdensity_projection}. In this regard, advanced techniques that include colour and redshift information, for example red IRAC colours \citep{Wylezalek2013,Wylezalek2014,Mei2023}, narrowband emitters \citep{Koyama2013cluster,Shimakawa2014HAE,Hu_WD2021NatAs}, and photometric or spectroscopic redshifts \citep[e.g.][]{Cucciati2018,Sillassen2022_HPC1001,Helton2023jwst,Jin2023_Cosmic_Vine}, have significantly improved the efficiency and robustness of the selection. 
Second, far-infrared (FIR) and (sub-)millimetre-bright dusty star-forming galaxies have been used to detect high-$z$ protocluster candidates, which were later confirmed by follow-up observations, for example GN20 \citep{Daddi2009GN20}, AzTEC3 \citep{Capak2011Nature}, the SCUBA2-selected HDF850.1 \citep{Walter2011CI}, the \textit{Herschel}-selected DRC \citep{Oteo2018cluster} and HerBS-70 \citep{Bakx2024_HerBS70}, and the millimetre-selected SPT2349-56 \citep{Miller2018cluster_z4}.
Third, extended X-ray emission from hot plasma in the intracluster medium (ICM) can be used to trace mature clusters \citep[e.g.][]{Stanford2006_XMM_cluster}. 
Fourth, Sunyaev-Zel'dovich (SZ) emission \citep[][]{SunyaevZeldovich1970} is a sign of a collapsed massive structure, where cosmic microwave background photons are accelerated by the host ICM in clusters via inverse Compton scattering \citep[e.g.][]{Staniszewski2009_SZ_survey}. 
Finally, extended Ly$\alpha$ emission (Ly$\alpha$ blobs), stemming from cool gas in dense regions associated with overdensities of galaxies, can be used to trace actively star-forming massive groups \citep[e.g.][]{Prescott2008_Lya_blob_protocluster}.
A new technique that uses Ly$\alpha$ absorption has recently been exploited to search for galaxy groups missed by other surveys: Ly$\alpha$ tomography \citep{Lee2018_Lya_tomography,Newman2022_Lya_tomography}. 

Overdensity mapping and dusty star-forming galaxy (DSFG) tracers are widely used to search for massive structures at $z>2$, for which the other methods are either infeasible (e.g. SZ and X-ray) or require prohibitively time-consuming observations. 
Therefore, a combined search of FIR-luminous objects and overdensities of optical/NIR galaxies provides an efficient and powerful tool for selecting massive galaxy groups and protoclusters. However, this method still suffers from projection effects, and high quality photometric redshifts are required to associate the FIR emission with the candidate overdensities.
Recently, the advent of deep, wide, and panchromatic extragalactic survey fields, such as the Cosmic Evolution Survey (COSMOS), and the construction of state-of-the-art, multi-band photometric catalogues (e.g. COSMOS2020; \citealt{Weaver2022COSMOS2020}) have made this cluster selection method more viable than ever before. The efficiency of this combined search method has been verified by some pilot projects; they successfully selected the cluster CLJ1001, which was later spectroscopically confirmed to reside at $z=2.51$ \citep{Wang_T2016cluster}, as well as three massive groups at $z\sim3$ \citep{Daddi2021Lya,Daddi2022Lya}. These discoveries demonstrated the potential of building large, homogeneously selected samples of massive groups and protoclusters in large and deep survey fields.

The dark matter halo mass is a vital parameter of these identified groups and protoclusters, and an accurate estimate of halo masses is crucial to assessing their dynamical and evolutionary states \citep{Chiang2013cluster,Ata2022cluster,Montenegro-Taborda2023BCG}.
To date, various methods have been exploited to estimate the dark matter halo mass of galaxies and galaxy groups, including (1) the stellar-to-halo mass relation  \citep[SHMR; e.g.][]{Behroozi2013Mhalo,Shuntov2022}, (2) dynamical constraints of member galaxies \citep[e.g.][]{Wang_T2016cluster,Miller2018cluster_z4}, (3) X-ray emission from the hot ICM \citep[e.g.][]{Gobat2011,Wang_T2016cluster}, and (4) the SZ effect from the inverse Compton scattering of cosmic microwave background photons \citep[e.g.][]{Gobat2019_CLJ1449,DiMascolo2023_SZ_spiderweb}.
However, constraining the dark matter halo mass of structures in the early Universe remains a challenging task.
This is because (1) optical data are often too shallow to allow for accurate stellar mass measurements \citep{Daddi2021Lya}, (2) spectroscopic surveys of high-$z$ structures are too incomplete for membership identification, (3) X-ray observations from current facilities are not deep enough to probe the hot ICM \citep{Overzier2016}, and (4) the SZ effect is a powerful tool but one that is very demanding in terms of observing time, and only a couple of structures have been detected at $z\gtrsim2$ \citep{Gobat2019_CLJ1449,DiMascolo2023_SZ_spiderweb}.
Recently, \cite{Wang_T2016cluster} found that the stellar mass density profile of the $z=2.5$ X-ray-detected cluster CLJ1001 can be fitted by a projected Navarro-Frenk-White (NFW) profile \citep*{Navarro1997_NFW_profile}, suggesting that the structure is virialized and that the halo has a relatively high concentration \citep{Sun2024_CLJ1001}. The similar shape of the two profiles implies that the halo mass can be estimated from the stellar mass density profile, which provides an efficient and powerful approach to constraining the halo mass of collapsed structures.
However, this novel method has yet to be tested with a statistically robust sample at high redshifts and still suffers from limitations (1) and (2).


Another open question concerns the growth of galaxies residing in massive structures.
Theoretically, structures with dark matter halo masses above $M_{\rm shock}\simeq10^{12}\,{\rm M_{\odot}}$ would generate shocks and heat the infalling gas of the intergalactic medium to the temperature of the ICM \citep{Birnboim2003_hot_accretion}, preventing further star formation and the growth of cluster galaxies. 
However, advanced simulations predict a mechanism of gas inflow in which cold streams travelling along intersections of dense filaments are able to penetrate the halo without being shock-heated \citep{Dekel2006,Dekel2009Nature,RosdahlBlaizot_2012_cold_accretion,Mandelker2020_cold_streams}. This regime of cold gas inflow in a hot environment is defined by a redshift-dependent theoretical upper limit of the host halo mass where these streams can efficiently occur, $M_{\rm stream}$, and a theoretical lower limit of the host halo mass where shock heating occurs, $M_{\rm shock}$ \citep{Dekel2013}. 
Nevertheless, observational evidence for this picture is lacking. 
The model from \cite{Goerdt2010core} predicted that this mode of cold accretion would be detectable via collision-driven Ly$\alpha$ emission. This idea is supported by the recent discovery of giant Ly$\alpha$ nebulae in massive galaxy groups at $1.9<z<3.3$ by \cite{Daddi2021Lya,Daddi2022Lya}. 
The rate of accretion of both cold and warm gas can be quantified with the baryonic accretion rate (BAR), which is predicted to scale with the dark matter halo mass \citep{Goerdt2010core} and the total SFR it feeds \citep{Daddi2022Lya}. To further verify and observationally constrain this picture of cold accretion, a large sample of homogeneously selected massive galaxy groups and protoclusters is needed. 

Finally, the fate of massive structures in the early Universe remains unresolved. Whether massive groups of galaxies at high-$z$ will form present-day clusters, a proposition that is fundamental for precisely defining the term `protocluster', remains an open question.
Currently, there are no direct observables that can offer a robust characterization and accurate classification of protoclusters \citep{Overzier2016}. Instead, comparisons between the observables and cosmological simulations provide a feasible approach to inferring the evolutionary stage and final fate of structures \citep[e.g.][]{Chiang2013cluster,Miller2018cluster_z4,Ata2022cluster,Jin2023a,Jin2023_Cosmic_Vine}. 

In this paper we report spectroscopic confirmation of eight galaxy groups from the NOrthern Extended Millimeter Array (NOEMA) formIng Cluster survEy (NICE) in the COSMOS field and study their integrated stellar, gas, dust, and dark matter properties.
The paper is organized as follows: We describe the sample and selection in \cref{sec:target-selection}. In \cref{sec:Data} we describe the observations and data reduction. In \cref{sec:methods} we explain in detail our analysis methods. 
We present our results in \cref{sec:results}, discuss the corresponding science in \cref{sec:discussion}, and summarize this study in \cref{sec:summary}.
%
We adopt a flat cosmology with parameters $H_0=70\,{\rm km\,s^{-1}\,Mpc^{-1}}$, $\Omega_{m}=0.27$, and $\Omega_\Lambda=0.73$, and use a \citet{Chabrier2003} initial mass function. Magnitudes are in the AB system \citep{Oke1974_AB_sys}.

\section{NOEMA formIng Cluster survEy (NICE)}
\label{sec:target-selection}

NICE is a 159 hours NOEMA large programme (ID:M21AA, PIs: E. Daddi and T. Wang) targeting 48 massive galaxy group candidates in the COSMOS, Lockman Hole, Elais-N1, Boötes, and XMM-LSS fields. 
This programme is complemented by a 40 hours Atacama Large submillimeter Array (ALMA) programme (ID: 2021.1.00815.S, PI: E. Daddi) targeting 25 candidates in the southern sky in the ECDFS, COSMOS, and XMM-LSS
fields. Four candidates are observed with both NOEMA and ALMA. 
Overall, 69 targets are selected in a total of 46 deg$^2$ sky area.
The first discovery of NICE is a $z=3.95$ star-bursting group in the Lockman Hole field, which was recently reported in \citet{Zhou2023_LHSBC3_NICE}. 
In this paper, we focus on the eight candidates in the COSMOS field (\cref{tab:obs_table}), of which four  are observed with NOEMA, three are observed with ALMA, and one is observed with both facilities. 



In Table~\ref{tab:obs_table} we list the eight massive galaxy group candidates of this study: seven of them (i.e. COS-SBC3, COS-SBC4, COS-SBC6, COS-SBCX1, COS-SBCX3, COS-SBCX4, and COS-SBCX7) were selected as a significant overdensity of red IRAC sources with red \textit{Herschel} colours, following the identical selection method utilized by \cite{Zhou2023_LHSBC3_NICE}. The selection is described in detail in \citet{Zhou2023_LHSBC3_NICE}, and we briefly list the selection criteria below:


(1) Overdensity of red IRAC sources:
\begin{align}
   \hspace{1cm} & [3.6]-[4.5]>0.1 \nonumber \\ 
   \hspace{1cm} & 20< [4.5] < 23 \label{eq:IRAC_fluxes} \\
   \hspace{1cm} & {\Sigma_{N}>5\sigma,} \nonumber
\end{align}

\noindent where [3.6] and [4.5] are IRAC channel 1 and 2 magnitudes from the COSMOS2020 catalogue \citep{Weaver2022COSMOS2020}. 
We used the distance to the $N$-th nearest neighbour $r_N$ to quantify the local galaxy density, and constructed a galaxy surface density map of $\Sigma_N=N/(\pi r^2_N)$. We selected overdensities with either $\Sigma_5$ or $\Sigma_{10}$ that are $5\sigma$ above the field levels in log scale (e.g. \citealt{Wang_T2016cluster}).

(2) \textit{Herschel} detection with red colours:
\begin{align}
   \hspace{1cm} & S_{500\,{\rm \mu m}}  >30\,{\rm mJy} \nonumber \\
   \hspace{1cm} & S_{350\,{\rm \mu m}}/S_{250\,{\rm \mu m}}  >1.07  \label{eq:Herschel_fluxes} \\ 
   \hspace{1cm} & S_{500\,{\rm \mu m}}/S_{350\,{\rm \mu m}}  >0.72. \nonumber
\end{align}

Additionally, the target HPC1001 was originally selected by \cite{Sillassen2022_HPC1001} using the overdensity of COSMOS2020 sources at $z_{\rm phot}\sim3.7$ and an overdensity of ALMA sources. 
We note that HPC1001 has red IRAC colours as in \cref{eq:IRAC_fluxes} but was not selected as an IRAC overdensity. This is because HPC1001 is extremely compact, and the members are severely blended in the low-resolution IRAC images. 
However, the $\Sigma_5$ of $z\sim3.7$ sources in HPC1001 is 6.8$\sigma$ above the average, which is one of the strongest overdensities in the COSMOS field at $z>3$ \citep{Sillassen2022_HPC1001}.
Furthermore, HPC1001 satisfies the criteria of \textit{Herschel} selection (\cref{eq:Herschel_fluxes}) and shows even redder colours. 
Therefore, the eight candidates share a consistent selection, constituting a homogeneous sample. Benefiting from the rich multi-wavelength datasets and well-established catalogues in the COSMOS field, the eight high-$z$ group candidates are entitled to the best data quality in the NICE sample, constituting an ideal sample for this initial statistical study of the NICE programme.



\begin{table*}
    \centering
    \setlength{\tabcolsep}{2pt}
    \caption{Observations presented in this work.}
    \begin{tabular}{c c c c c c c c c c c}
    \hline\hline
        Target name & RA & Dec. & $z_{\rm spec}$ & Programme ID & Sensitivity$^a$ & Beam & Ancillary ID & Sens.$^b$ & Ang. Res.$^c$\\
                  & [deg] & [deg] & & & [mJy/beam] & [arcsec] & & [mJy/beam] & [arcsec]\\
        \hline       
        (NOEMA) \\
        HPC1001 & 150.4656 & 2.6359 & $3.613$  & M21AA & $0.13$ & $4.7\times3.7$ & 2013.1.00034.S & 0.06 & 0.4\\
        COS-SBCX3 & 150.3113 & 2.4511 & $3.031$ & M21AA & $0.15$ & $5.4\times4.5$ & 2016.1.00463.S & 0.26 & 0.8\\
        COS-SBCX4 & 150.7509 & 2.4132 & $2.646$ & M21AA & $0.13$ & $4.2\times1.8$ & 2016.1.00463.S & 0.27 & 0.8\\ 
        COS-SBCX7 & 149.9898 & 1.7978 & $2.415$ & M21AA & $0.14$ & $4.7\times3.8$ & 2015.1.00137.S & 0.12 & 1.0\\
        \hline
        (ALMA) \\
        COS-SBC3 & 150.7196 & 2.6995 & $2.365$ & 2021.1.00815.S & $0.13$ &$0.84\times0.55$ & 2021.1.00246.S & 0.02 & 1.6\\
        COS-SBC6 & 149.7057 & 2.2160 & $2.323$ & 2021.1.00815.S & $0.13$ & $0.87\times0.53$ & 2016.1.00463.S & 0.27 & 0.8\\
        COS-SBC4 & 150.0364 & 2.2177 & $1.65^b$ & 2021.1.00815.S & $0.13$ & $0.72\times0.57$ & 2016.1.00463.S & 0.27 & 0.8\\
        \hline
        (NOEMA + ALMA)\\
        COS-SBCX1 & 150.3492 & 2.7619 & $2.422$ & M21AA & 0.14 & $4.8\times3.8$ & 2016.1.00463.S & 0.26 & 0.8\\
         & & & & 2021.1.00815.S & $0.23$ & $0.83\times0.66$ \\
         \hline
    \end{tabular}
    {Notes: $^a$line sensitivity over $500\,{\rm km s^{-1}}$. $^b$The redshift is measured from Subaru/FMOS observations \citep{Kashino2019_FMOS_COSMOS}. $^b$Continuum sensitivity of the ancillary ALMA data. $^c$Angular resolution of the ancillary ALMA data.
    \label{tab:obs_table}}
\end{table*}

\section{Data}
\label{sec:Data}
The eight galaxy group candidates in the COSMOS field are observed with the NOEMA and ALMA interferometers. A summary of our observations is provided in \cref{tab:obs_table}, and we describe them in detail below.

\subsection{NOEMA}
As a part of the NICE large programme, five targets were observed with NOEMA (Table~\ref{tab:obs_table}). The observations were designed with two frequency setups of two sidebands each in NOEMA Band 1, covering CO(3-2) and CO(4-3) in the redshift ranges $2<z<4$ and $z>3$, respectively. The first setup covered the frequencies of the most probable CO lines. If lines were detected with the first setup, the second setup was not executed.
The observations were conducted in October 2022 with a total on-source time of 10.8h for the five targets in track sharing mode with array configurations C and D. The data were reduced and calibrated using the institut de radioastronomie millimétrique (IRAM) software package\footnote{\url{https://www.iram.fr/IRAMFR/GILDAS}}. The final data products were generated in uv tables, reaching an average continuum sensitivity of $17\,{\rm \mu Jy/beam}$ and line sensitivity ${\rm 0.13~mJy/beam}$ over a 500~km~s$^{-1}$ line width at $\sim$100~GHz, with an average angular resolution of $\sim4.1"$ (see details in Table 1).

\subsection{ALMA}
ALMA Band 4 and 5 observations of four targets were carried out in Cycle 8. We designed four frequency tunings that cover one or multiple CO and CI lines based on the estimated redshift of the groups. The four tunings cover the frequency range $135-183\,{\rm GHz}$, but leave two gaps at 157.5--168.5~GHz and 170--178~GHz.
The observations were conducted in August 2022, reaching an rms sensitivity of ${\rm 0.13~mJy/beam}$ over a 500~km~s$^{-1}$ line width at $\sim150$~GHz. The raw data were reduced and calibrated using the Common Astronomy Software Application (\texttt{CASA}; \citealt{McMullin2007CASA}) pipeline. Following our standard pipeline (e.g. \citealt{Coogan2018,Jin2019alma,Jin2022,Zhou2023_LHSBC3_NICE}), we converted the calibrated measurement sets to \texttt{uvfits} format for further analysis with the \texttt{GILDAS/mapping} software. In each ALMA tuning, we achieved an average continuum sensitivity of $16\,{\rm \mu Jy/beam}$ and an angular resolution of $\sim0.7"$.

\subsection{Spectrum extraction pipeline}
The extraction of NOEMA and ALMA spectra is carried out with the pipeline adopted in \citet{Zhou2023_LHSBC3_NICE}. Briefly, we extracted spectra in the $uv$ space (visibility) using the \texttt{uvfit} routine of \texttt{GILDAS}. The \texttt{uvfit} run was performed on the original uv tables at all frequencies, where we adopted a point source model for the sources, using prior positions from the COSMOS2020 catalogue or ALMA continuum images.
To enhance the sensitivity in overlapping frequency ranges, we combined all spectral windows of ALMA into a single 1D spectrum for each source.

\subsection{Ancillary data}

We utilized the rich multi-wavelength data and catalogues in the COSMOS field. In the optical and NIR, we adopted the COSMOS2020 photometric catalogue \citep{Weaver2022COSMOS2020}, along with the provided  photometric redshifts and stellar masses that were crucial for the identification of spectral lines in our data and for characterization of the physical properties of the member galaxies.
At FIR and sub-millimetre wavelengths, we made use of the MIPS 24~$\mu$m \citep{LeFloch2009}, \textit{Herschel} 100--500~$\mu$m \citep{Lutz2011}, SCUBA-2 850~$\mu$m \citep{Simpson2019}, and AzTEC 1.1~mm maps \citep{Aretxaga2011}, and we measured the integrated fluxes of this sample by performing the super-deblending technique \citep{Liu2018_super_deblending,Jin2018cosmos}. 
For radio bands, we used the low-resolution MeerKAT-DR1 image with a beam size of $8.9''\times8.9''$ and frequency range 1.15--1.35~GHz (\citealt{Jarvis2016mightee,Heywood2022_Meerkat}; Hale et al. in prep.).
We also used the archival ALMA data (ID: 2016.1.00463.S, PI: Y. Matsuda; ID: 2021.1.00246.S, PI: C. Chen; ID: 2015.1.00137.S, PI: N. Scoville; 2013.1.00034.S, PI: N. Scoville) for the dust continuum imaging in Band 6 and 7, and ALMA photometry from the A$^3$COSMOS catalogue \citep{Liu2019A3COSMOS} where available.

\section{Methodology}
\label{sec:methods}
In this section we describe the methods adopted to determine the redshifts and group membership, as well as the integrated FIR properties, the BAR, and the halo mass of the groups.

\subsection{Line detection and redshift determination}

Following \cite{Zhou2023_LHSBC3_NICE}, we ran a line-searching algorithm as in \citet{Coogan2019_CO} and \citet{Jin2019alma} to search for the highest integrated signal-to-noise ratio (S/N) emission lines in the 1D NOEMA/ALMA spectra. We fit the line-free continuum emission as a power-law with a fixed slope of 3.7 in frequency, assuming modified blackbody emission with $\beta\sim1.7$ \citep{Magdis2012SED}, by masking out channels where significant ($P_{\rm chance}<5\%$, \citealt{Jin2019alma}) 
emission lines are detected. The continuum-subtracted spectra are then fitted with a Gaussian line-profile using \texttt{MPFIT}\footnote{\url{http://cow.physics.wisc.edu/~craigm/idl/idl.html}}, at the frequencies identified by the line searching algorithm. 
The redshift can be robustly identified if two or more lines are detected on one source. For sources only detected with a single line, we determined their redshifts by comparing them to the photo-z redshift probability density function (PDF(z)) from the COSMOS2020 catalogue.
As an example, in Fig.~\ref{fig:pdfz-hpc1001}, we determined the best $z_{\rm spec}$ solution by comparing all possible redshift solutions for each object with optical/NIR photometric PDF(z) from the Classic \texttt{LePhare} version of COSMOS2020. 
For galaxies with a broad PDF(z) and an emission line detection close to the frequency of the structure as determined by other secure members, it is assumed that its redshift is closest to the redshift of the structure. 
Such cases only occur on two sources (HPC1001 ID1272853 and SBCX4 ID1049929), and one of them (ID1272853) has been recently confirmed with our ALMA [CI] observation (ID: 2023.1.00652.S, PI: N. Sillassen), which further validates our identification method.

\subsection{Candidate member selection}
With the photometric redshifts in the COSMOS2020 catalogue \citep{Weaver2022COSMOS2020}, we selected candidate group members with $|z_{\rm phot}-z_{\rm spec,group}|<0.1(1+z_{\rm spec,group})$ within the virial projected radius from the group centres (see \cref{tab:dark_matter}), where $z_{\rm spec,group}$ is the spectroscopic redshift of the central galaxy. This redshift range is defined by the representative uncertainty at the faint end of the COSMOS2020 catalogue of $\sim10\%$ \citep{Weaver2022COSMOS2020}. The virial radius limit ensures that we were only probing the inner region of these structures. We note there is a radial- and stellar-mass-dependent expected interloper fraction (\cref{fig:2d-interlopers}-right); that is, low stellar mass candidate members far from the centre have a higher chance of being interlopers. We discuss implications of interlopers in detail in \cref{sec:interlopers}. 


\subsection{Integrated FIR spectral energy distributions}

A deblended FIR catalogue of the COSMOS field is publicly available \citep{Jin2018cosmos}; however, it is not directly applicable for individual galaxies in crowded group and protocluster environments.
Due to the overdense nature and the large \textit{Herschel} and SCUBA2 beams (15--36$''$), in the group centre tens of galaxies can be present in one beam, which leads to severe blending of sources. 
Given the high compactness of the eight groups, we applied the super-deblending technique \citep{Jin2018cosmos,Liu_DZ2017} with improved priors to measure the integrated fluxes in \textit{Herschel} and SCUBA2 images, following the same method applied in \citet{Daddi2021Lya} and \citet{Zhou2023_LHSBC3_NICE}.

In detail, for each group, we defined one prior at the peak position of the SCUBA2 850~$\mu$m detection \citep{Simpson2019} to represent the whole structure. In the prior list, we excluded sources within a $10''$ radius of the prior to reduce the crowding \citep{Liu_DZ2017}. Then we subtracted faint foreground and background sources by convolving the point spread function (PSF) of each instrument to the corresponding image, and ran PSF fitting on the fixed prior positions together with other sources from COSMOS2020 that are beyond the $10''$ radius of the group centre. 
We note that nearly all ALMA and NOEMA continuum sources in the groups are spectroscopically confirmed to be group members; hence, the contamination of dusty interlopers is negligible, and the fitting is straightforward.
The PSF fitting is performed for \textit{Herschel} 100-500$\mu$m, SCUBA2 850$\mu$m, and MeerKAT images, and the measured fluxes and their uncertainties are then calibrated by Monte Carlo simulations performed on the maps (see \citealt{Jin2018cosmos} for details). Finally, we checked the residual images and find that a few groups are resolved in the MeerKAT map (beam size $\sim9''$); hence, we adopted aperture photometry for these. 
Due to its relatively high resolution (6$''$), the MIPS 24$\mu$m image cannot be fit with a single PSF. We thus adopted the total weighted 24$\mu$m photometry of individual members in the \citet{Jin2018cosmos} catalogue.
The resultant photometry is summarized in Table~\ref{tab:FIR-photo}.
 
To infer the integrated physical properties of these groups, we fit the integrated FIR spectral energy distribution (SED) using \texttt{STARDUST} \citep{Kokorev2021} with the above FIR photometry, as well as the integrated ALMA and NOEMA continuum fluxes where available. The fitting was performed at the spectroscopic redshift of each group (\cref{tab:obs_table}). Only photometric points with a significant detection ($>3\sigma$) were considered in the fitting, and the rest were treated as $3\sigma$ upper limits \citep{Kokorev2021}.
We note that the radio photometry is not included in the fitting. Instead, we extrapolated a radio component based on the IR luminosity using the stellar mass-dependent IR-radio relation from \cite{Delvecchio2021}, assuming the average stellar mass of spectroscopically confirmed ALMA continuum-detected sources. This allows us to identify potential radio excess by comparing the IR-derived radio model with the measured radio flux.



\subsection{Estimate of dark matter halo mass}
\label{sec:method-dmh}
Based on literature studies \citep[e.g.][]{Wang_T2016cluster,Daddi2021Lya,Daddi2022Lya,Sillassen2022_HPC1001}, we adopted and developed in total six methods for estimating the dark matter halo mass\footnote{We use $M_{\rm h}(n)$ as a shorthand to refer to the n-th method for halo mass estimation.}. Our six methods are split into three main techniques; the first three methods employ the simple SHMR with the peak stellar mass and the total stellar mass, respectively, the fourth method employs overdensity with clustering bias, and the final two use projected stellar mass surface density profile fitting. 
We briefly summarize the six halo mass methods in \cref{tab:mhalo_summary}, and refer the details as follows:

\begin{table*}[ht]
\caption{Halo mass estimate methods in this work.}
    \centering
    \setlength{\tabcolsep}{3pt}
    \renewcommand{\arraystretch}{1.25}
    \begin{tabular}{c c c c c c c}
      \hline\hline
        Method & Input & SHMR & SMF & Overdensity & NFW fit & Assumption\\
        \hline
        \multicolumn{7}{c}{\rm (Stellar mass scaling with SHMR)}\\
        $M_{\rm h}(1)$ & $M_{\rm \ast,BCG}$ & \citealt{Behroozi2013Mhalo} & -- & -- & -- & -- \\
        $M_{\rm h}(2a)$ & $M_{\rm \ast,tot}$ & \citealt{Shuntov2022} & \citealt{Muzzin2013} & -- & -- & Virialization\\
        $M_{\rm h}(2b)$ & $M_{\rm \ast,tot}$ & \citealt{van_der_Burg2014} & \citealt{Muzzin2013} & -- & -- & Virialization\\
        \hline
        \multicolumn{7}{c}{\rm (Galaxy overdensity-based)}\\
        $M_{\rm h}(3)$ & -- & -- & -- & This work & -- & Virialization \\
        \hline
        \multicolumn{7}{c}{\rm ($M_*$ surface density profile fitting)}\\
        $M_{\rm h}(4)$ & $\Sigma_{M_*}(R)$ & -- & -- & -- & This work & NFW profile\\
        $M_{\rm h}(5)$ & $\Sigma_{M_*}(R)$ & -- & -- & -- & This work & NFW profile \& fixed $c^*$\\
        \hline\hline
    \end{tabular}\\
    {Note: The details are described in Sect.~\ref{sec:method-dmh}, and the results are presented on Table~\ref{tab:dark_matter}. $^* c$ is the concentration parameter of the dark matter halo.}
    \label{tab:mhalo_summary}
\end{table*}

(1) We derived a halo mass by scaling the stellar mass of the most massive central group member with the SHMR in \citet{Behroozi2013Mhalo};

(2a) Following the methodology in \citet{Daddi2021Lya,Daddi2022Lya} and \citet{Sillassen2022_HPC1001}, we computed the mass-complete total stellar mass of members within the radius $R<{R_{\rm vir}}$. $R$ selected from mass-complete COSMOS2020 \citep{Weaver2022COSMOS2020}, where the expected contamination from the background is $<10\%$ (e.g. \cref{fig:2d-interlopers}-left). We then extrapolated the mass down to $10^7\,{\rm M_{\odot}}$ assuming the field stellar mass function (SMF) from \citet{Muzzin2013}. We subtracted the contamination of the background from the recovered total mass. This background-corrected total stellar mass was then scaled to the redshift-dependent central and satellite stellar mass ($M_{\rm \ast,cent}+M_{\rm \ast,satellite}$) SHMR in the COSMOS2020 catalogue from \citet{Shuntov2022}. This recovered halo mass was corrected for the measuring stellar mass within a radius smaller than virial, by assuming a NFW density profile \citep*{Navarro1997_NFW_profile}.


(2b) Using the total stellar mass in $M_{\rm h}(2a)$, we derived a halo mass by scaling the total stellar mass with the SHMR of $z\sim1$ clusters from \citet{van_der_Burg2014}.

 (3) We estimated halo mass based on the overdensity of the groups above the field level, following the methodology presented in \citet{Sillassen2022_HPC1001}. A stellar mass cut, selected where the interloper fraction at the virial radius of the group was $<10\%$ (e.g. \cref{fig:2d-interlopers}-right), was applied to the entire catalogue. First, we measured the average number density of the photo-z-selected galaxies ($|z_{\rm phot}-z_{\rm spec,group}|<0.1(1+z_{\rm spec,group})$)  in the mass cut COSMOS2020 Classic \texttt{LePhare} catalogue, and obtained the group core density by measuring the number density within the virial radius centred on the core. 
The depth difference between the field and the group selection was accounted for by assuming a group velocity dispersion $\Delta v=400\,{\rm km/s}$, which is the velocity dispersion of a virialized group with $M_{\rm h}\sim2\times10^{13}\,{\rm M_\odot}$ at $z\sim2.5$ \citep{Ferragamo2021_veldisp}. 
We then calculated the mass of a sphere with comoving virial radius $R_{\rm 200}$, defined as the radius within which the average density is 200 times the critical density of the universe with initial halo mass from $M_{\rm h}(2a)$:\begin{equation}
    M_{\rm h}=\frac{\rho~{\rm \delta}}{b}\frac{4}{3}~\pi(R_{\rm 200}(1+z))^3
    \label{eq:mh4}
.\end{equation}
Here $\delta=(\Sigma_{\rm group}-\Sigma_{\rm field})/\Sigma_{\rm field}$ is the overdensity 
 and $\rho$ is the average matter density at $z$ $\left(\rho=\rho_c\Omega_m,~ \rho_c=\frac{3H(z)^2}{8\pi G}, ~H(z)=H_0\sqrt{\Omega_m(1+z)^3+\Omega_{\Lambda}}\right)$, where $\rho_{c}$ is the critical density of the universe at $z$, $H(z)$ is the Hubble parameter at $z$, $H_0$ is the Hubble parameter at $z=0$,  $\Omega_m$ and $\Omega_\Lambda$ are the fraction of matter and vacuum energy of the total energy in the Universe, respectively, and $b$ is a clustering bias calculated with the \citet{Tinker2010haloBias} formalism. Using the halo mass from $M_{\rm h}(2a)$ as an initial guess on the halo mass, we recalculated the bias in an iterative fashion from the result of this method until convergence.

 (4) Inspired by the stellar mass density profile of the $z=2.5$ X-ray detected cluster CLJ1001 being consistent with a NFW profile \citep{Wang_T2016cluster}, we explored fitting the projected stellar mass density of each group using a NFW dark matter halo profile \citep*{Navarro1997_NFW_profile}. We assumed that the structures are virialized, their stellar mass density profiles can be described by a NFW profile, and the stellar mass profile follows the underlying dark matter profile. 
We first sampled the projected NFW profiles with a range of halo mass and concentration parameter values at a fixed redshift (see \cref{sec:nfw_profiles}). The radial bins are annuli with increasing radius $R$ from $R\sim10$~pkpc out to $\sim2\,{\rm pMpc}$. We calculated a characteristic mass, $M_{0,i}=(\delta M_{{\rm \ast,total},i})^2/M_{{\rm \ast,total,}i}$, and background stellar mass density in each annulus, for the mass-complete $z$-phot selected COSMOS2020 catalogue in 10000 randomly placed annuli. Combining the model and background, we estimated the expected number of galaxies in each radial bin, $N_{{\rm model},i}$. The observed characteristic number in each radial bin was calculated as $N_{{\rm obs,}i}=M_{{\rm \ast,total},i}/M_{0,i}$. We found the best model by maximizing the combined probability of finding $N_{\rm obs}$ when expecting $N_{\rm model}$; $p_{\rm combined}=\Pi_i(p_i)$ where $p_i=p(N_{\rm obs,i}\leq N_{{\rm model,}i})$ when $N_{{\rm model,}i} > N_{{\rm obs,}i}$ and $p_i=p(N_{{\rm obs,}i}\geq N_{{\rm model,}i})$ when $N_{{\rm model,}i} \leq N_{{\rm obs,}i}$ assuming a Poisson distribution. 
The projected stellar mass surface density was calculated within the annuli (i.e. $\Sigma(R)$) centred on the mass and distance to FIR peak weighted barycentre of spectroscopically confirmed member galaxies. 
The best-fit profile yields a total baryonic mass $M_{\rm b,total}$. We adopted a dark matter to baryonic mass ratio of $\Omega_{m}/\Omega_{b}-1\approx5.4$ from PLANCK cosmology \citep{Planchk2020_cosmo_params} to convert $M_{\rm b,total}$ to halo mass $M_{\rm h}(4)$.

(5) Assuming that the stellar mass density profiles largely follow NFW models, we further explored deriving halo mass solely based on the shape of the density profiles, independent of a scaling with stellar mass.
We fit the stellar mass surface density profiles and their background levels using a NFW model with scale radius ($R_{\rm s}$) as a variable. 
Unlike with $M_{\rm h}(4)$, we fixed the concentration parameter, $c,$ of the halos using the predictions from \citet{Ludlow2016_concentrations}, assuming a prior halo mass from the average of $M_{\rm h}(2a)$, $M_{\rm h}(3)$, and $M_{\rm h}(4)$. The best-fit $R_s$ value allowed us to obtain a virial radius $R_{200}=R_{s}c$. This virial radius is directly correlated with the halo mass, and hence we obtained the halo mass using the redshift-dependent $M_{\rm h}-R_{\rm 200}$ relation from \citet{Goerdt2010core}.




\section{Results}
\label{sec:results}
\begin{figure*}[!htbp]
    \centering
    \includegraphics[width=0.85\textwidth]{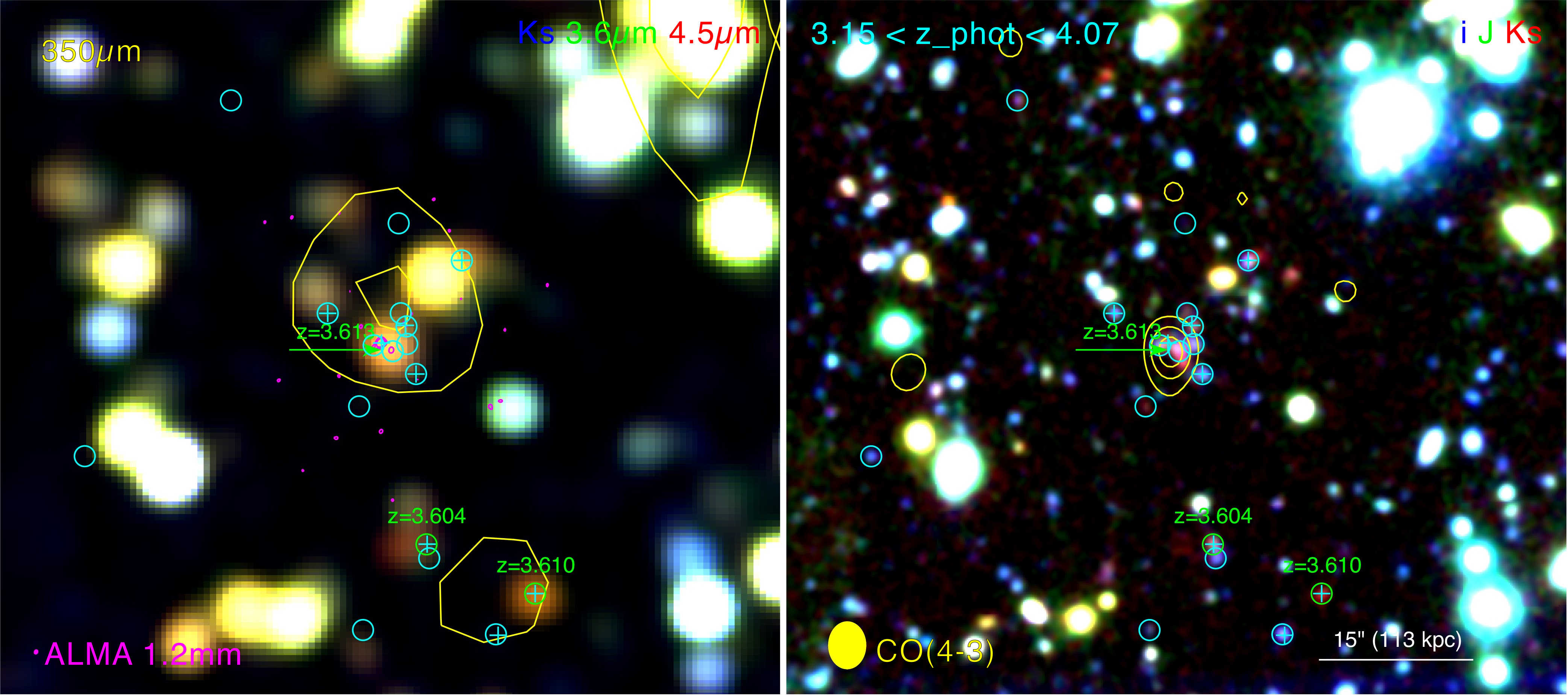}
    \includegraphics[width=0.85\textwidth]{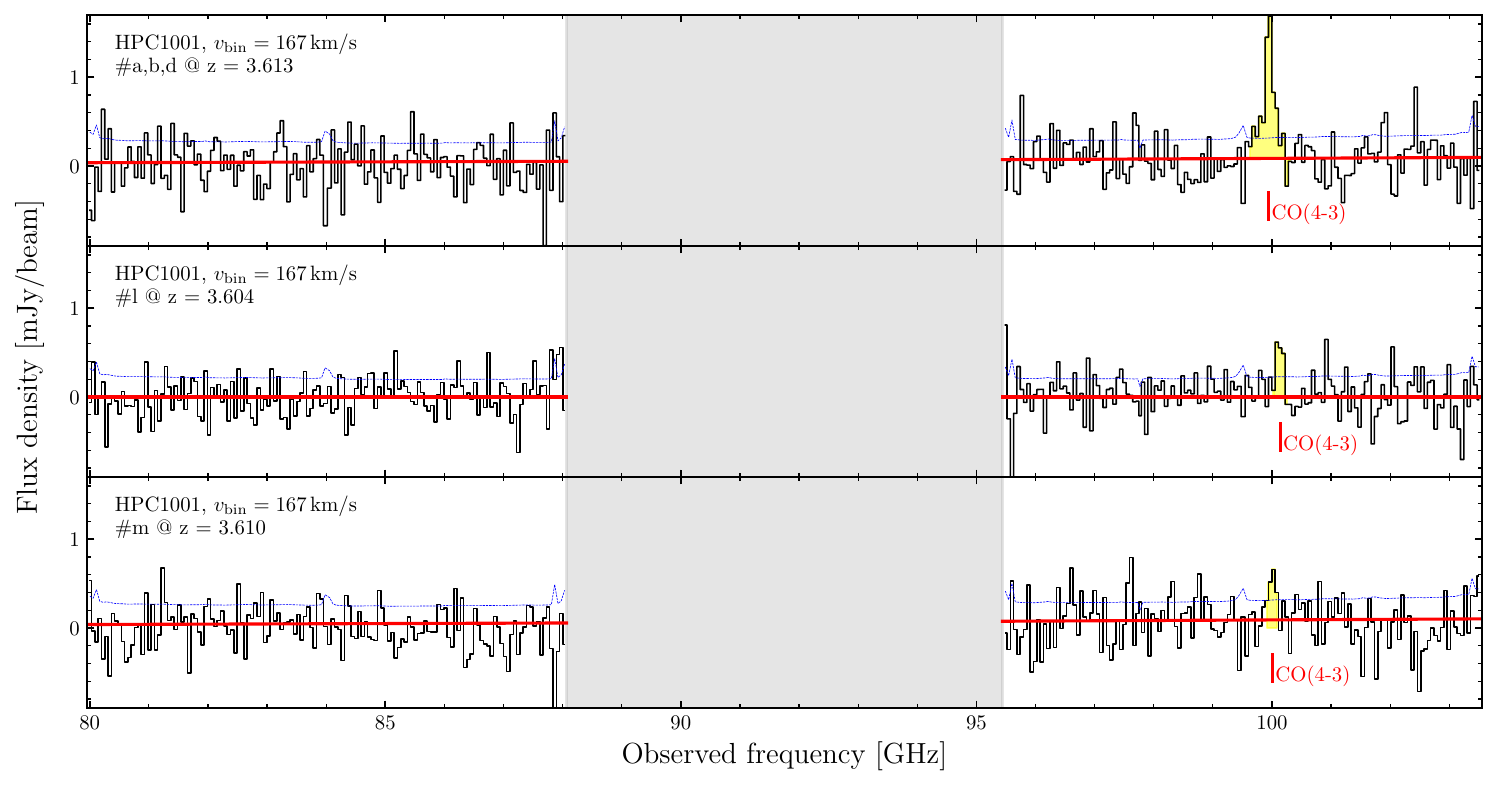}
    \caption{Colour images and NOEMA spectra of HPC1001.\ Photometrically selected galaxies with $3.15<z_{\rm phot}<4.07$ are shown in cyan circles. Spectroscopically confirmed galaxies are shown in green circles, with the spec-z labelled. Cyan plus signs indicate where 1D 3mm spectra have been extracted. {\bf Top left:} Red-green-blue image corresponding to IRAC/4.5 and 3.6$\,\mu$m, and UVISTA/{\it Ks} overlaid with yellow SPIRE/350${\rm \mu m}$ contours at $10,20, \ {\rm and} \ 30\,{\rm mJy}$ levels. Magenta contours show ALMA Band 6 1.2 mm dust continuum at $3,5, \ {\rm and} \ 7\sigma$ levels. {\bf Top right:} Red-green-blue image corresponding to UVISTA/{\rm Ks, J}, and ACS/{\rm i}. The yellow contours show CO(J=4-3) line emission at $3,5, \ {\rm and} \ 7\,\sigma$ levels. {\bf Bottom:} NOEMA Band 1 spectra in the HPC1001 core. The continuum level is marked with a red line, and significant ($P_{\rm chance}<5\%$) emission lines are marked in yellow. The dotted blue line shows the flux uncertainty (1$\sigma$) per channel. The structure name, bin velocity width, ID, and redshift are provided. Corresponding plots for the seven other fields are shown in \cref{fig:SBC3-color-spectra,fig:SBC4-color-spectra,fig:SBC6-color-spectra,fig:SBCX1-color-spectra,fig:SBCX3-color-spectra,fig:SBCX4-color-spectra,fig:SBCX7-color-spectra}.}
    \label{fig:Sillassen-color-spectra}
\end{figure*}

In this section we report the results of our analyses adopting the methodologies presented in \cref{sec:methods}. We produced colour images and spectra for confirmed and candidate members (\cref{fig:Sillassen-color-spectra,fig:SBC3-color-spectra,fig:SBC4-color-spectra,fig:SBC6-color-spectra,fig:SBCX1-color-spectra,fig:SBCX3-color-spectra,fig:SBCX4-color-spectra,fig:SBCX7-color-spectra}) and multi-wavelength cutouts (\cref{fig:SBC3_cutouts,fig:SBC4_cutouts,fig:SBC6_cutouts,fig:SBCX1_cutouts,fig:SBCX3_cutouts,fig:SBCX4_cutouts,fig:SBCX7_cutouts,fig:Sillassen_cutouts}). Coordinates and physical properties of confirmed and candidate members are shown in Tables C1 to C8.

\subsection{Emission lines and redshifts}
We detect robust ($\sigma>7$,$p_{\rm chance}<0.001$) emission lines in seven out of the eight pointings, with no significant lines detected in COS-SBC4. To identify the lines and determine the redshift, we calculated all possible redshift solutions of CO and [CI] lines allowed by the observed frequencies, and compared them with the PDF(z) in the COSMOS2020 catalogue. As shown in Fig.~\ref{fig:pdfz-hpc1001} to \ref{fig:pdfz-sbc6}, the best redshift solution is determined as the one that is closest to the peak of PDF(z). In cases where the group frequency is significantly offset from the peak of the PDF(z), for example ID1272853 in \cref{fig:pdfz-hpc1001}, we reject solutions closer to the peak as one or more bright lines would be covered but none are detected. In total, we identify 22 emission lines of CO(3-2), CO(4-3), CO(5-4), and [CI](1-0) for 20 sources. A summary of the detected lines is presented in \cref{tab:emission-lines}. We detect a single emission line from 18 galaxies, and two sources are detected with two or more lines.
For 58\% (95\%)  of the sources, the best $z_{\rm spec}$ solution lies within the 1$\sigma$ (2$\sigma$) uncertainty of the $z_{\rm phot}$ estimates (\cref{fig:pdfz-hpc1001,fig:pdfz-sbc3,fig:pdfz-sbc6,fig:pdfz-sbcx1,fig:pdfz-sbcx3,fig:pdfz-sbcx4,fig:pdfz-sbcx7}). The mean $z_{\rm spec}$ solution is $-0.57\sigma$ in the PDF(z).

\begin{table*}[!htbp]
    \centering
    \renewcommand{\arraystretch}{1.25}
    \caption{Significant emission lines detected with NICE.}
    \begin{tabular}{c c c c c c c c c c}
    \hline\hline
        Structure & ID & RA & Dec. & $\nu_{\rm obs}$ & Line & $S_{\rm line}$ & Width$_{\rm FWZI}$ & S/N & ${P_{\rm chance}}^{a}$ \\
         & & [deg] & [deg] & [GHz] & & [${\rm Jy\,km/s}$] & [${\rm km/s}$] & & \\ 
         \hline
         HPC1001 & -- & 150.4659 & 2.6362 & 99.957 & CO(4-3) & $1.03\pm0.09$ & $1824\pm167$ & 10.9 & <0.001\\
         HPC1001 & 1272853 & 150.4618 & 2.6294 & 100.004 & CO(4-3) & $0.38\pm0.08$ & $1036\pm266$ & 3.9 & 0.001$^b$ \\
         HPC1001 & 1274387 & 150.4647 & 2.6307 & 100.130 & CO(4-3) & $0.36\pm0.08$ & $1288\pm300$ & 4.3 & 0.019$^b$\\
         \hline
         COS-SBCX3 & 1088787 & 150.3105 & 2.4515 & 85.803 & CO(3-2) & $0.90\pm0.08$ & $571\pm49$ & 11.6 & <0.001 \\
         COS-SBCX3 & 1088927 & 150.3117 & 2.4510 & 85.768 & CO(3-2) & $0.62\pm0.07$ & $428\pm45$ & 9.5 & <0.001 \\
         \hline
         COS-SBCX4 & 1050531 & 150.7504 & 2.4129 & 94.805 & CO(3-2) & $1.71\pm0.06$ & $653\pm27$ & 26.8 & <0.001\\
         COS-SBCX4 & 1049510 & 150.7512 & 2.4124 & 94.947 & CO(3-2) & $0.38\pm0.07$ & $653\pm112$ & 5.8 & <0.001\\
         COS-SBCX4 & 1049929 & 150.7522 & 2.4140 & 94.789 & CO(3-2) & $0.30\pm0.05$ & $392\pm62$ & 6.3 & <0.001\\
         \hline
         COS-SBCX1 & 1408110 & 150.3480 & 2.7611 & 168.415 & CO(5-4) & $0.59\pm0.08$ & $554\pm70$ & 7.9 & <0.001\\
          & " & " & " & 143.797 & [CI](1-0) & $0.43\pm0.06$ & $557\pm82$ & 6.8 & <0.001\\
          & " & " & " & 134.686 & CO(4-3) & $0.80\pm0.13$ & $1145\pm182$ & 6.3 & <0.001\\
          & " & " & " & 101.046 & CO(3-2) & $0.98\pm0.15$ & $859\pm133$ & 6.5 & <0.001\\
         \hline
         COS-SBCX7 & 394609 & 149.9896 & 1.7977 & 101.228 & CO(3-2) & $0.79\pm0.09$ & $675\pm74$ & 9.1 & <0.001\\
         COS-SBCX7 & 394944 & 149.9882 & 1.7980 & 101.221 & CO(3-2) & $0.57\pm0.07$ & $491\pm63$ & 7.8 & <0.001\\
         COS-SBCX7 & 392257 & 149.9910 & 1.7967 & 101.247 & CO(3-2) & $0.54\pm0.09$ & $675\pm113$ & 6.0 & <0.001\\
         COS-SBCX7 & 392639 & 149.9816 & 1.7960 & 101.310 & CO(3-2) & $0.75\pm0.20$ & $491\pm132$ & 3.7 & 0.028$^b$\\
         \hline
         COS-SBC3 & 1345246 & 150.7193 & 2.6998 & 154.812 & CO(5-4) & $1.00\pm0.08$ & $725\pm55.3$ & 13.1 & <0.001\\
         COS-SBC3 & 1340799 & 150.7226 & 2.6963 & 137.021 & CO(4-3) & $0.72\pm0.11$ & $308\pm49$ & 6.4 & <0.001\\
          & " & " & " & 146.305 & [CI](1-0) & $0.62\pm0.15$ & $390\pm95$ & 4.1 & 0.022$^b$\\
         \hline
         COS-SBC6 & 835289 & 149.7053 & 2.2153 & 138.757 & CO(4-3) & $0.94\pm0.10$ & $909\pm98$ & 9.3 & <0.001\\
         COS-SBC6 & 839791 & 149.7053 & 2.2171 & 138.718 & CO(4-3) & $1.05\pm0.12$ & $1044\pm116$ & 9.0 & <0.001\\
         COS-SBC6 & 838104 & 149.7074 & 2.2140 & 168.253 & CO(4-3) & $0.39\pm0.07$ & $305\pm57$ & 5.4 & <0.001\\
         \hline 
    \end{tabular}
    {\\ Notes: $^a$chance probability over the entire spectrum as defined in \citet{Jin2019alma}. $^b$chance probability when limiting the line search to $\pm2000\,{\rm km/s}$ from the brightest line \citep{Zhou2023_LHSBC3_NICE}.}
    \label{tab:emission-lines}
\end{table*}

The $z=3.61$ HPC1001 is the most distant group in this sample, in which we detect three CO lines. Notably, the emission line from the core region has a velocity width of $1824\pm167\,{\rm km/s}$. This large width suggests a blending of multiple sources in the low resolution NOEMA beam ($\sim4''$, \cref{fig:Sillassen-color-spectra}), consistent with the fact that three ALMA 1.2~mm continuum sources are detected (at 0.4'' resolution) within the NOEMA beam.
As the photometric redshifts of the two i, J, and Ks-detected galaxies in the core are consistent with the CO redshift (see \cref{fig:pdfz-hpc1001}), we suspect the three galaxies to be at the same redshift $z=3.613$. Similarly, \citet{Gomez-Guijarro2019} studied \textit{Herschel} selected sources blended at similar scales, and found all blended candidate member galaxies at the same redshift, if this is also true for HPC1001, this would result in a total of five $z_{\rm spec}$ confirmed members. Our ongoing high-resolution ALMA observations (ID: 2023.1.00652.S, PI: N. Sillassen) will reveal the members in the compact core.
Interestingly, we spectroscopically confirm the nearby DSFG (HPC1001.m) at $z=3.610$ (i.e. at a velocity difference of just $-141\,{\rm km/s}$ from the core). At a projected distance of $\sim204\,{\rm kpc}$, the nearby DSFG is outside the expected virial radius of the dark matter halo (\cref{tab:dark_matter}), indicating that the DSFG HPC1001.m is in a larger structure at the same redshift.

In the other groups, we can separate individual members without blending issues. We confirm two members in COS-SBCX3, three members in COS-SBCX4, four members in COS-SBCX7, one member in COS-SBC3, and two members in COS-SBC6. In the pointing COS-SBCX1, we confirmed a member with four emission lines CO(3-2), CO(4-3), [CI](1-0), and CO(5-4) all with S/N$>$6. Though no line was detected in COS-SBC4 with ALMA, we confirm three members with Subaru/FMOS \citep{Kashino2019_FMOS_COSMOS}. All emission lines at the redshift of COS-SBC4, $z=1.65$, would fall out of the frequency coverage of our ALMA observations.

\subsection{Integrated FIR properties}\label{sec:fir-results}
\begin{figure*}[!htbp]
    \centering
    \includegraphics[width=0.49\textwidth]{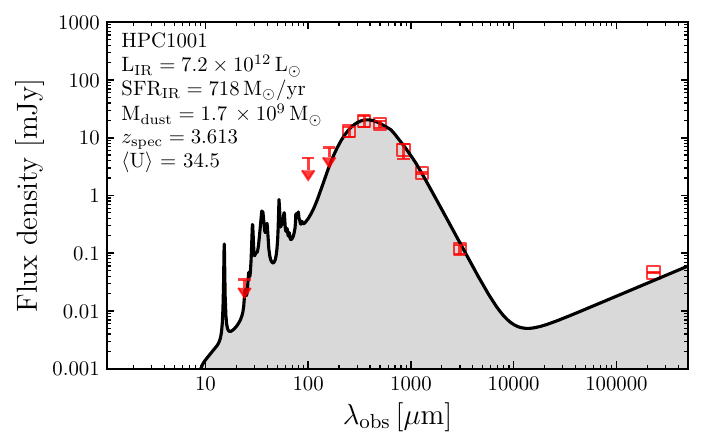}
    \includegraphics[width=0.49\textwidth]{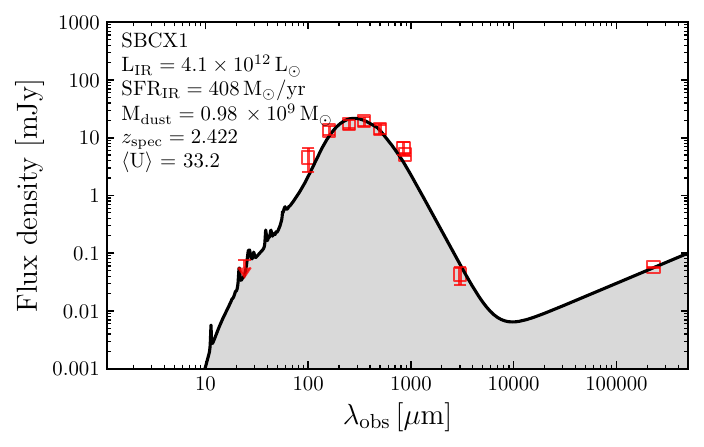}
    \includegraphics[width=0.49\textwidth]{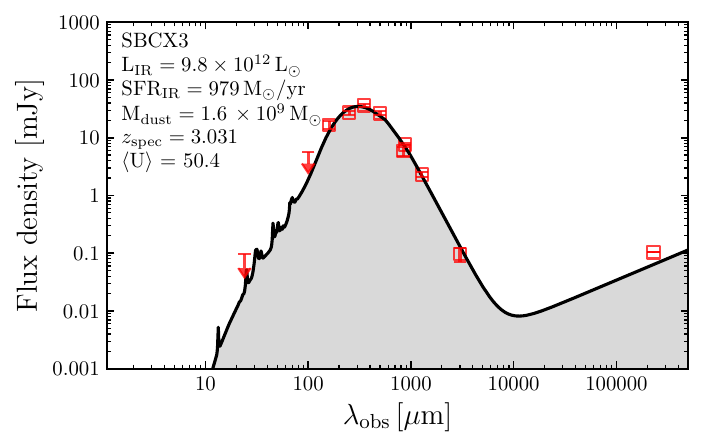}
    \includegraphics[width=0.49\textwidth]{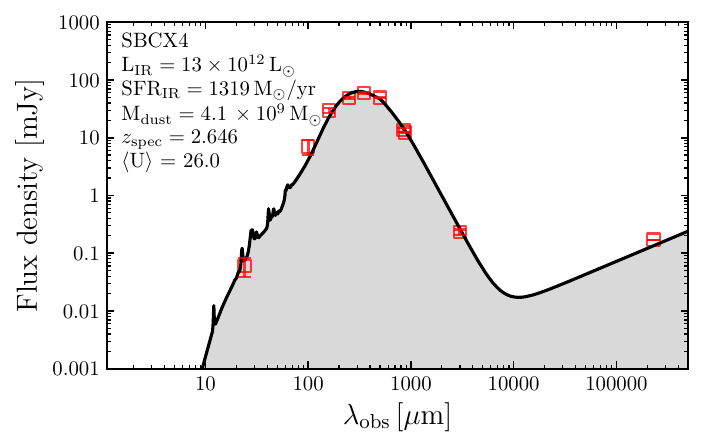}
    \includegraphics[width=0.49\textwidth]{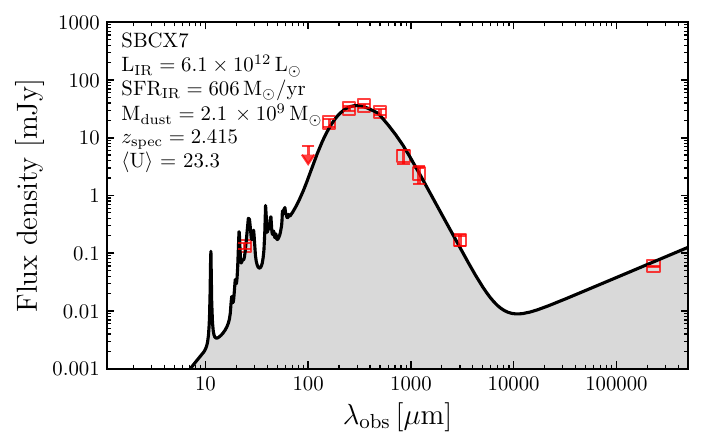}
    \includegraphics[width=0.49\textwidth]{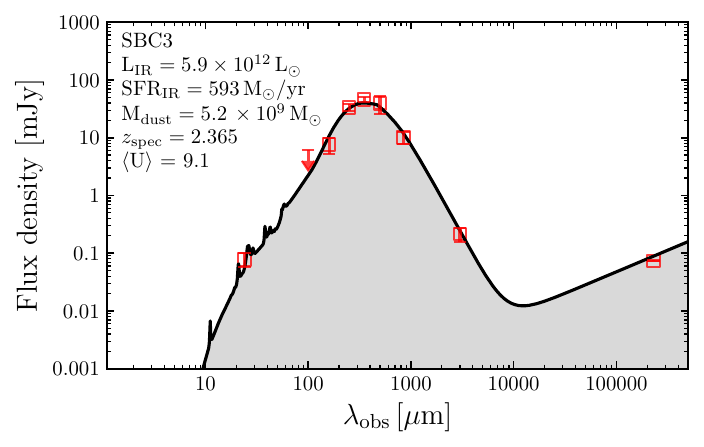}
    \includegraphics[width=0.49\textwidth]{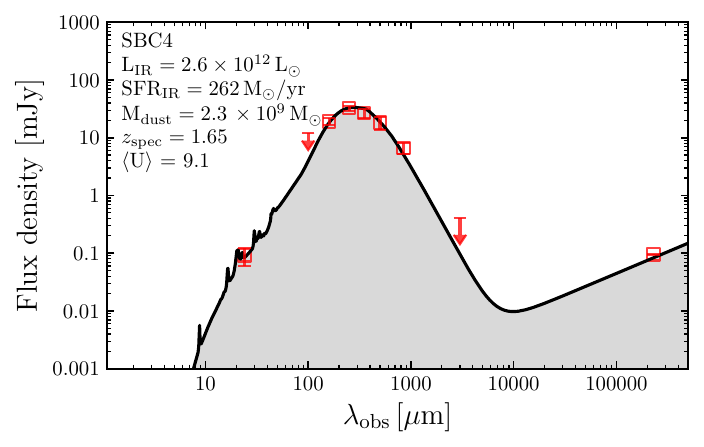}
    \includegraphics[width=0.49\textwidth]{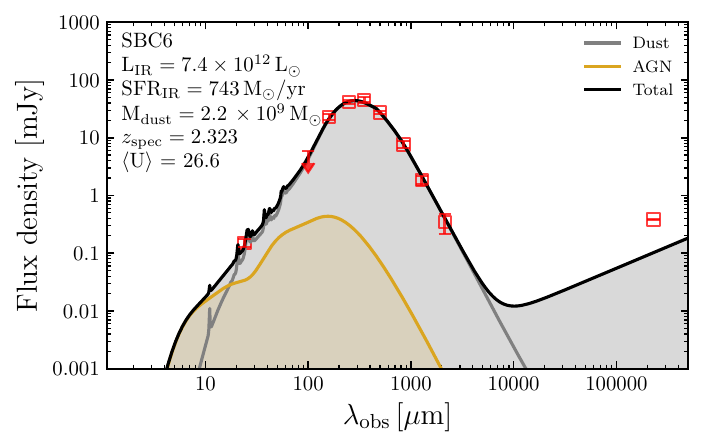}
    \caption{Integrated FIR SEDs of our sample of high-$z$ structures, which are fitted with dust models (grey curve; \citealt{Magdis2012SED}) and, in the case of COS-SBC6, a mid-IR AGN template (yellow; \citealt{Mullaney2011}) (see \cref{sec:fir-results}). We show the best-fit SFR, dust mass, and radiation field $\langle {\rm U}\rangle$ in each panel. 
    We note that the radio photometry is not included in the fitting; instead we extrapolated a radio component from IR luminosity using the IR-radio relation from \citet{Delvecchio2021}.}
    \label{fig:fir-seds}
\end{figure*}

In Fig.~\ref{fig:fir-seds} we present the best-fit of the integrated FIR photometry with \texttt{STARDUST} based on dust templates from \cite{Magdis2012SED}. 
We obtain FIR properties of the integrated dust emission of the groups, including total IR luminosity ($L_{\rm IR}$),  SFR, dust mass ($M_{\rm dust}$) and mean radiation field ($\langle U \rangle$). 
As summarized in Table~\ref{tab:BAR}, we obtain integrated SFRs in the range ${\rm SFR_{IR}}=(262-1319)\,{\rm M_{\odot}/yr}$, dust masses in the range ${\rm M_{\rm dust}}=(1.0-5.2)\times10^9\,{\rm M_{\odot}}$ (optically thin model), and dust-weighted
mean starlight intensity scale factors in the range $\langle {\rm U}\rangle=(9-50)$ (\cref{fig:fir-seds}).

Remarkably, seven of the eight FIR SEDs are well fit by pure dust emission, that is, they need no mid-IR or radio active galactic nucleus (AGN) contribution to fit the data. As shown in Fig.~\ref{fig:fir-seds}, they closely follow the IR-radio correlation with a dispersion of 0.11 dex (inter quartile range), indicating that they are powered predominantly by star formation. However, COS-SBC6 exhibits clear radio excess that suggests the presence of a radio-loud AGN. Adding a mid-IR AGN component to the fit of COS-SBC6 yields a mid-IR AGN contribution of $\sim10\%$ to the total IR luminosity. We recall that the radio part of the model is not part of the fit but extrapolated using the \citet{Delvecchio2021} IR-radio correlation. By including the radio excess found in the $z=3.95$ LH-SBC3 presented in \citet{Zhou2023_LHSBC3_NICE}, this gives a radio-loud AGN fraction of 2/9 (i.e. 22\%) in the NICE sample of massive galaxy groups at $1.6<z<4$. This low fraction of radio-loud AGNs implies that this massive group population is primarily in star formation mode, and they are not significantly affected by AGN activity. 


\newpage

\subsection{Dark matter halos}
\begin{figure}[!htbp]
    \centering
    \includegraphics[width=\columnwidth]{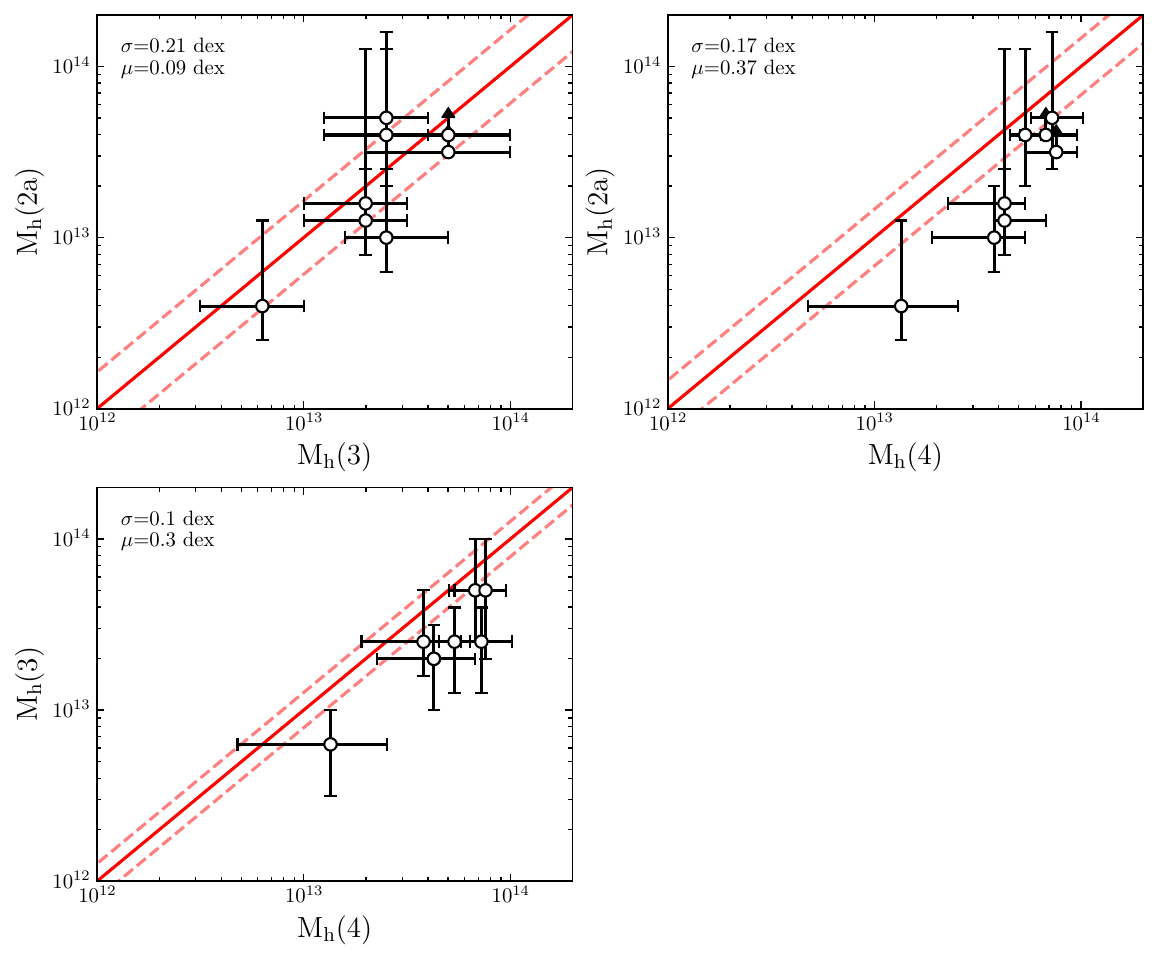}
    \caption{Comparison between dark matter halo estimate methods $M_{\rm h}(2)$, $M_{\rm h}(4)$, $M_{\rm h}(5)$, and $M_{\rm h}(6)$. The standard deviation, $\sigma,$ and the mean offset from the 1:1 relation, $\mu,$ are provided in each panel. The solid and dashed lines correspond to the 1:1 relation and 1$\sigma$, respectively.}
    \label{fig:Mh-comp}
\end{figure}

Using the methods described in \cref{sec:method-dmh}, we estimated halo masses of $\log(M_{\rm h}/{\rm M_\odot})=12.8-13.7$ as listed in \cref{tab:dark_matter}.
In \cref{fig:Mh-comp} we compare $M_{\rm h}(2a)$, $M_{\rm h}(3)$, and $M_{\rm h}(4),$ which represent the three main techniques of stellar mass scaling, projected galaxy overdensity, and NFW profile fitting, respectively. 
We calculated the standard deviation and mean offset from a 1:1 relation between each two methods (\cref{fig:Mh-comp}). We find that the results from different methods agree within 0.2--0.3 dex with systematic offsets $\lesssim 0.4$ dex, which are consistent with the expected uncertainty of lower-mass halos ($M_{\rm h}<4\times10^{13}\,{\rm M_\odot}$)  at high redshifts \citep[$z>2$;][]{Looser2021,Daddi2021Lya,Daddi2022Lya}. The mean offset is in all cases $\lesssim2\sigma$, and thus with this limited sample size we are unable to detect significant systematic offsets. We adopted an average of methods $M_{\rm h}(2a)$, $M_{\rm h}(3)$, and $M_{\rm h}(4)$ as the median of the best estimate, and the uncertainty is conservatively adopted as 0.3 dex, based on the scatter between methods (see \cref{fig:Mh-comp}).
\label{sec:dark matter-halo-results}

\begin{figure*}[!htbp]
    \centering
    \includegraphics[width=0.85\columnwidth]{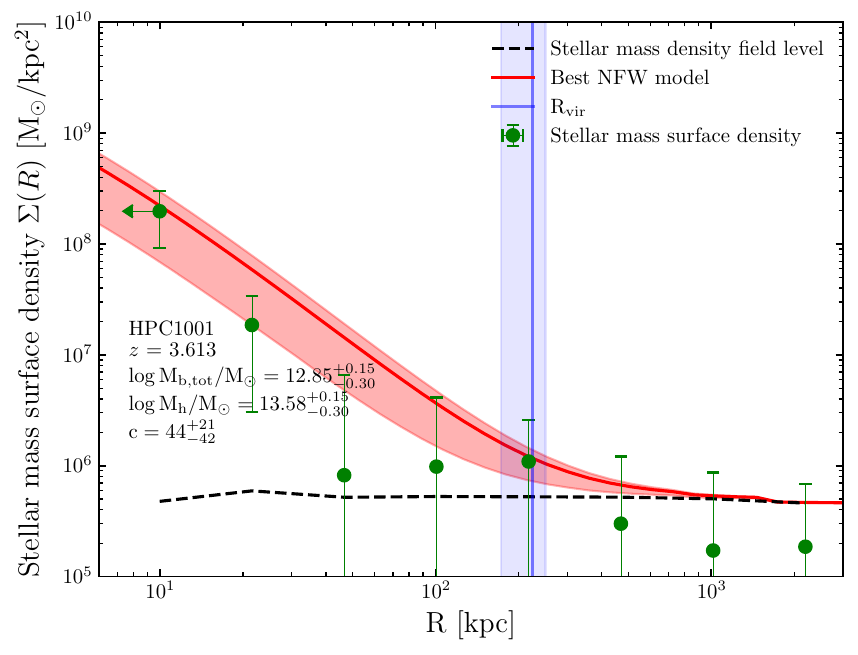}
    \includegraphics[width=0.85\columnwidth]{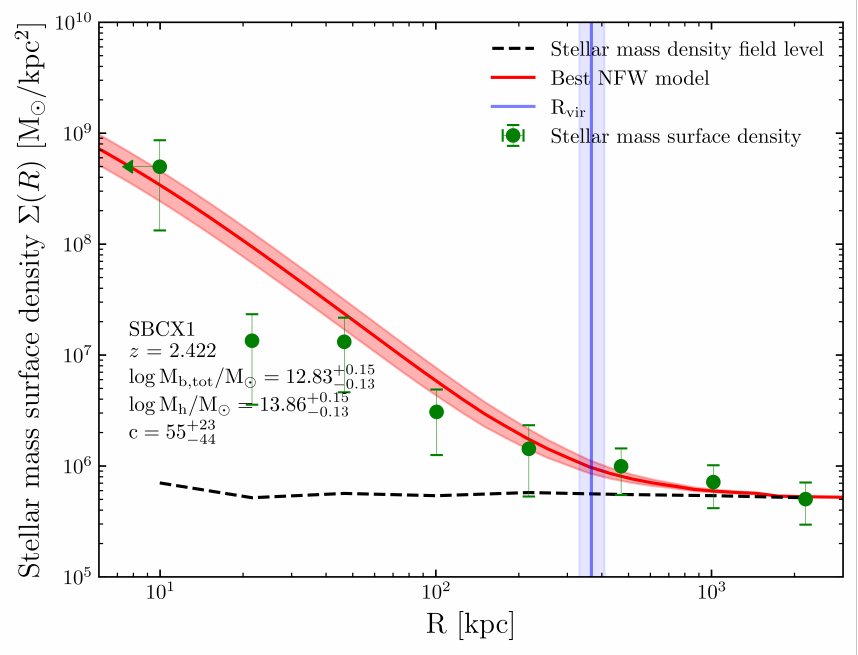}
    \includegraphics[width=0.85\columnwidth]{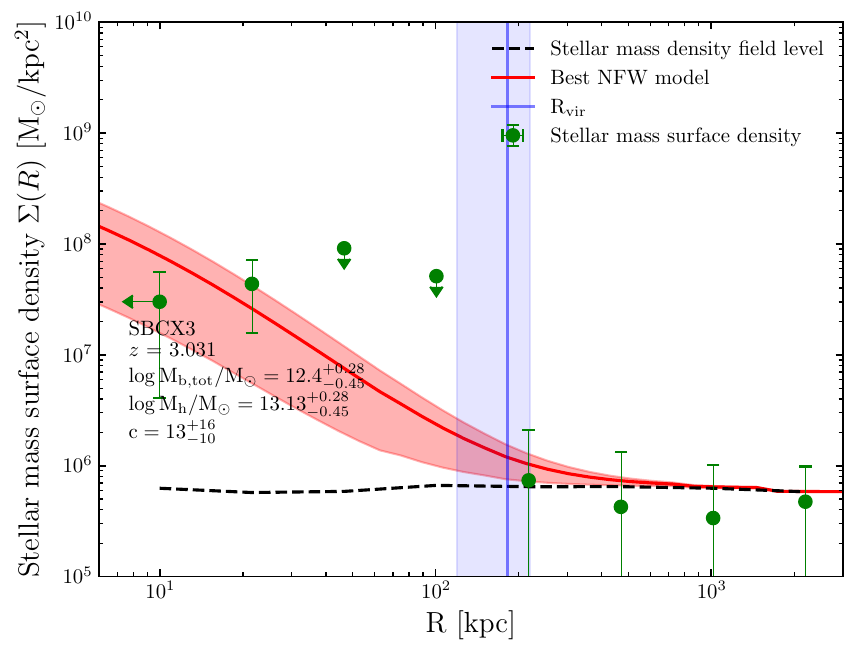}
    \includegraphics[width=0.85\columnwidth]{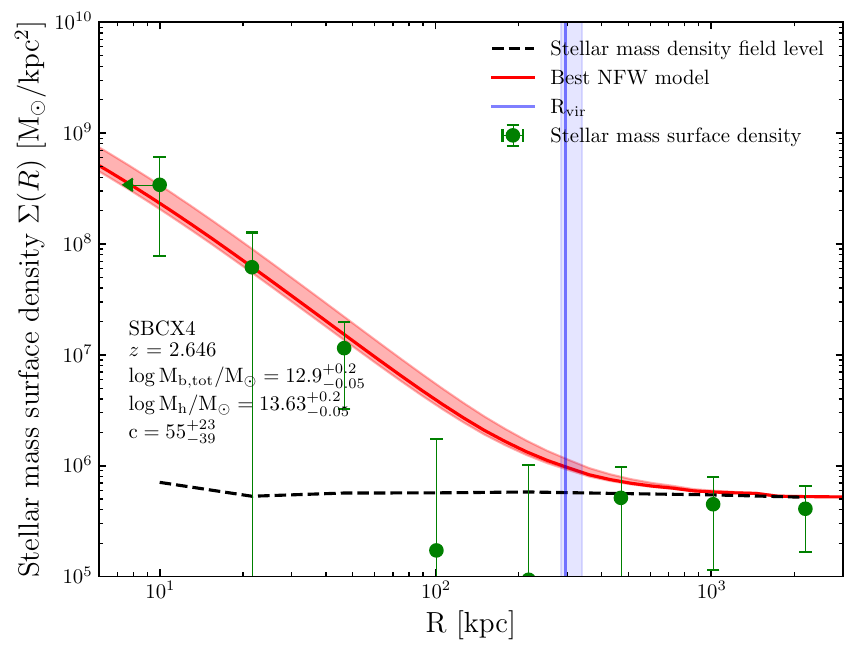}
    \includegraphics[width=0.85\columnwidth]{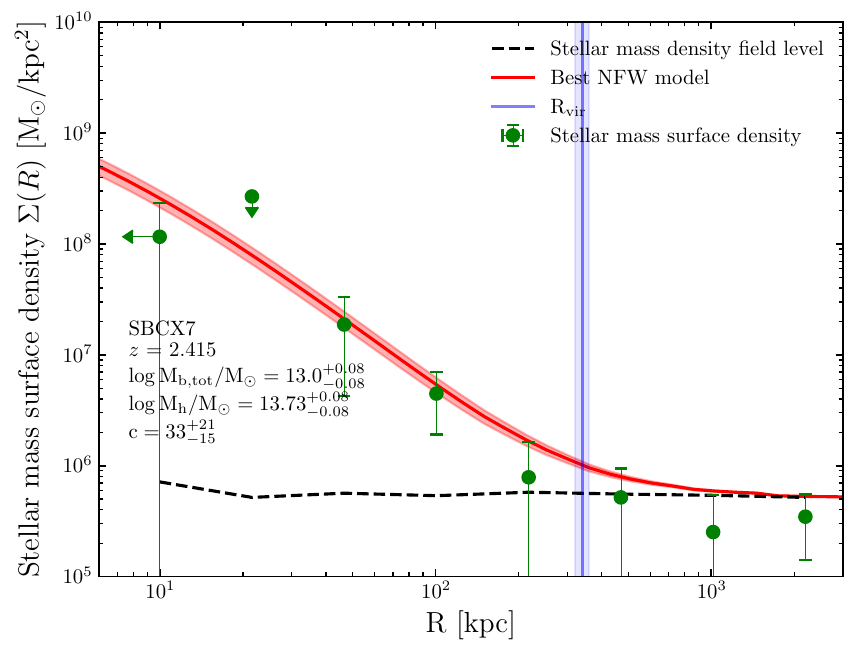}
    \includegraphics[width=0.85\columnwidth]{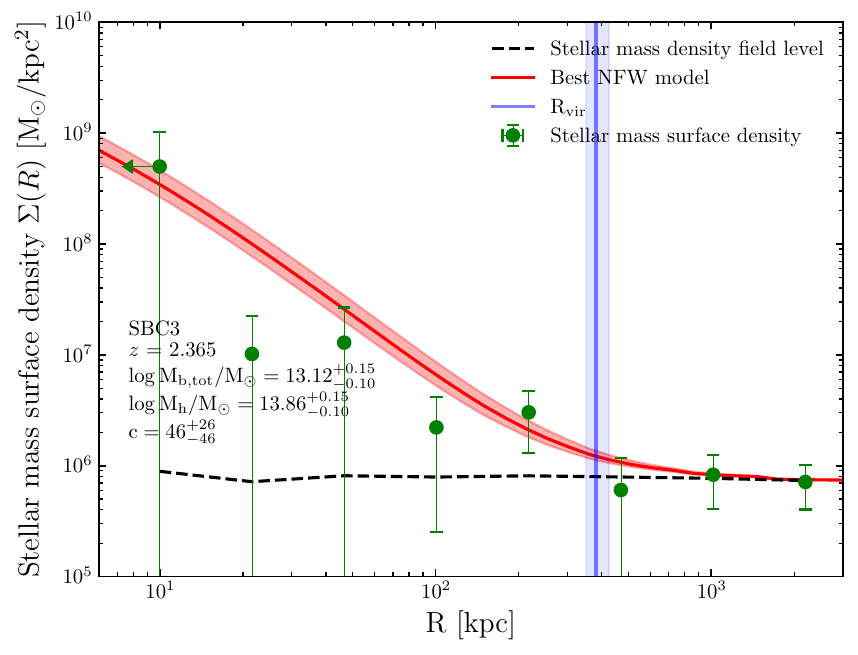}
    \includegraphics[width=0.85\columnwidth]{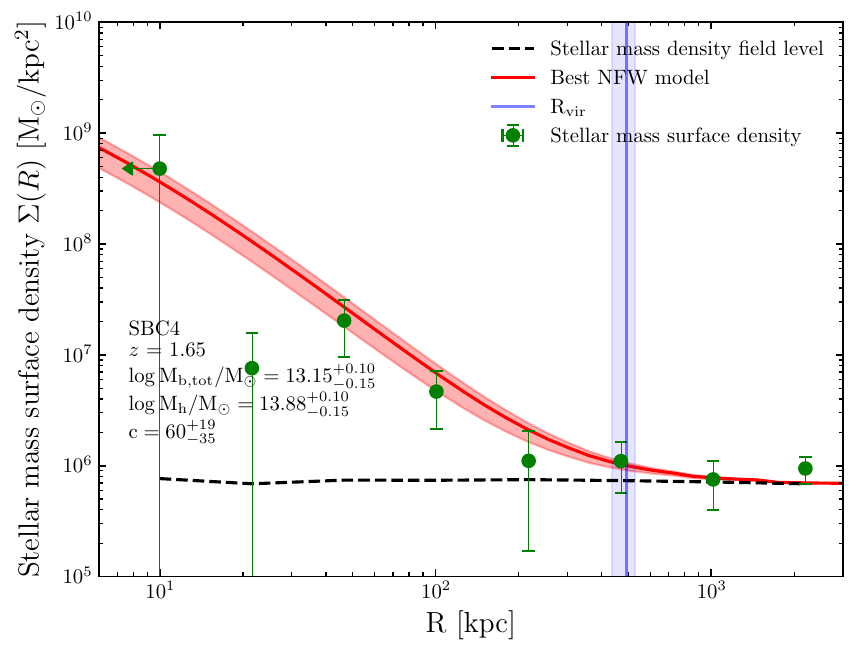}
    \includegraphics[width=0.85\columnwidth]{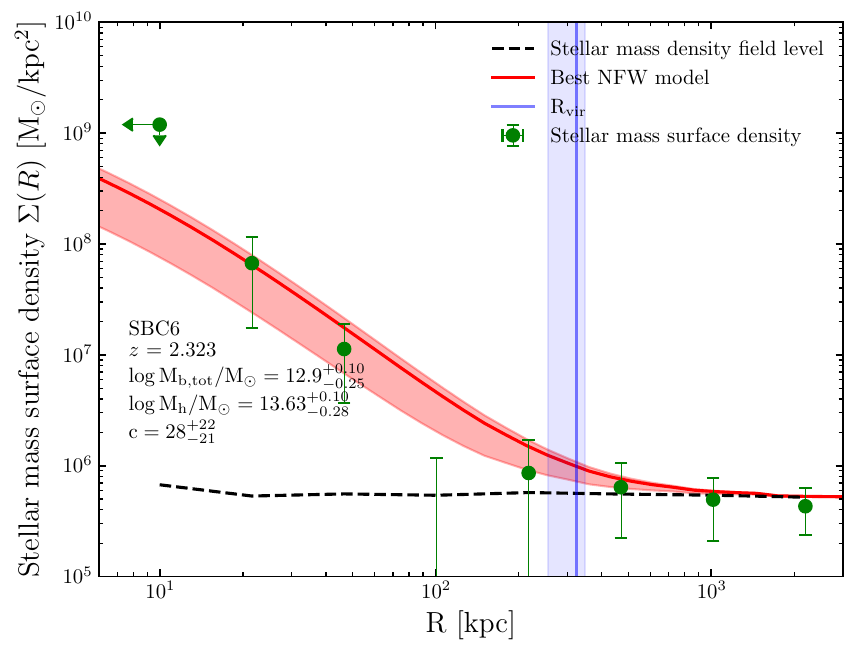}
    \caption{Projected density profiles of the eight structures. We fit a NFW profile to the radial stellar mass density. 
    Green circles mark the projected stellar mass density, with the field level shown as a dashed black line. The red curve shows the best-fit NFW profile with uncertainty as the red shaded area, and the blue line indicates the derived virial radius \citep{Goerdt2010core} with uncertainty as the blue shaded area.
    The name of the structure, $z_{\rm spec}$, fitted halo mass, and fitted concentration parameters are indicated in each panel. Upper limits are at $3\sigma$ significances.}
    \label{fig:NFW-profiles}
\end{figure*}

\begin{table*}[!htbp]
    \centering
    \setlength{\tabcolsep}{2.5pt}
    \renewcommand{\arraystretch}{1.3}
    \caption{Dark matter halo mass results for all groups.}
    \begin{tabular}{c c c c c c c c c c c c}
    \hline\hline
        Structure & $\sigma_{\rm OD}$$^{a}$ & $M_{\rm \ast, max}$$^b$ & Total $M_{\ast}$  & $M_{\rm h}(2a)$  & $M_{\rm h}(3)$ & $M_{\rm h}$(4) & $M_{\rm h}$ Best$^c$ & $R_{\rm vir}$$^d$ & $M_{\rm h}(1)$ & $M_{\rm h}(2b)$ & $M_{\rm h}$(5)\\
        & & [$10^{11}{\rm M_\odot}$] & [$10^{11}{\rm M_\odot}$] & ${\rm log\,M_\odot}$ & ${\rm log\,M_\odot}$ & ${\rm log\,M_\odot}$  & ${\rm log\,M_\odot}$ & [${\rm pkpc}$] & ${\rm log\,M_\odot}$  & ${\rm log\,M_\odot}$ & ${\rm log\,M_\odot}$ \\
            \hline
        HPC1001 & 7.7 & $1.0\pm0.2$ & $2.4\pm0.2$ & $13.0^{+0.3}_{-0.2}$ & $13.4^{+0.3}_{-0.2}$ & $13.6^{+0.2}_{-0.3}$ & $13.3\pm0.3$ & $181\pm42$ & $>12.8$ &  $12.8\pm0.3$ & $12.9^{+0.9}_{-1}$\\
        COS-SBCX3 & 6.0 & $0.5\pm0.1$ & $1\pm0.1$ & $12.6^{+0.5}_{-0.2}$ & $12.8^{+0.2}_{-0.3}$ & $13.1^{+0.3}_{-0.5}$  & $12.8\pm0.3$ & $141\pm33$ & $12.3^{+0.4}_{-0.1}$ & $12.2\pm0.4$ & $12.8^{+2.4}_{-1.3}$\\
        COS-SBCX4 & 6.0 & $1.6\pm0.4$ & $4.4\pm0.6$ & $13.1^{+0.3}_{-0.2}$ & $13.3^{+0.2}_{-0.3}$ & $13.6^{+0.2}_{-0.1}$  & $13.3\pm0.3$ & $229\pm53$ & $>13.1$ & $13.2\pm0.3$ & $13.2^{+1.3}_{-1}$\\
        COS-SBCX1 & 7.0 & $2.5\pm0.6$ & $8.6\pm0.6$ & $>13.6$ & $13.7^{+0.3}_{-0.3}$ & $13.9^{+0.1}_{-0.2}$  & $13.7\pm0.3$ & $331\pm70$ & $>13.8$ & $13.7\pm0.2$ & $13.4^{+2.6}_{-1.3}$\\
        COS-SBCX7 & 5.6 & $1.3\pm0.3$ & $4.5\pm0.6$ & $13.6^{+0.5}_{-0.3}$ & $13.4^{+0.2}_{-0.3}$ & $13.7^{+0.1}_{-0.1}$ & $13.6\pm0.3$ & $308\pm71$ & $13.5^{+1.4}_{-0.4}$ & $13.3\pm0.2$ & $13.3^{+0.8}_{-0.8}$\\
        COS-SBC3 & 4.6 & $3.2\pm0.7$ & $5.2\pm0.7$ & $13.7^{+0.5}_{-0.3}$ & $13.4^{+0.2}_{-0.3}$ & $13.9^{+0.2}_{-0.1}$ & $13.7\pm0.3$ & $337\pm71$ & $>14.1$ & $13.3\pm0.2$ & $13.3^{+1.7}_{-1.3}$\\
        COS-SBC6 & 4.6 & $1.3\pm0.3$ & $2.5\pm0.2$ & $13.2^{+0.9}_{-0.3}$ & $13.3^{+0.2}_{-0.3}$ & $13.6^{+0.1}_{-0.3}$ & $13.4\pm0.3$ & $271\pm57$ & $13.5^{+1.4}_{-0.4}$ & $12.9\pm0.3$ & $13.0^{+1.7}_{-1.3}$\\
        COS-SBC4 & 7.4 & $1.6\pm0.6$ & $7.9\pm0.4$ & $>13.5$ & $13.7^{+0.3}_{-0.4}$ & $13.9^{+0.1}_{-0.2}$ & $13.7\pm0.3$ & $429\pm90$ & $>13.0$ & $13.8\pm0.2$ & $13.4^{+0.7}_{-0.6}$\\
    \hline
    \end{tabular}
    {\\Notes: Lower limits are at $1\sigma$ significance. $^a$Peak significance of overdensity. $^b$$M_{\rm \ast,max}$ is the stellar mass of the most massive central spectroscopically confirmed member galaxy. $^c$The best estimate halo mass is the average of $M_{\rm h}(2a)$, $M_{\rm h}(3)$, and $M_{\rm h}(4)$, and the uncertainty is estimated by the average scatter between the methods (\cref{fig:Mh-comp}). $^d$Virial radius of $M_{\rm h}$ Best, using the $M_{\rm h}-R_{\rm vir}$ relation from \citet{Goerdt2010core}.}
    \label{tab:dark_matter}
\end{table*}


Uniquely, as shown in \cref{fig:NFW-profiles}, the best fit of $M_{\rm h}(4)$ shows that all stellar mass density profiles can be fitted by a NFW profile. This suggests that these structures are likely already collapsed and hosted by a single dark matter halo.
Furthermore, $M_{\rm h}(4)$ allows us to constrain the concentration parameter $c$ of the dark matter halos (see \cref{tab:concentration}). In \cref{fig:concentration} we compare the measured concentrations with the predicted values $c(M, z)$ from simulations \citep{Ludlow2016_concentrations}.
The comparison shows a standard deviation from the 1:1 relation of 0.28 dex, and our measured mean concentration is 0.4 dex higher than the predictions from simulations. 

\subsection{Baryonic accretion rate}
Following the procedure in \citet{Daddi2022Lya} and using Eq. (5) in \citet{Goerdt2010core}, we estimated the total BAR: 
\begin{equation}
    {\rm BAR}\simeq 137\left(\frac{M_{\rm h}}{10^{12}{\rm M_\odot}}\right)^{1.15}\left(\frac{1+z}{1+3}\right)^{2.25}\,{\rm M_\odot\,yr^{-1}}
    \label{eq:bar}
.\end{equation}

Based on the best estimate of halo masses (\cref{tab:dark_matter}), we calculated the BAR using \cref{eq:bar}, yielding ${\rm BAR}=(1200-8400)\,{\rm M_{\odot}/yr}$ (\cref{tab:BAR}). 
The SFR arising from cold accretion can be generalized as follows:
\begin{equation}
    {\rm SFR}=C_{\rm SFR}\times\left\{\begin{array}{l l}
       \left(\frac{M_{\rm stream}}{M_{\rm h}}\right)^{\alpha_{\rm SFR}}{\rm BAR} & M_{\rm h} > M_{\rm stream}  \\
       {\rm BAR} & M_{\rm h} \lesssim M_{\rm stream}
       \end{array}\right.
    \label{eq:SFR_BAR}
,\end{equation}
where $M_{\rm stream}$ is the theoretical upper limit halo mass, where cold streams efficiently occur \citep{Dekel2006,Daddi2022Lya}.

Using the BAR and the fitted values of $\log C_{\rm SFR}=-0.54\pm0.23$ and $\alpha=0.78\pm0.28$ from \citet{Daddi2022Lya} with \cref{eq:SFR_BAR}, the expected SFR derived by the BAR are ${\rm SFR}=(100-1700)\,{\rm M_{\odot}/yr}$ (\cref{tab:BAR}).
In six of the eight groups, the BAR-derived SFRs are consistent with our measured IR SFRs within the uncertainties, only the least massive COS-SBCX3 and the lowest redshift COS-SBC4 groups have a higher SFR than expected from the BAR. Although tentative, this rough consistency supports the baryon accretion models.

\begin{figure*}[!htbp]
    \centering
    \includegraphics[width=0.95\columnwidth]{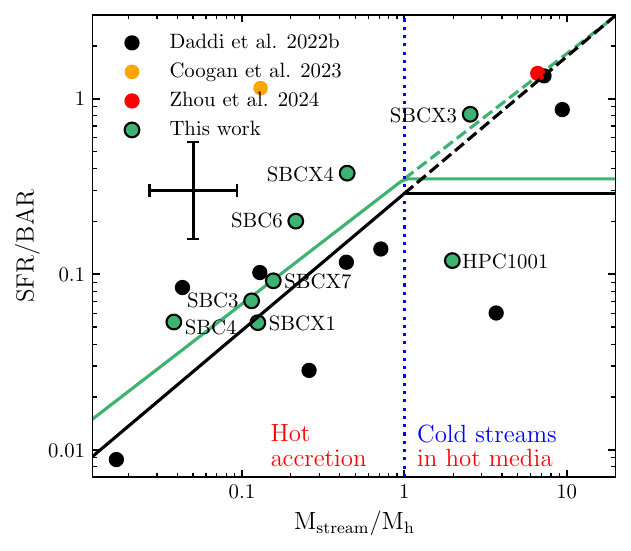}
    \includegraphics[width=0.95\columnwidth]{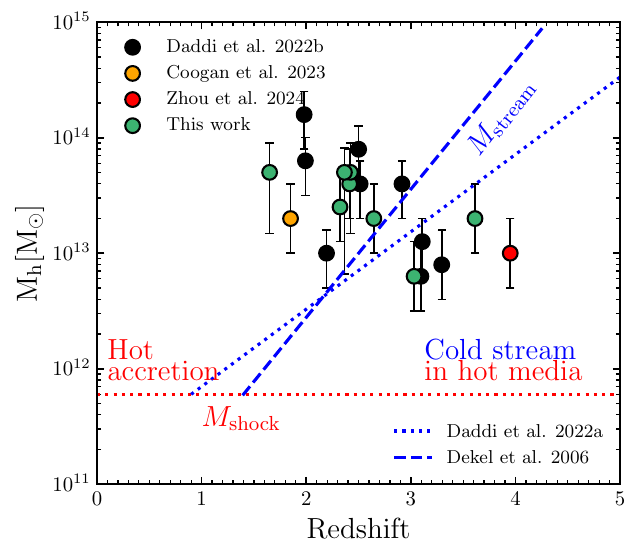}
    \caption{Baryonic accretion in massive galaxy groups and (proto-)clusters. {\bf Left:} SFR$_{\rm IR}$/BAR fraction versus ${\rm M_{stream}/M_{\rm h}}$ ratio for massive galaxy groups in this work (green dots), \citet[black dots]{Daddi2022Lya}, \citet[red dot]{Zhou2023_LHSBC3_NICE}, and \citet[orange dot]{Coogan2023_EGS_Group}. Black lines mark the fit from \citet{Daddi2022Lya}. We fit the SFR/BAR$-{\rm M_{stream}/M_{\rm h}}$ relation with all data points and show the best-fit as a green line. The black cross in the top-left corner shows representative errors. The solid lines show a model with bending at $M_{\rm stream}>M_{\rm h}$, and the dashed lines show a model with a single trend. The dotted blue line shows the separation between the hot accretion and cold stream regimes.
    {\bf Right:} Dark matter halo mass versus redshift for this sample and the samples from \citet{Daddi2022Lya}, \citet{Zhou2023_LHSBC3_NICE}, and \citet{Coogan2023_EGS_Group}, overlaid with dotted and dashed lines separating the three regimes of gas inflow from \citet{Dekel2006}, \citet{Mandelker2020_cold_streams}, and \citet{Daddi2022_mstream}.}
    \label{fig:Mdm_Mstream_znice}
\end{figure*}

In \cref{fig:Mdm_Mstream_znice} we show this sample on the diagrams of gas accretion from \cite{Daddi2022Lya} and \cite{Dekel2013}.
Interestingly, the two most distant groups, HPC1001 and COS-SBCX3, occupy the regime of cold streams in hot media, suggesting cold gas inflow. As HPC1001 has a similar halo mass to the group RO-1001 \citep{Daddi2021Lya}, the cold streams should be detectable via Ly$\alpha$ emission. Given that the dark matter halo estimates are prone to large uncertainties, future observations of Ly$\alpha$ will be ideal to robustly constrain the BAR and reveal possible cold gas accretion.

As shown in \cref{fig:Mdm_Mstream_znice}, our sample enlarges the sample size of massive groups in \citet{Daddi2022Lya} by a factor of two.
By including the sample in \citet{Daddi2022Lya} and \citet{Zhou2023_LHSBC3_NICE}, we fit a quasi-linear model to the SFR/BAR-${\rm M_{stream}/M_{\rm h}}$ relation, where the theoretical upper limit halo mass of cold gas streams ${\rm M_{stream}}$ is adopted from Eq. (2) in \citet{Daddi2022_mstream}. The best fit gives $\alpha_{\rm SFR}=0.71\pm0.16$, $\log C_{\rm SFR}=-0.46\pm0.22$ (\cref{fig:Mdm_Mstream_znice}) with a scatter of $0.38\,{\rm dex}$ assuming a linear relation, and $0.40\,{\rm dex}$ assuming flattening at ${\rm M_{stream}/M_{\rm h}}>1$. As discussed in \citet{Coogan2023_EGS_Group}, this group in the CEERS field is a strong outlier, and we therefore excluded it from the fitting. Our fit result is in excellent agreement with the result from \citet{Daddi2022Lya}, and is improved with $2\times$ better statistics. Adopting instead the definition of $M_{\rm stream}$ from \citep{Dekel2006}, does not significantly change the fitting results, but slightly increases the scatter to 0.43 dex assuming a single relation.

In the hot-accretion regime, where the estimate of $M_{\rm DM}$ is less important \citep{Daddi2022Lya}, we find a 0.30 dex scatter from the relation. In the cold-stream regime, we find a scatter of $0.53$ dex and mean offset of $0.12$ dex from the flattening model, and scatter of $0.45$ dex and mean offset of $-0.35$ dex from the linear relation. Adopting the linear relation would result in two structures with a $\sim1.1$ dex deviation from the model, where the bending model results in one $\sim0.8$ dex outlier. 


\begin{table*}[!htbp]
    \centering
    \caption{Baryonic accretion rate (BAR) and integrated FIR properties (${\rm L}_{\rm IR}$,${\rm SFR}_{\rm IR}$,$\langle {\rm U}\rangle$, and dust mass $M_{\rm dust}$).}
    \renewcommand{\arraystretch}{1.25}
    \begin{tabular}{c c c c c c c c c}
    \hline \hline
        Name & $z$ & $M_{\rm h}$ Best & BAR & SFR$_{\rm BAR}$ & ${\rm L_{\rm IR}}$ & ${\rm SFR_{IR}}$ & $\langle {\rm U}\rangle$ & ${\rm M_{dust}}$\\
         & & ${\rm log~M_\odot}$ & [${\rm M_\odot\,yr^{-1}}$] & [${\rm M_\odot\,yr^{-1}}$] & [${\rm10^{12}L_\odot}$] & [${\rm M_\odot\,yr^{-1}}$] & & [$10^9{\rm M_\odot}$]\\
         \hline 
         HPC1001 & 3.613 & $13.3\pm0.3$ & $6000\pm5000$ & $1700\pm1600$ & $7.18\pm0.04$ & $718\pm4$ & $35\pm1$ & $1.7\pm0.2$\\
         COS-SBCX3 & 3.031 & $12.8\pm0.3$ & $1200\pm900$ & $300\pm300$ & $9.79\pm0.02$ & $979\pm2$ & $50\pm1$ & $1.6\pm0.2$\\
         COS-SBCX4 & 2.646 & $13.3\pm0.3$ & $3500\pm2800$ & $800\pm900$ & $13.2\pm0.4$ & $1319\pm45$ & $26\pm1$ & $4.1\pm0.5$\\
         COS-SBCX1 & 2.422 & $13.7\pm0.3$ & $7700\pm6900$ & $600\pm700$ & $4.1\pm0.1$ & $408\pm13$ & $33\pm1$ & $1.0\pm0.7$\\
         COS-SBCX7 & 2.415 &  $13.6\pm0.3$ & $6600\pm5300$ & $600\pm600$ & $6.1\pm0.2$ & $606\pm21$ & $23\pm1$ & $2.1\pm0.7$\\
         COS-SBC3 & 2.365 & $13.7\pm0.3$ & $8400\pm6700$ & $500\pm600$ & $5.93\pm0.03$ & $593\pm3$ & $9\pm1$ & $5.2\pm1.6$\\
         COS-SBC6 & 2.323 & $13.4\pm0.3$ & $3700\pm2900$ & $400\pm400$ & $7.4\pm0.1$ & $743\pm2$ & $27\pm1$ & $2.2\pm0.3$\\
         COS-SBC4 & 1.65 & $13.7\pm0.3$ & $4900\pm3900$ & $100\pm100$ & $2.62\pm0.01$ & $262\pm1$ & $9\pm1$ & $2.3\pm1.0$\\
         \hline
    \end{tabular}
    \label{tab:BAR}
\end{table*}

\section{Discussion}
\label{sec:discussion}


\subsection{Dark matter halo mass estimates: Prospects and caveats}

Accurate estimates of halo masses are vital to inferring the evolutionary state and eventual fate of massive structures; however, it remains a challenging task especially at high redshifts. 
In this work, we applied three methods based on SHMR $M_{\rm h}(1), (2a), (2b)$, a method based on galaxy overdensity $M_{\rm h}(3)$, and developed two methods based on NFW profile fitting $M_{\rm h}(4), (5)$. Every method has its strengths and limitations, we discuss the details in the following. 

$M_{\rm h}(1)$ is widely used in the literature \citep[e.g.][]{Ito2023,Brinch2023_pcz06,Helton2024_overdensity,Jin2023_Cosmic_Vine}, and is solely based on the stellar mass of the most massive member galaxy. However, it is a very simplified assumption that the stellar mass of the central galaxy is solely correlated with halo mass, with a large dispersion between BCG mass and halo mass \citep[e.g.][]{Montenegro-Taborda2023BCG}. The accuracy of the halo mass derived with this approach can be severely affected by uncertainties in the stellar mass and redshift of the central galaxy, where the situation is even worse if the central galaxies are DSFGs, as the stellar mass estimate can be very uncertain \citep[e.g.][]{Long2023_DSFG_Mstar}. Furthermore, since the relation is derived for establishing expected stellar masses given a halo mass, this results in stellar mass saturation, that is, the SHMR is almost constant at $M_{\rm h}\gtrsim3\times10^{12}\,{M_\odot}$ \citep{Behroozi2013Mhalo}. Comparing with results from other methods, the resultant mass is inconsistent with the other methods, yielding mainly lower limits, which is rather risky when it is applied to high redshift groups. With these points in mind, we caution against the use of this method for massive galaxy groups, and massive individual galaxies. 

Method $M_{\rm h}(2a)$ scales to the total stellar mass and is clearly more advanced than $M_{\rm h}(1)$. It assumes that all members are in a collapsed structure hosted by a single dark matter halo. However, it is difficult to assess whether the structure is already collapsed. The SHMR scaling is redshift dependent and has been calibrated for a high redshift \citep[$z\sim5$;][]{Shuntov2022}. Nevertheless, it is not calibrated to real clusters as no massive clusters exist in the COSMOS field, where this SHMR is measured.
$M_{\rm h}(2b)$ shares similar benefits and risks as $M_{\rm h}(2a)$. This method has been applied in several studies \citep[e.g.][]{Daddi2022Lya,Ito2023,Coogan2023_EGS_Group}. However, the scaling relation is calibrated to $z\sim1$ clusters, which might not persist, and might evolve, at higher redshift, and might not be applicable to lower-mass halos. 
In this work, the most massive members are all spectroscopically confirmed, and only low-mass members are identified using photometric redshifts. Although some foreground and background interlopers could be included, their low masses are not expected to significantly impact the inferred total stellar mass, and by extension the halo mass (see \cref{sec:interlopers}).

Unlike the above discussed methods, $M_{\rm h}(3)$ is not directly dependent on the stellar mass of the cluster galaxies, instead relying solely on their spatial density. A big advantage is that it is independent of uncertainties in stellar mass measurements. On the other hand, the clustering bias of the halo is not well constrained, and we simply assumed a bias value from the \citet{Tinker2010haloBias} formalism, which adds another layer of uncertainty in the estimate. Because the bias value is dependent on the halo mass, and we used the bias to calculate a halo mass, this method is circularly defined; however, we avoided this circularity by iteratively calculating the bias using $M_{\rm h}(2a)$ as an initial guess. The bias is not 1:1  with halo mass, and the iterative approach converges quickly. 

Method $M_{\rm h}(4)$ assumes that the stellar mass density profile traces the dark matter density profile, and that they both follow a NFW model at high redshifts. With this method, we derived a total baryonic mass and inferred a halo mass using a dark matter-to-baryon mass ratio. This assumption has been validated in low redshift clusters, for example \citet{Annuziatella2014} and \citet{Palmese2016cluster} find both the stellar mass density and number density profiles of the clusters MACS J1206.2-0847 \citep{Biviano2013_nfw_fitting} and RXC J2248.7-4431 can be fitted by a NFW profile, and further at cosmic noon in CLJ1449 \citep{Strazzullo2013}.
\cite{Andreon2015cluster} find that the stellar-to-total mass ratio is radially constant in three $z\sim0.45$ clusters.
\cite{Caminha2017cluster} find that dark matter and hot gas asymmetry of the $z=0.44$ MACS 1206 closely follows the asymmetric distribution of the stellar component. 
Accordingly, if the stellar mass density profile of a high-$z$ group is found consistent with a NFW profile, it would suggest that the structure is likely collapsed. Uniquely, this method can constrain the halo mass and its concentration parameter simultaneously, which is a significant advance in characterizing dark matter halos of collapsed structures. 
As a proof of this method at high redshifts ($z>2$), we applied it to the most distant X-ray detected cluster CLJ1001 \citep[][see \cref{fig:CLJ1001Mh5}]{Wang_T2016cluster}. CLJ1001 is uniquely suited for this, as it is at a high redshift of $z=2.5$ and has both an X-ray and velocity dispersion calibrated halo mass. Our NFW fitting yields a halo mass of $M_{\rm h}=(8\pm2)\times10^{13}\,M_\odot$, with concentration $c=54^{+25}_{-46}$ and scale radius $R_s=7^{+3}_{-6}\,{\rm kpc}$ (\cref{fig:CLJ1001Mh5}). The recovered halo mass is in excellent agreement with the X-ray and velocity dispersion halo masses of $M_{\rm h}=(5.0\pm2.3)\times10^{13}\,M_\odot$, and, the scale radius is consistent with the one reported in \citet{Wang_T2016cluster} $R_s=0^{+8}_{-0}\,{\rm kpc}$.
Nevertheless, similarly to $M_{\rm h}(1)$, (2a) and (2b), this method suffers from the uncertainty on stellar masses and membership, and requires an accurate constraint on the background density level.
Moreover, $M_{\rm h}(4)$ is also impacted by the uncertainty on the dark matter-to-baryonic mass ratio and possible inconsistency between stellar mass density profile and dark matter halo profile at high-$z$. We also tested two methods for finding the exact centre of the group: (1) using the position of the most massive central galaxy; and (2) randomly shifting the centre position by up to a few arcseconds, instead of calculating the distance to FIR-peak-weighted barycentre. We find no significant changes in either the recovered halo masses or recovered concentrations, nor in the uncertainties of either values.

Method $M_{\rm h}(5)$ is a variant of $M_{\rm h}(4)$. It shares with $M_{\rm h}(4)$ the assumption of the consistency between stellar mass density profile and halo mass density profile. However, it is totally independent of the stellar masses, as it purely relies on the shape of the profile. This is a significant advantage with respect to other methods. Notably, adopting this method with the number density profile would further reduce the bias introduced by the stellar mass estimate of the central galaxy, as the halo mass would be solely based on the distribution of member galaxies. 
However, this method has two major drawbacks: (1) it must assume a fixed concentration parameter to break the degeneracy between scale radius and concentration, while the fixed concentration is based on simulation results with a prior halo mass, and neither are fully tested by observations; (2) the profile fitting to the stellar mass density is dominated by the central galaxy, fitting to the number density can remove this bias but requires a larger sample of confirmed memberships. Another problem with both $M_{\rm h}(4)$ and $M_{\rm h}(5)$ is the degeneracy between $M_{\rm h}$, $c$, and $R_{\rm s}$ as they are all co-dependent and influence the shape and scaling of the density profile (\cref{sec:nfw_profiles}). We tried to limit this degeneracy by fixing the concentration and purely fitting the scale radius without directly incorporating the halo mass; however, this adds another layer of uncertainty as the true concentration is unknown.

Overall, the self-consistency between different methods shows encouraging prospects in halo mass estimates for high redshift structures like these. Among these methods, the NFW profile fitting $M_{\rm h}(4)$ and (5) are exploited for high-$z$ groups for the first time, showing potential for state-of-the-art in characterizing dark matter halos.
Nevertheless, thorough analyses and tests with a large sample of low-z clusters are required to fully explore the prospects of the profile fitting techniques, and then exploit it towards high redshift, which is beyond the scope of this paper.




\subsection{Nature and fate: Are these structures clusters in formation?}

\begin{figure*}[!htbp]
    \centering
    \sidecaption
    \includegraphics[width=12cm]{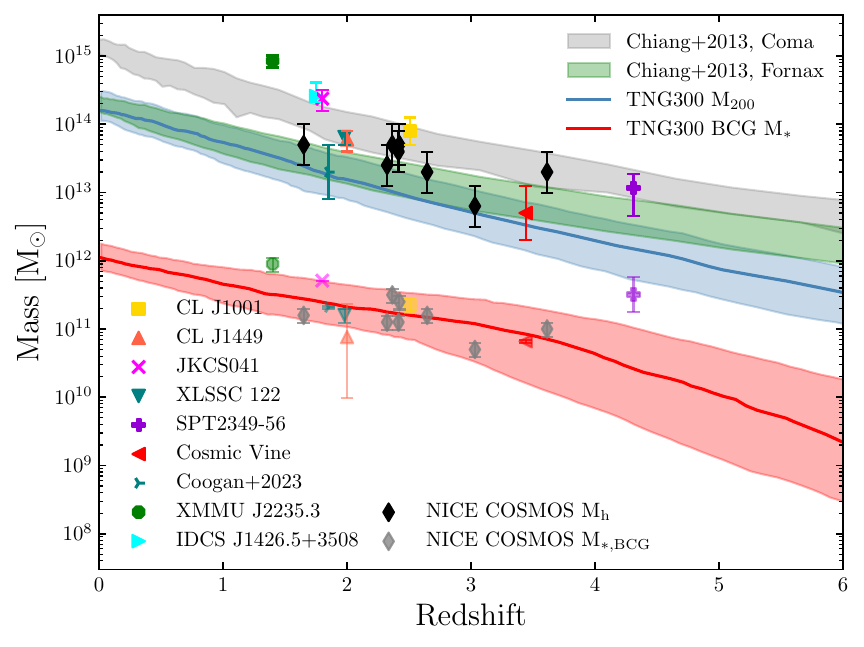}
    \caption{Dark matter halo mass (upper black points) and stellar mass of BCG (lower grey points) versus redshift. Overlaid are the halo mass of simulated progenitors of Fornax- and Coma-type clusters from \citet{Chiang2013cluster} and halo and BCG masses from the TNG300 simulation \citep{Montenegro-Taborda2023BCG}. The black and grey diamonds show halo masses and (proto-)BCG stellar masses from the NICE COSMOS sample. Other points show halo and (proto-)BCG masses from literature (proto-)clusters as identified in the figure \citep{Jee2009_XMMU_J22353,Rosati2009_XMMUJ22353,Brodwin2016_IDCS_J14,Newman2014_JKCS041,Trudeau2022_XLSSC122,Gobat2019_CLJ1449,Strazzullo2018_CLJ1449,Wang_T2016cluster,Miller2018cluster_z4,Rotermund2021_SPT2349_56,Coogan2023_EGS_Group}}
    \label{fig:Mass_z_nice}
\end{figure*}

To understand the nature and fate of these structures, following \cite{Jin2023_Cosmic_Vine}, we compared our observations with cosmological simulations. First, we compared the estimated halo masses with simulations of Coma- and Fornax-like clusters from \citet{Chiang2013cluster}. We find all groups to have halo masses consistent with progenitors of clusters with present day halo masses of at least $M_{\rm h}(z=0)>10^{14}\,{\rm M_\odot}$ (\cref{fig:Mass_z_nice}). We also compared the halo masses of this sample to clusters from the TNG300 simulation \citep[][\cref{fig:Mass_z_nice}]{Montenegro-Taborda2023BCG}; again, we find that the halo masses of all groups are consistent with those of the progenitors of $z=0$ clusters ($M_{\rm h}(z=0)>10^{14}~M_\odot$). 
Furthermore, we compared the stellar masses of the most massive group galaxies with that of BCG progenitors in the TNG300 simulation \citep{Montenegro-Taborda2023BCG}, and again find they are consistent (\cref{fig:Mass_z_nice}). 
This supports the idea that the eight groups are real forming clusters, and their central galaxies are likely proto-BCGs \citep{Jin2023_Cosmic_Vine}.

Another approach assessing the fate of massive structures is the observationally constrained large-scale simulations, which has been recently performed by \cite{Ata2022cluster} on the $z=2.5$ large-scale structure Hyperion \citep{Cucciati2018}. This simulation requires a highly complete spectroscopic survey of cluster members \citep{Ata2022cluster}; hence, deep spectroscopic follow-up observations on both the cores and large-scale structure are essential. On the other hand, dedicated cluster simulations have been achieved, for example Cluster-EAGLE \citep{Barnes2017ClusterEAGLE}, Magneticum \citep{Remus2023a_simu}, FLAMINGO \citep{Schaye2023FLAMINGO}, and TNG-Cluster \citep{Nelson2023TNG-Cluster}. Further work comparing a large sample with these simulations would provide new insights into galaxy and cluster formation.


\subsection{Baryonic accretion: Flattening or not?}
In \cref{fig:Mdm_Mstream_znice} we fit the BAR using both a quasi-linear single trend and assuming flattening at $M_{\rm stream}>M_{\rm h}$. With double sample size compared to \citet{Daddi2022Lya}, we now find a significant ($4.5\sigma$) slope compared with their $2.5\sigma$ fit. We excluded the group from \citet{Coogan2023_EGS_Group} as it remains a strong outlier from the expected SFR from the BAR. The group is also a strong outlier in terms of the proportions of stellar mass and SFR in the BCG, suggesting that it might be caught in a short-lived starburst connected to the formation of the BCG, or might signal a cooling flow \citep{Coogan2023_EGS_Group}. 
The scatter of our fit ($0.40$ dex) is somewhat lower from $0.45$ dex in \citet{Daddi2022Lya}. The data do not prefer the bending model or the linear model, with nearly identical scatter of $0.40$ dex and $0.38$ dex, respectively. Instead of using the updated $M_{\rm stream}$ definition in \citet{Daddi2022_mstream}, adopting the $M_{\rm stream}$ definition from \citet{Dekel2006}, does not change the scatter significantly. Because the scatter in the cold-stream regime is more dependent on the halo mass \citep{Daddi2022Lya}, we calculated the scatters in the hot (\cref{fig:Mdm_Mstream_znice}-left, $M_{\rm stream}/M_{\rm h}<1$) versus cold-stream (\cref{fig:Mdm_Mstream_znice}-left, $M_{\rm stream}/M_{\rm h}>1$) regimes and find a scatter of 0.30 dex in the hot and 0.53 dex in the cold-stream regime. The expected scatter in the BAR at fixed halo mass is 0.2 dex \citep{Correa2015_BAR_scatter}, and the scatter arising from the fraction of cold gas to total gas, $f_{\rm cold}$, is 0.1 dex \citep{Correa2018_f_cold}, adding these in quadrature yields 0.3 dex in excellent agreement with our scatter in the hot regime. In the cold-stream regime, the scatter from $f_{\rm cold}$ is negligible as all accreted gas is cold, leaving a scatter from the BAR of 0.2 dex and from $M_{\rm h}$ of 0.3 dex, adding these in quadrature yields 0.36 dex, significantly lower than the observed scatter. This increased observed scatter is only from six points and would require a larger sample to confirm; however, if real, this could indicate that there is some intrinsic stochasticity, inefficiencies in converting accreted baryonic matter into SFR, feedback mechanisms, or a combination of these processes. 

\subsection{Interlopers in candidate members}
\label{sec:interlopers}
Given the majority of candidate members are selected with photometric redshifts, it is common that foreground and background interlopers can be included in the candidate members.
We quantified the contamination of possible interlopers in two ways: (1) we measured the background number count in the mass-complete photo-$z$ selection in COSMOS2020 Classic LePhare and calculated the expected number of interlopers that can appear by chance in an aperture around the groups. We find possible mass-complete interloper fractions of $4-36\%$ (${\rm median}=11.9\%$) within a $r=10''$ aperture, and the fractions increase with larger aperture, which is $23-73\%$ (${\rm median}=45.8\%$) within 30" (see \cref{fig:2d-interlopers}-left). (2) We tested our candidate selection method in CLJ1001 \citep{Wang_T2016cluster} using the COSMOS2020 photo-z, and compare with the highly complete spectroscopic redshift sample of \citet{Sun2024_CLJ1001}. We find two clear interlopers and a total interloper fraction of $17-35\%$ in a $r=30''$ aperture, which is largely consistent with the result of method (1). 
Furthermore, we also tested the interloper contamination in the total stellar mass budgets. We measured the background stellar mass in the COSMOS2020 catalogue and compared it with the total stellar mass of the groups, finding mass-complete fractions of $2-10\%$ (${\rm median}=3.7\%$) within $r=10''$ and $11-42\%$ (${\rm median}=17.3\%$) within $r=30''$, respectively (see \cref{fig:2d-interlopers}-right). 
We note that the interloper contamination were accounted for by subtracting them from the total stellar masses or overdensity when deriving halo masses.

\subsection{Line-of-sight projection: Chance probabilities}
\label{sec:los-prob}
To understand the probability of finding two line-of-sight projected groups, we calculated the chance probability of finding two halos of $M_{\rm h}/2$ within $R_{\rm vir}/2$ and indistinguishable $z$-phot distributions (i.e. the redshift selection for halos are overlapping), using halo mass functions calculated with \texttt{hmf}\footnote{\url{https://github.com/halomod/hmf}} \citep{Murray2013_hmf}. We multiplied this chance probability by the total number of expected halos of mass $M_{\rm h}/2$ in the entire COSMOS field. The highest chance probability of this sample is the lowest redshift COS-SBC4, with $p_{\rm chance,los}=2\times10^{-4}$, as seen in \cref{tab:pchance_los}. This is only assuming the case of twin halos, and possible combinations of triples or more halos are not considered; however, this would not change the ballpark probabilities considerably.

\begin{figure}
    \centering
    \includegraphics[width=0.95\columnwidth]{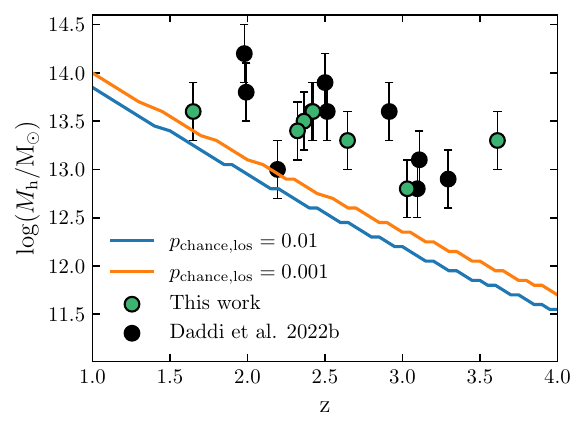}
    \caption{Halo mass limits as a function of redshift. The limits correspond to the chance probability of finding two overlapping halos of mass, $M_{\rm h}/2$: $p_{\rm chance,los}=1\%$ and $p_{\rm chance,los}=0.1\%$. We overplot the halo masses of this work and the sample from \citet{Daddi2022Lya}.}
    \label{fig:pchance_los}
\end{figure}

In general, we find decreasing chance probability with increasing redshift and halo mass. We explored the halo mass limits in the COSMOS field (assuming a full redshift dispersion in the structure of $\Delta z=0.1$), where the chance probability of being a line-of-sight projection is $<1\%$ and $<0.1\%$, as shown in \cref{fig:pchance_los}. At low redshift (z<1.5) only massive halos with $\log(M_{\rm h}/{\rm M_\odot})>13.5$ are unlikely to be line-of-sight projection of two lower mass halos, whereas at $z\sim3.5$ the mass limit decreases to $\log(M_{\rm h}/{\rm M_\odot})\sim12$. As shown in (\cref{fig:pchance_los}), the NICE sample and \citet{Daddi2022Lya} sample are all above the limits, with the majority of them significantly above the limits by $\sim$0.8 dex, indicating that they are unlikely line-of-sight projections, further supporting their massive group nature.

\begin{table}[!htbp]
    \centering
    \begin{tabular}{c c}
        Structure & $p_{\rm chance,los}$ \\
        \hline
        HPC1001 & $9\times10^{-16}$\\
        COS-SBCX3 & $3\times10^{-7}$\\
        COS-SBCX4 & $8\times10^{-8}$\\
        COS-SBCX1 & $1\times10^{-8}$\\
        COS-SBCX7 & $3\times10^{-8}$\\
        COS-SBC3 & $3\times10^{-6}$\\
        COS-SBC6 & $6\times10^{-6}$\\
        COS-SBC4 & $2\times10^{-4}$\\
        \hline
    \end{tabular}
    \caption{Chance probabilities of a line-of-sight projection of two $M_{\rm h}/2$ halos.}
    \label{tab:pchance_los}
\end{table}

\section{Summary and conclusions}
\label{sec:summary}
We selected eight high-$z$ group candidates based on the overdensity of red IRAC sources that have red \textit{Herschel} colours. As part of the NICE large programme, we followed up on these candidates with NOEMA and ALMA observations. We summarize the results as follows:

(1) We spectroscopically confirmed a sample of eight massive galaxy groups in the redshift range $1.65\leq z\leq3.61$ in the COSMOS field, in which 21 members were detected with CO lines or H$\alpha$ line emission \citep{Kashino2019_FMOS_COSMOS}. Using photometric redshifts in the COSMOS2020 catalogue, we selected a total of $\sim$250 candidate members in these structures.

2) Utilizing \textit{Herschel} and SCUBA2 data, we measured the integrated FIR-to-radio photometry of the eight structures and performed FIR SED fitting using dust and mid-IR AGN models. Seven of the eight group SEDs are well fitted by a single dust template. Only the group COS-SBC6 exhibits a significant radio excess and a mid-IR AGN component, yielding a radio excess fraction of 22\% in the total NICE sample. This indicates that these massive groups are presumably dominated by star formation and that strong AGN activity does not play a major role.

(2) We applied six methods for estimating the dark matter halo mass of each structure, including the SHMR, galaxy overdensity with bias, and the NFW profile fitting technique. The results from the different methods are overall consistent within 0.2--0.3 dex.
Adopting the average results from different methods, our best estimate of the halo masses is $\log(M_{\rm h}/{\rm M_\odot})=12.8-13.7$. Using the NFW profile fitting technique, we found that the stellar mass densities of the eight groups can be fitted by a NFW profile, suggesting they are likely collapsed structures. We tentatively constrained the concentration parameters of the halos, and the most massive groups are overall consistent with predictions from the simulations \citep{Ludlow2016_concentrations}, albeit with a large scatter. 

(3) Using scaling relations between dark matter halo masses, the BAR, and the SFR, we calculated the expected BAR for each structure. The expected SFRs from baryonic accretion in all structures are in good agreement with the FIR-measured integrated SFRs. Together with literature samples, we derive a quasi-linear relation between SFR/BAR and M$_{\rm stream}$/M$_{\rm h}$, with ${\rm SFR/BAR}=10^{-0.46\pm0.22}({\rm M_{stream}/M_{h}})^{0.71\pm0.16}$ and a scatter of $\sim0.4$ dex. This supports the idea that the star formation in these structures is fed by gas accretion. Specifically, HPC1001 and COS-SBCX3 occupy the regime of cold gas streams in hot media, which could make them ideal laboratories for studying cold gas accretion in dense, $z\gtrsim3$  environments.

(4) By comparing these massive groups with simulations \citep{Chiang2013cluster,Montenegro-Taborda2023BCG}, we find that the former have halo masses and  proto-BCG stellar masses that are consistent with them being the progenitors of $z=0$ clusters. This suggests that these structures are likely forming clusters in the early Universe and that their central galaxies are forming BCGs.

All these results point to these structures being forming clusters and groups in the early Universe that, in turn, are growing their masses via gas accretion. 
Future spectroscopic observations of the large-scale and central cores, combined with comprehensive complementary simulations, are essential to confirming their nature and further assessing their evolution. This work serves as a pilot study of the NICE sample. In the near future we will expand studies in this work with the full NICE sample and investigate the properties of individual group galaxies to shed light on the formation of structures and galaxies in dense environments. 

\begin{acknowledgements}
We thank the anonymous referee for helpful and constructive comments. We thank Camila Correa for useful discussions on baryonic accretion. The Cosmic Dawn Center (DAWN) is funded by the Danish National Research Foundation under grant DNRF140. 
SJ acknowledges the financial support from the European Union's Horizon Europe research and innovation program under the Marie Sk\l{}odowska-Curie Action grant No. 101060888. TW acknowledges support by National Natural Science Foundation of China (Project No. 12173017 and Key Project No. 12141301), National Key R\&D Program of China (2023YFA1605600), and the China Manned Space Project (No. CMS-CSST-2021-A07). LZ is supported by the National Natural Science Foundation of China (NSFC grant 13001103) and the National Key R\&D Program of China No. 2022YFF0503401. 
This work is based on observations carried out under project number M21AA with the IRAM NOEMA Interferometer. IRAM is supported by INSU/CNRS (France), MPG (Germany) and IGN (Spain).
We are grateful for the help received from IRAM staff during observations and data reduction.
This paper makes use of the following ALMA data: ADS/JAO.ALMA 2021.1.00815.S, 2021.1.00246.S, 2016.1.00463.S, 2015.1.00137.S, and 2013.1.00034.S. ALMA is a partnership of ESO (representing its member states), NSF (USA) and NINS (Japan), together with NRC (Canada), MOST and ASIAA (Taiwan), and KASI (Republic of Korea), in cooperation with the Republic of Chile. The Joint ALMA Observatory is operated by ESO, AUI/NRAO and NAOJ. 
\end{acknowledgements}

\bibliographystyle{aa}
\bibliography{biblio}

\onecolumn

\begin{appendix}
\section{Colour images and spectra}

\begin{figure*}[!h]
    \centering
    \includegraphics[width=\textwidth]{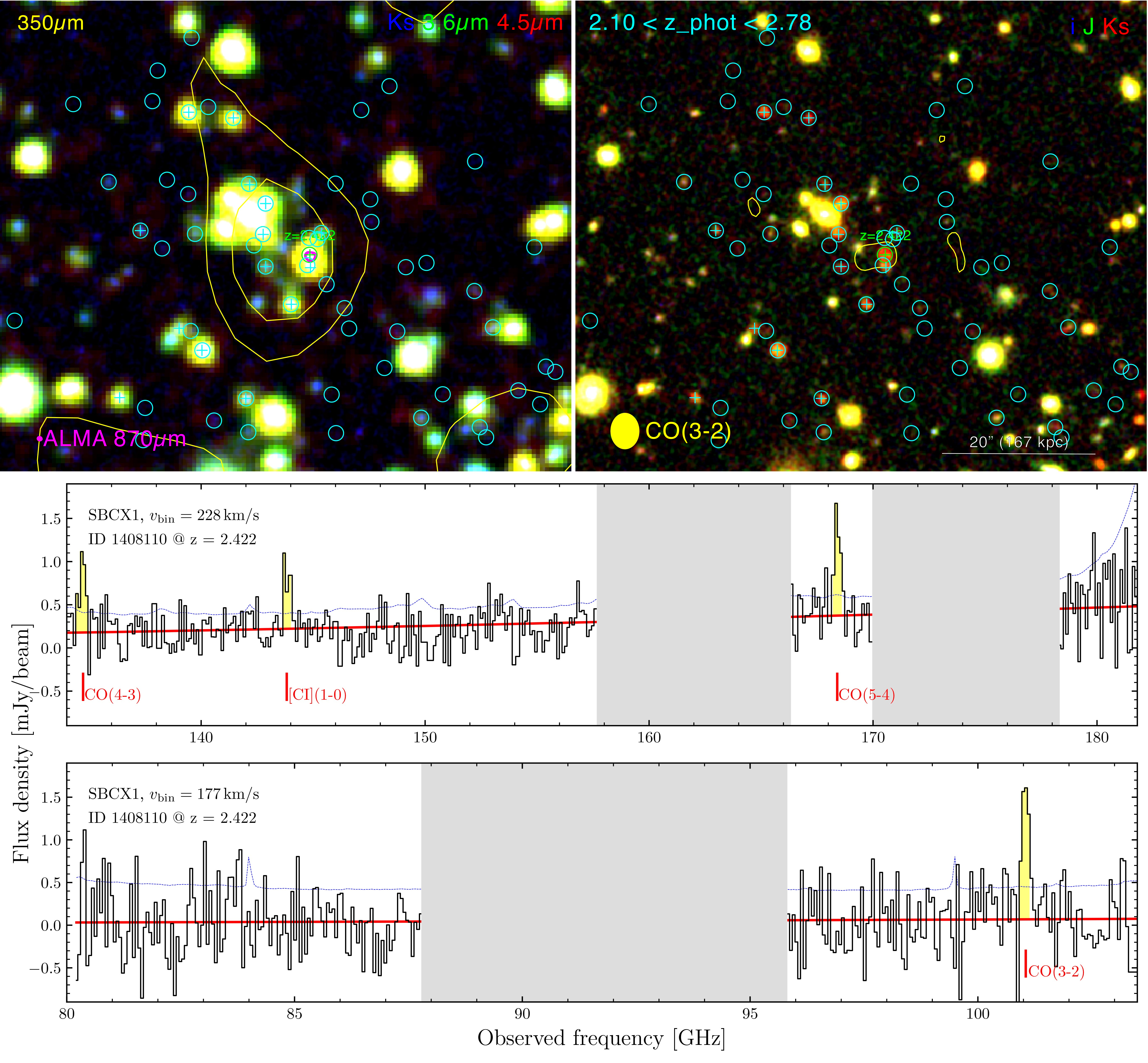}
    \caption{Colour images overlaid with ALMA Band 7 dust continuum, and ALMA (top) and NOEMA (bottom) spectra of COS-SBCX1.\ Photometrically selected galaxies with $2.10<z_{\rm phot}<2.78$ are marked with cyan circles. See the caption of \cref{fig:Sillassen-color-spectra}.}
    \label{fig:SBCX1-color-spectra}
\end{figure*}

\clearpage

\begin{figure*}[!ht]
    \centering
    \includegraphics[width=\textwidth]{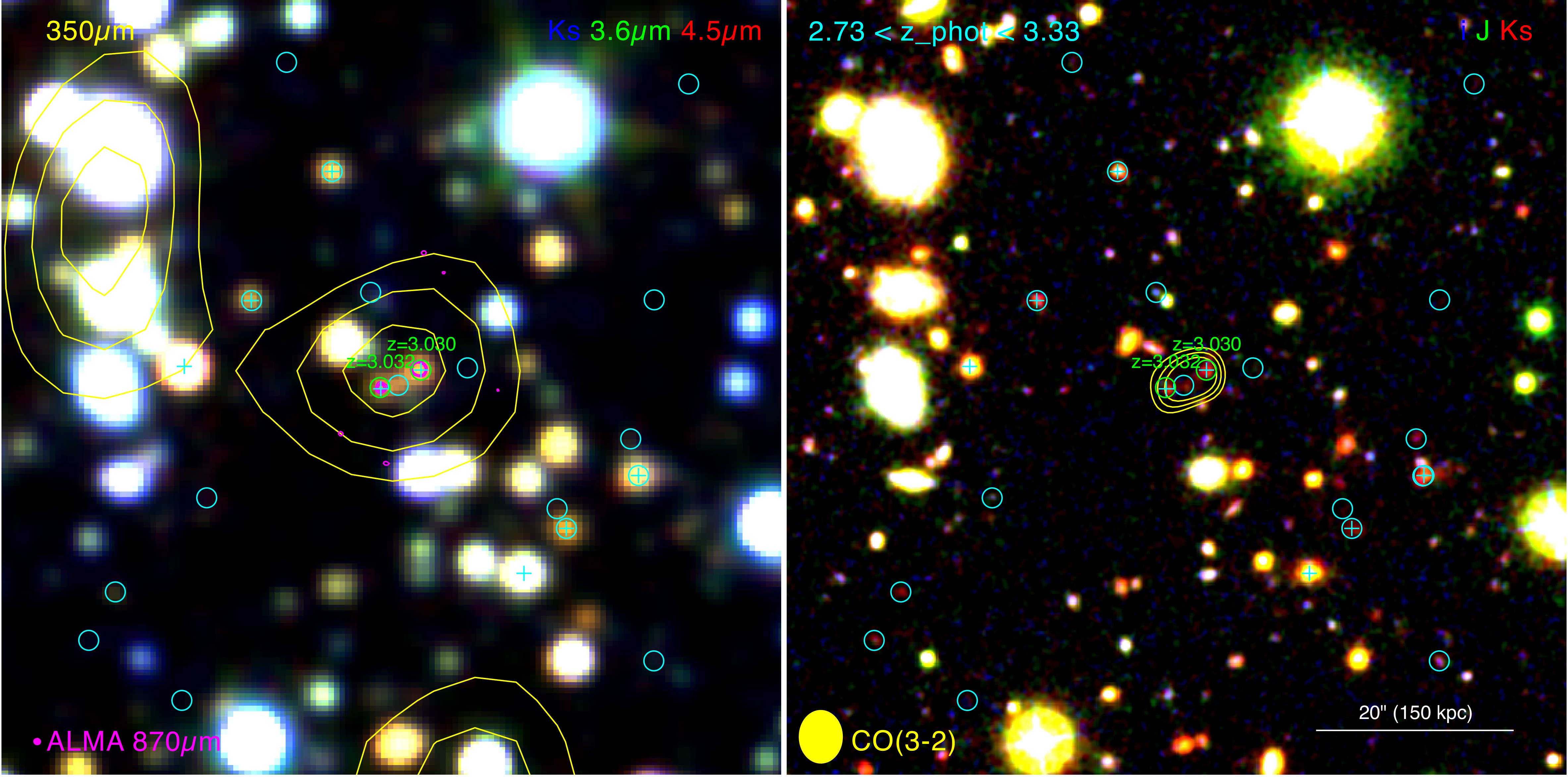}
    \includegraphics[width=\textwidth]{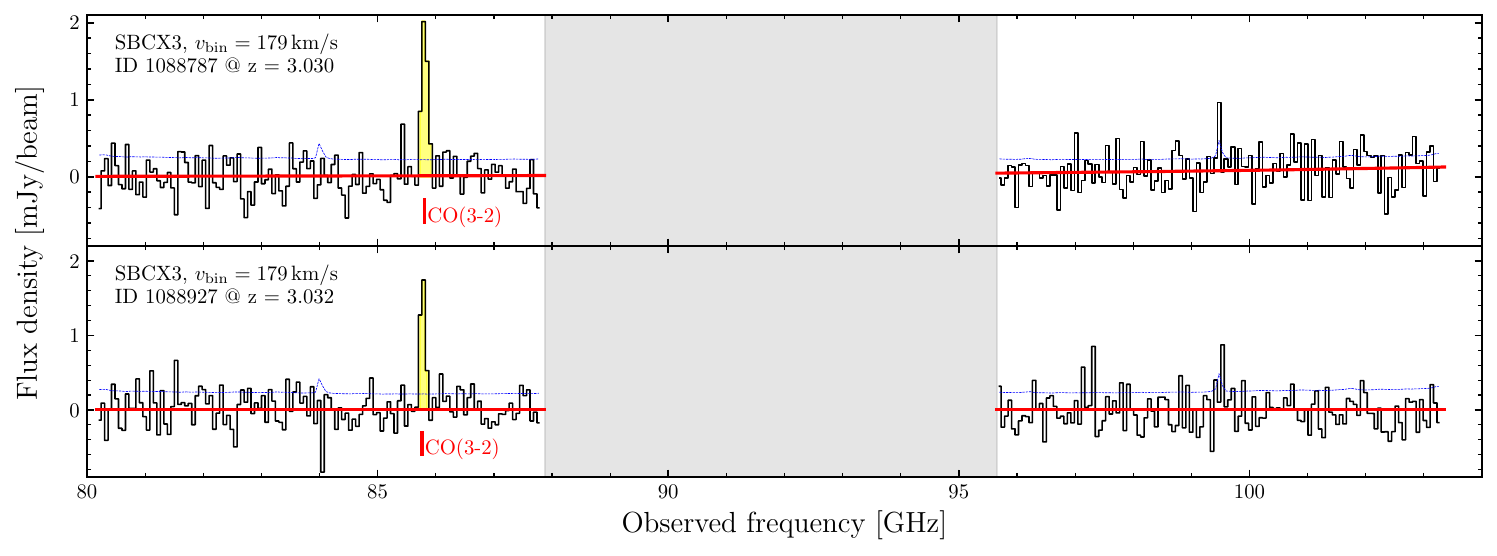}
    \caption{Colour images overlaid with ALMA Band 7 dust continuum and NOEMA spectra of COS-SBCX3. Photometrically selected galaxies with $2.73<z_{\rm phot}<3.33$ are shown as cyan circles. See the caption of \cref{fig:Sillassen-color-spectra}.}
    \label{fig:SBCX3-color-spectra}
\end{figure*}

\clearpage

\begin{figure*}[!ht]
    \centering
    \includegraphics[width=\textwidth]{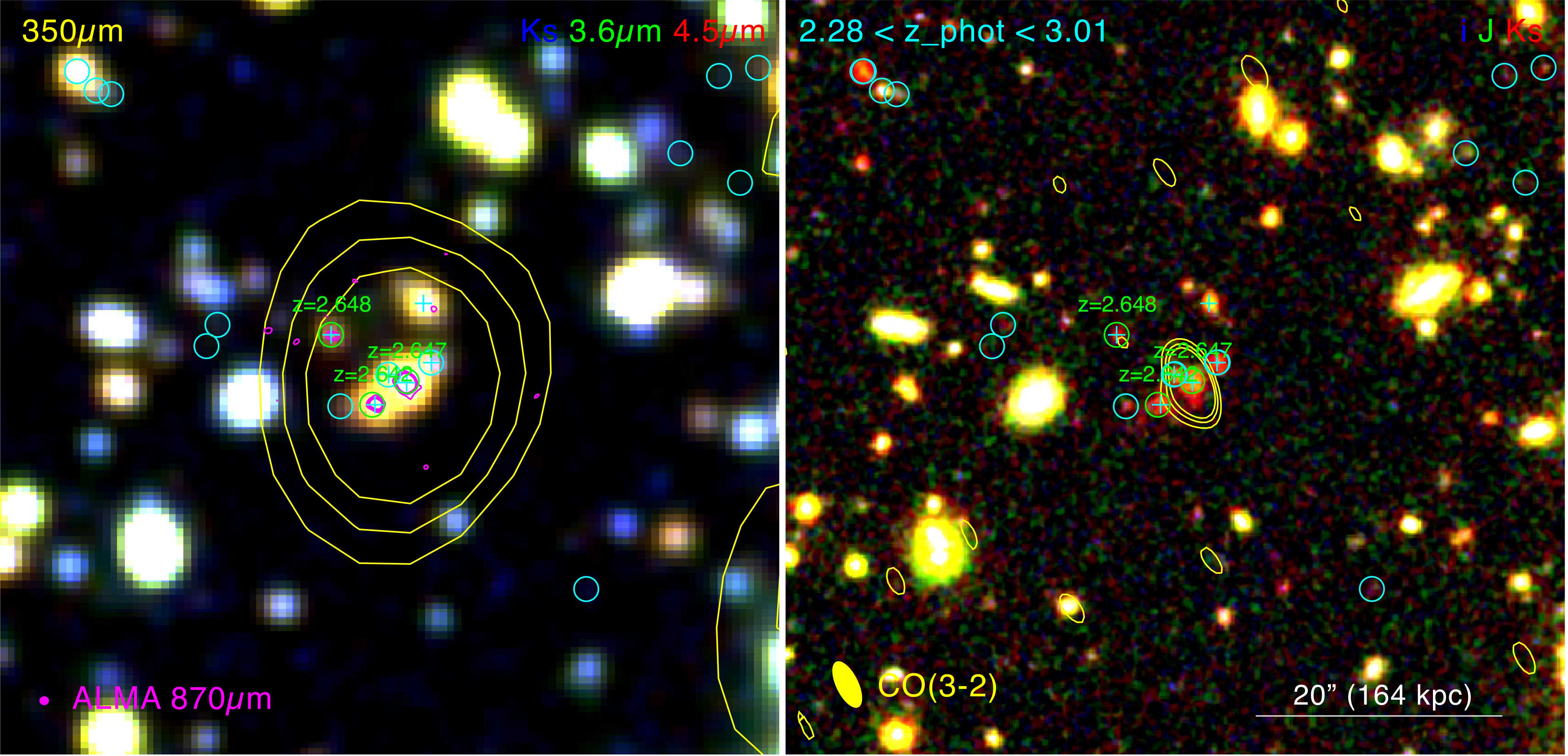}
    \includegraphics[width=\textwidth]{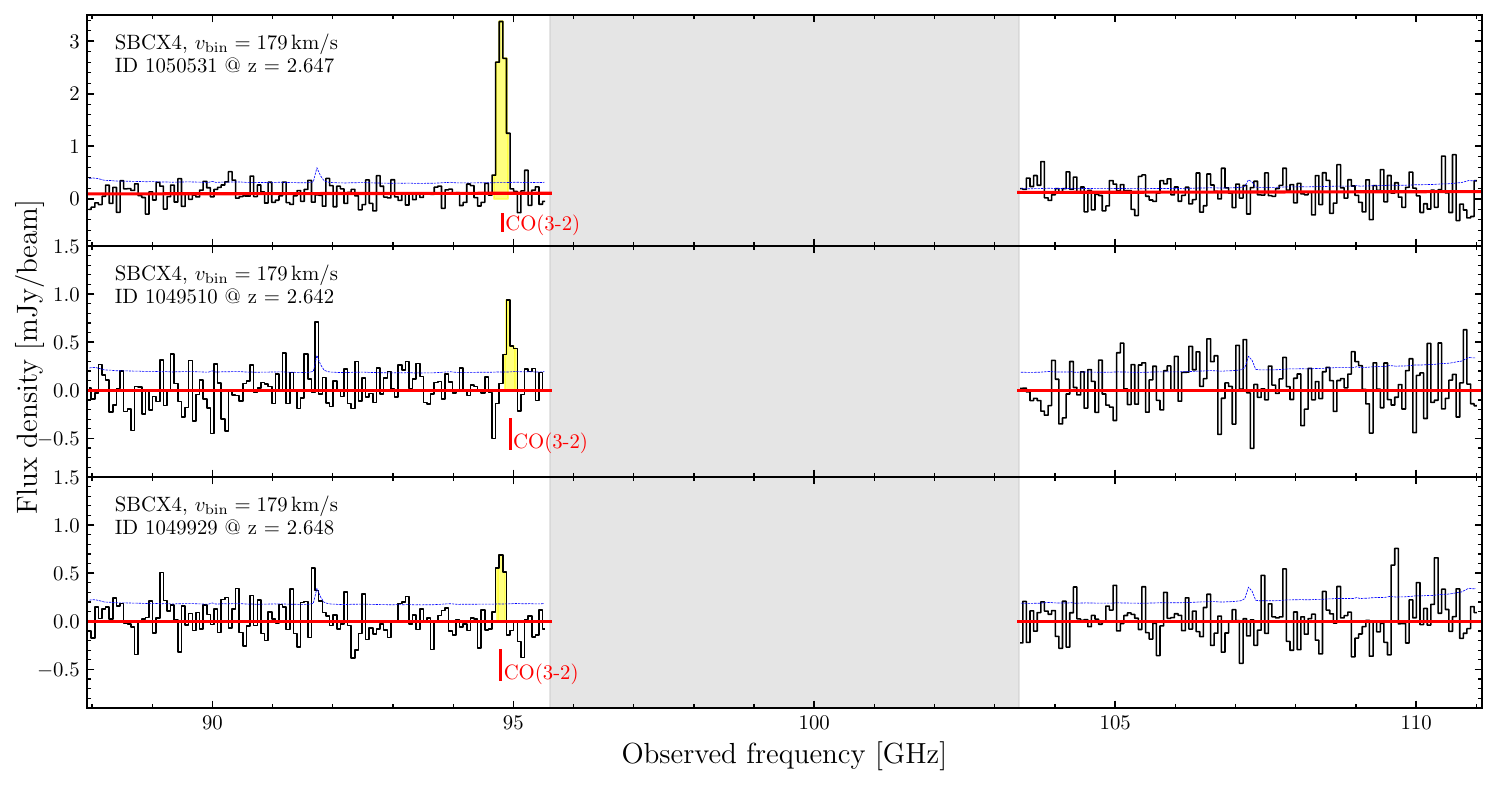}
    \caption{Colour images overlaid with ALMA Band 7 dust continuum and NOEMA spectra of COS-SBCX4.\ Photometrically selected galaxies with $2.28<z_{\rm phot}<3.01$ are marked by cyan circles. See the caption of \cref{fig:Sillassen-color-spectra}.}
    \label{fig:SBCX4-color-spectra}
\end{figure*}

\clearpage

\begin{figure*}[!ht]
    \centering
    \includegraphics[width=\textwidth]{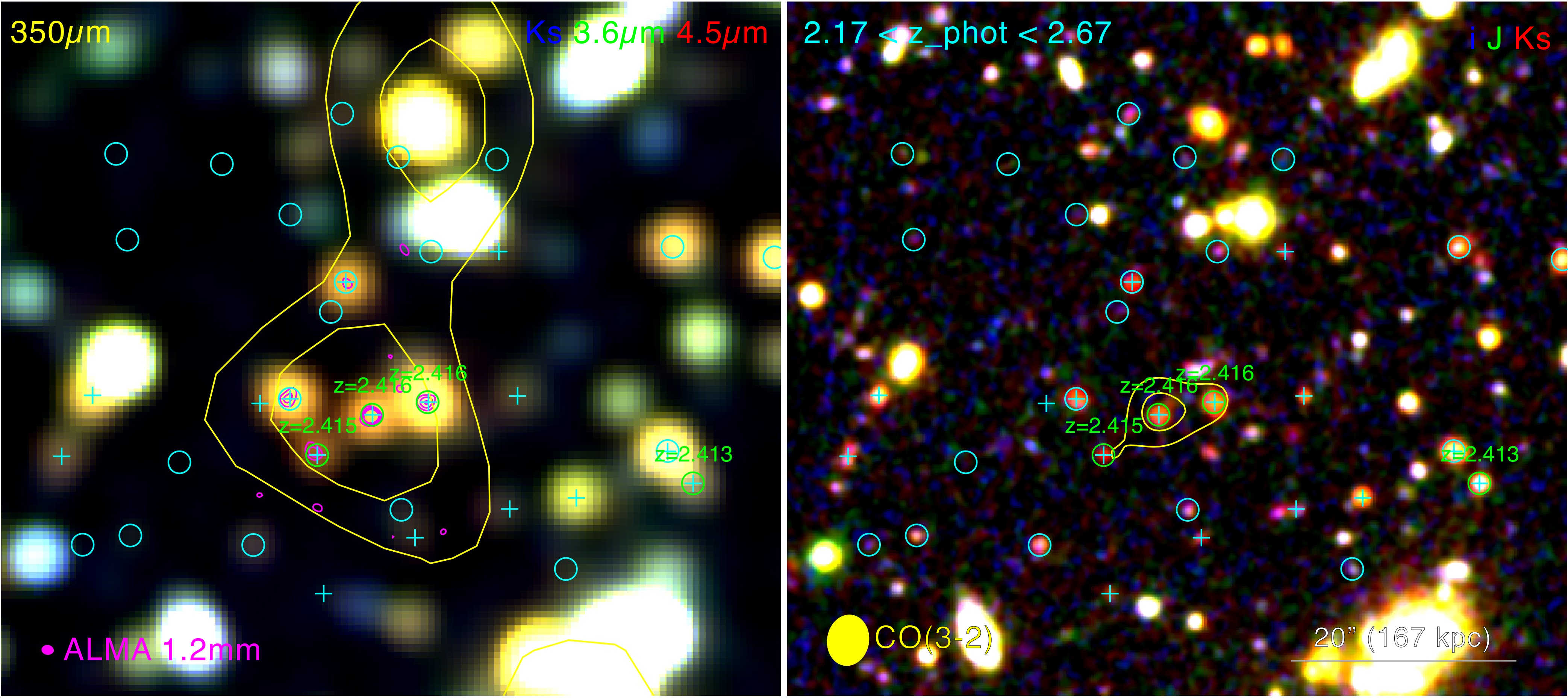}
    \includegraphics[width=\textwidth]{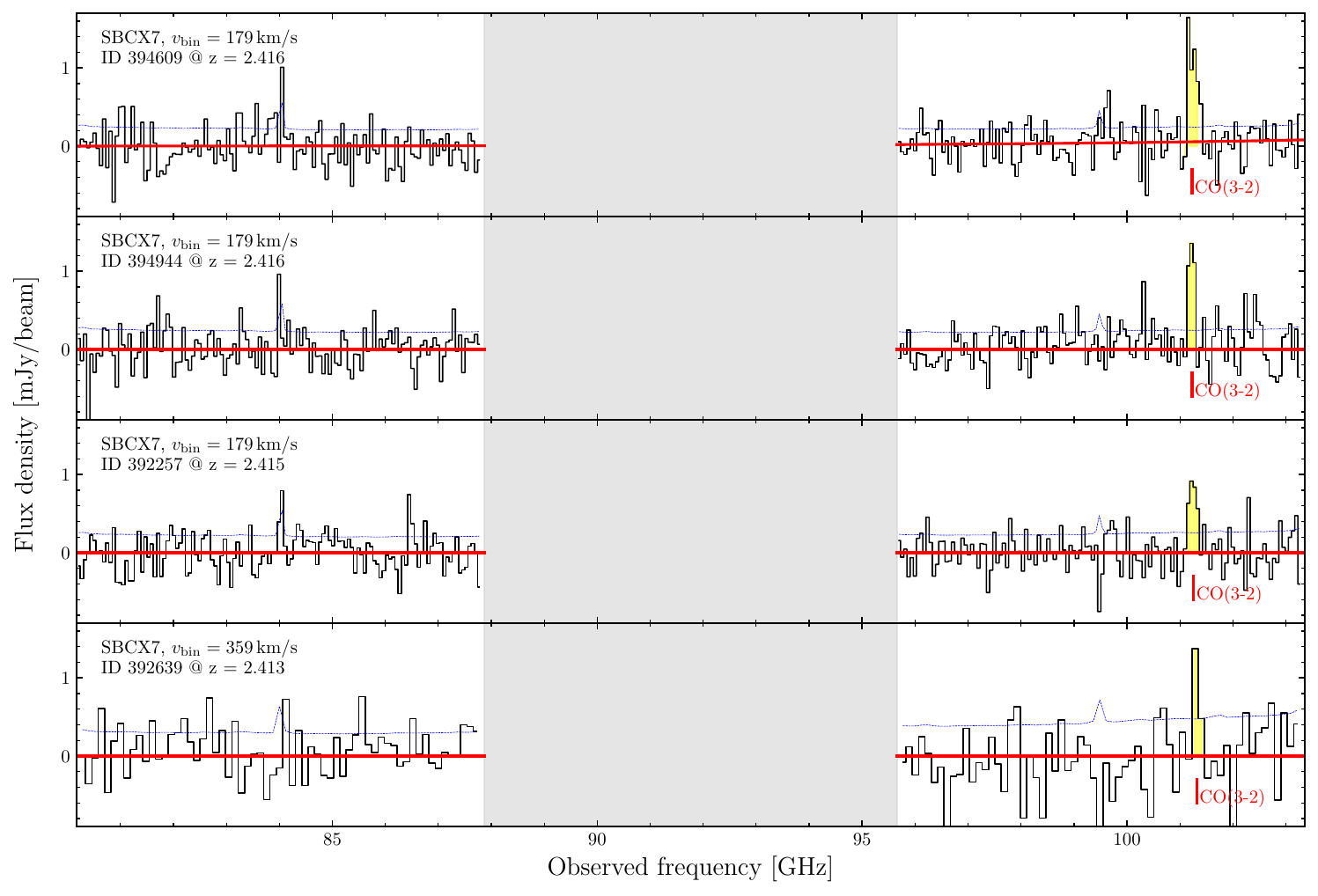}
     \caption{Colour images overlaid with ALMA Band 6 dust continuum and NOEMA spectra of COS-SBCX7.\ Photometrically selected galaxies with $2.17<z_{\rm phot}<2.68$ are marked by cyan circles. See the caption of \cref{fig:Sillassen-color-spectra}.}
     \label{fig:SBCX7-color-spectra}
\end{figure*}

\clearpage

\begin{figure*}[!ht]
    \centering
    \includegraphics[width=\textwidth]{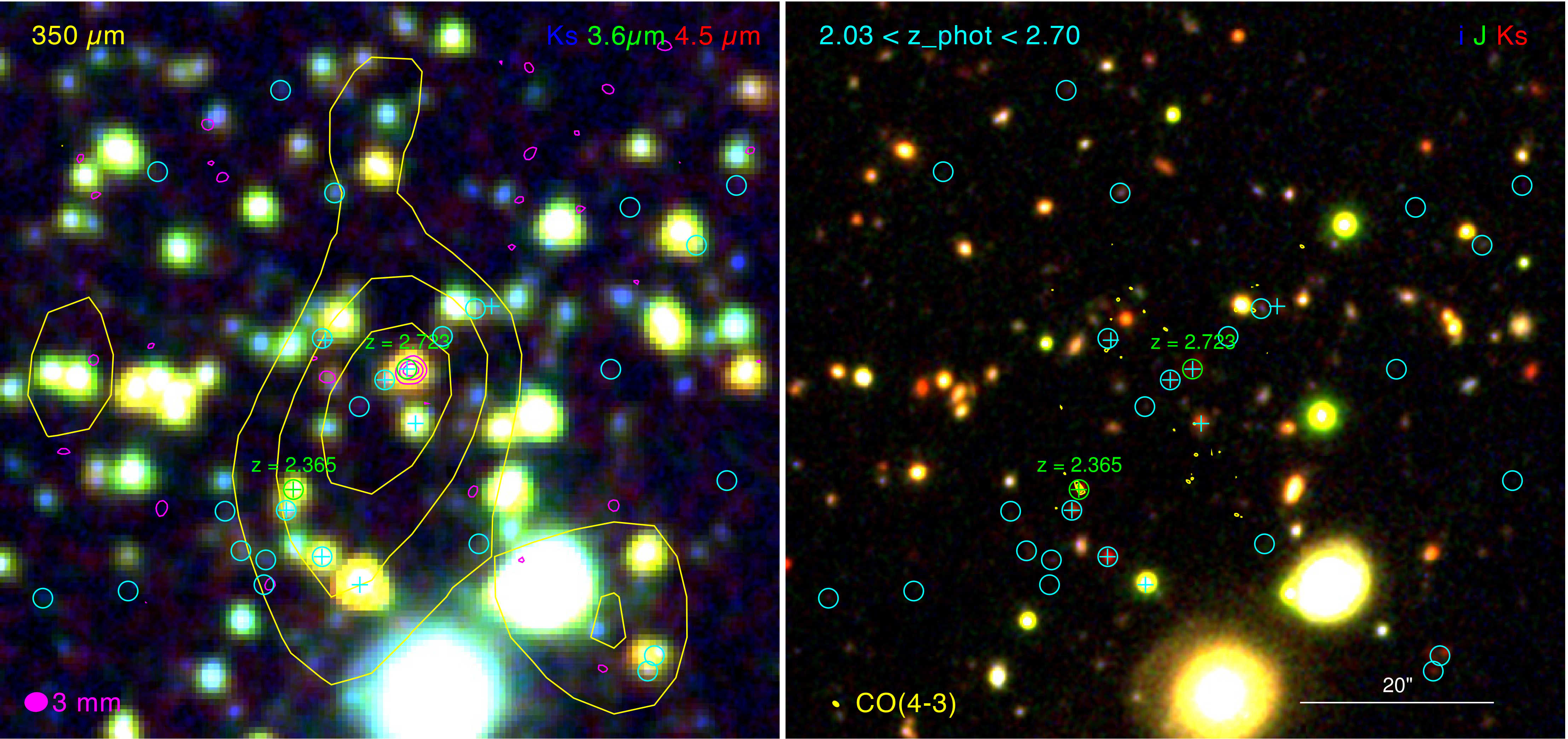}
     \includegraphics[width=\textwidth]{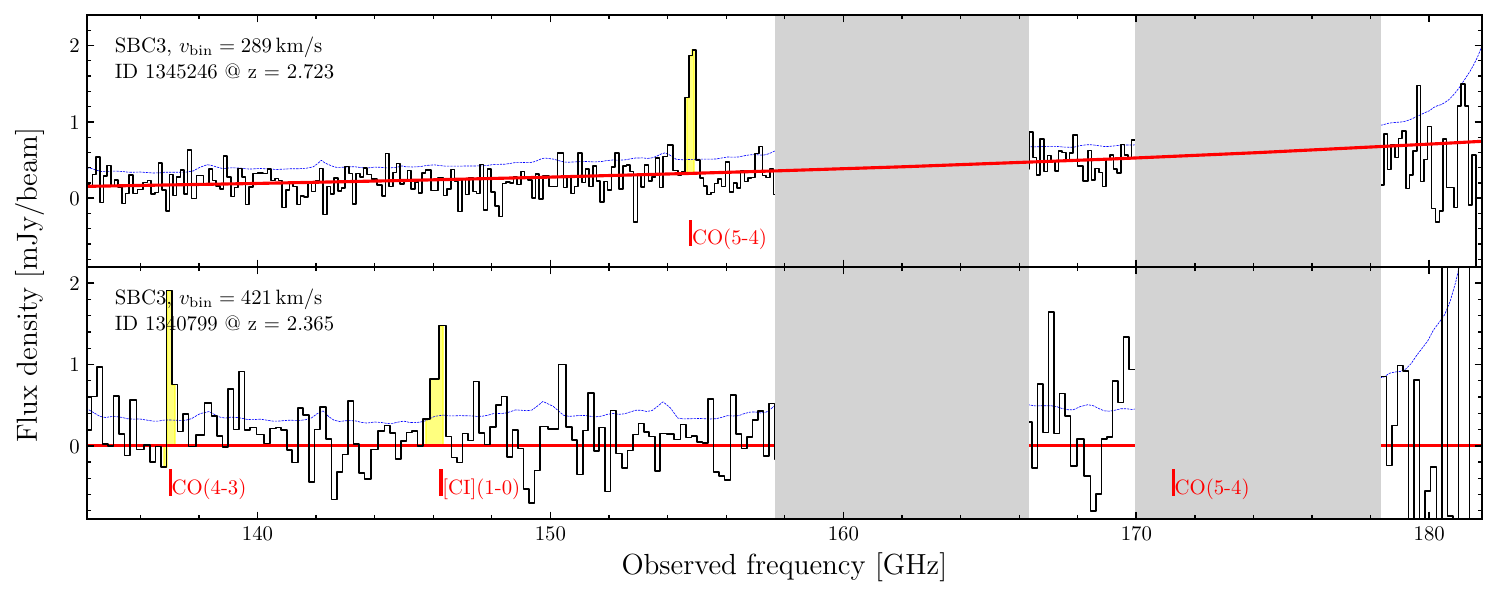}
     \caption{Colour images overlaid with ALMA Band 3 dust continuum and ALMA spectra of COS-SBC3.\ Photometrically selected galaxies with $2.03<z_{\rm phot}<2.70$ are marked by cyan circles. See the caption of \cref{fig:Sillassen-color-spectra}.}
     \label{fig:SBC3-color-spectra}
\end{figure*}

\clearpage

\begin{figure*}[!ht]
    \centering
     \includegraphics[width=\textwidth]{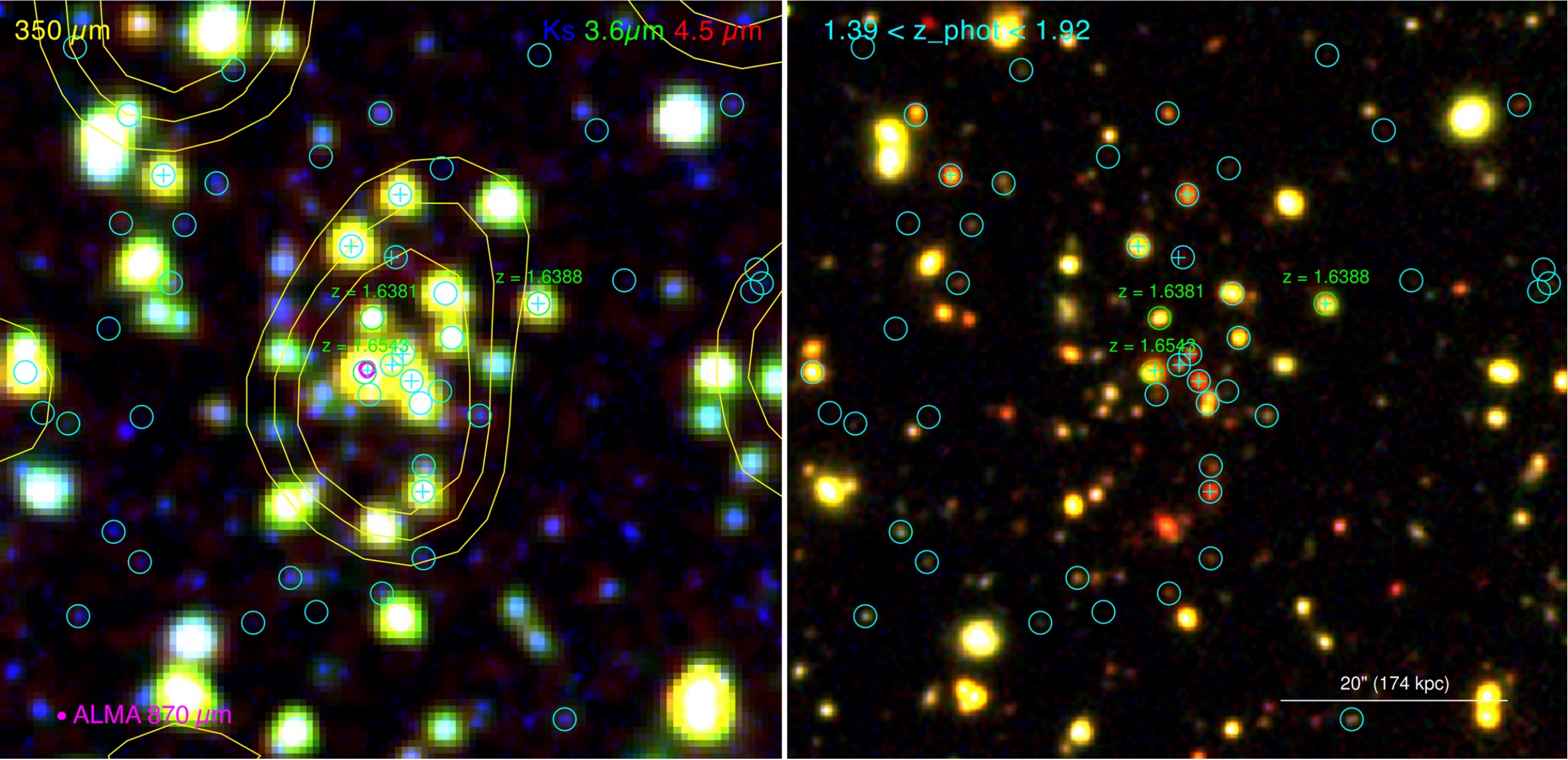}
     \caption{Colour images overlaid with ALMA Band 7 dust continuum of COS-SBC4.\ Photometrically selected galaxies with $1.39<z_{\rm phot}<1.92$ are marked with cyan circles. ALMA spectra are not shown, as no emission lines fall within the frequency coverage. See the caption of \cref{fig:Sillassen-color-spectra}.}
     \label{fig:SBC4-color-spectra}
\end{figure*}

\clearpage

\begin{figure*}[!ht]
    \centering
     \includegraphics[width=\textwidth]{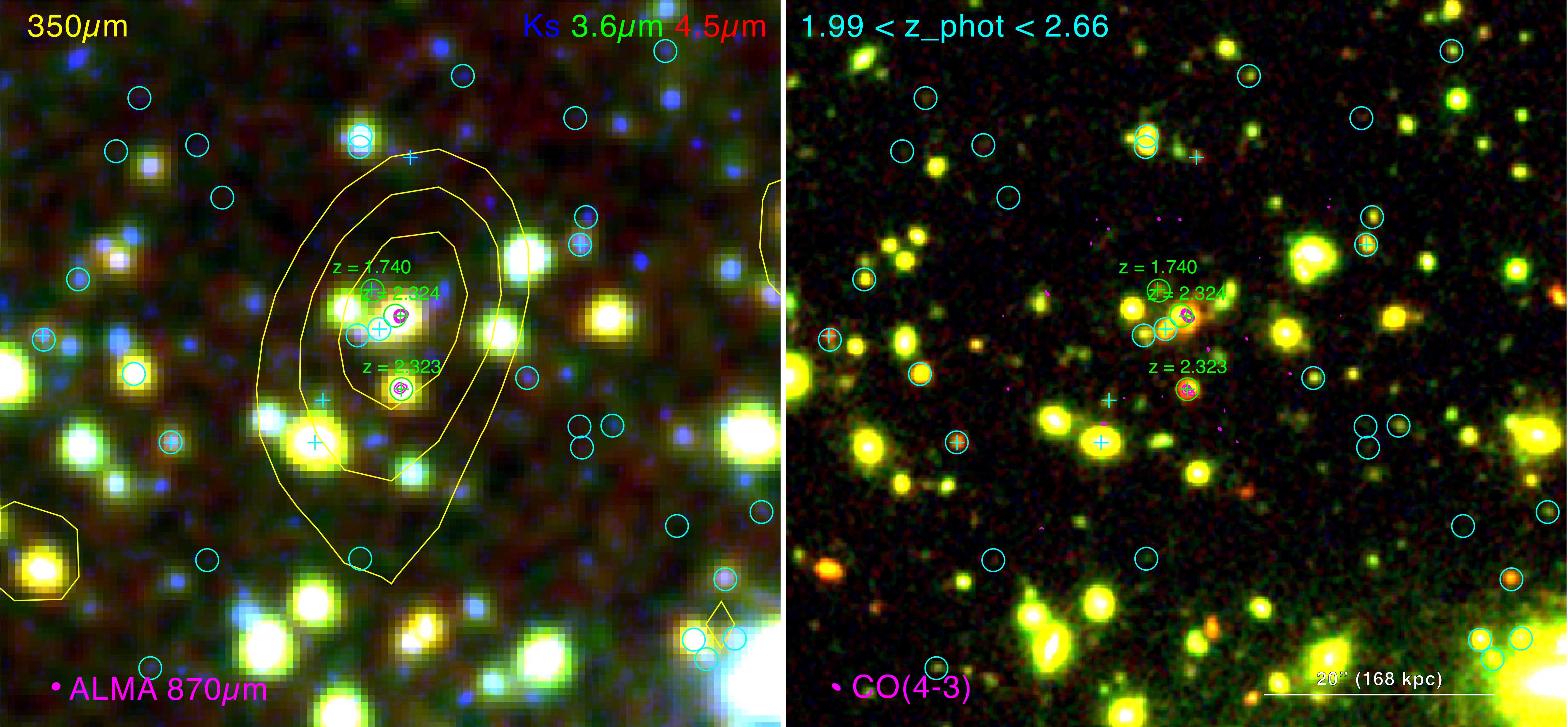}
     \includegraphics[width=\textwidth]{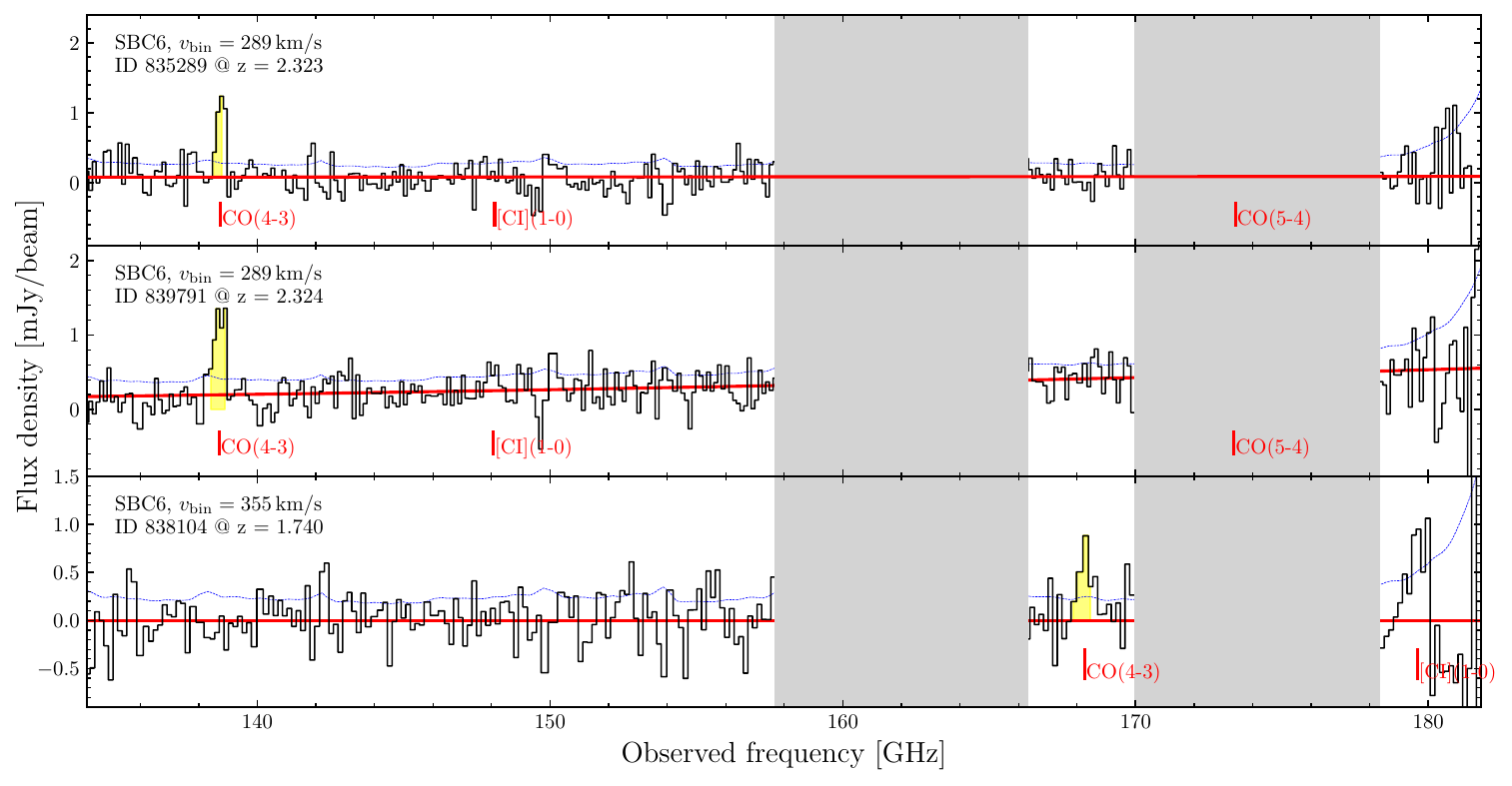}
     \caption{Colour images overlaid with ALMA Band 7 dust continuum and ALMA spectra of COS-SBC6.\ Photometrically selected galaxies with $1.99<z_{\rm phot}<2.66$ are marked with cyan circles. See the caption of \cref{fig:Sillassen-color-spectra}.}
     \label{fig:SBC6-color-spectra}
\end{figure*}

\clearpage

\twocolumn

\section{Multi-wavelength cutouts}
\begin{figure}[!htbp]
    \centering
    \includegraphics[width=\columnwidth]{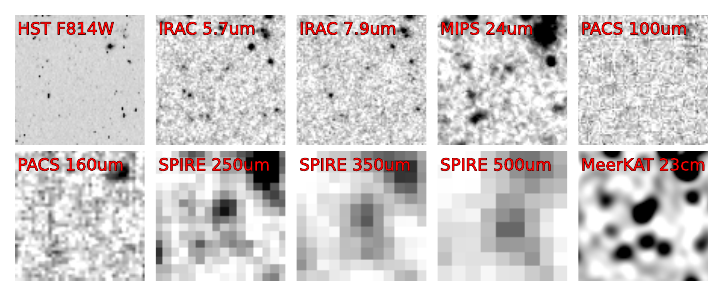}
    \caption{$90''\times90''$ size cutouts of HPC1001. }
    \label{fig:Sillassen_cutouts}
\end{figure}
\begin{figure}[!htbp]
    \centering
    \includegraphics[width=\columnwidth]{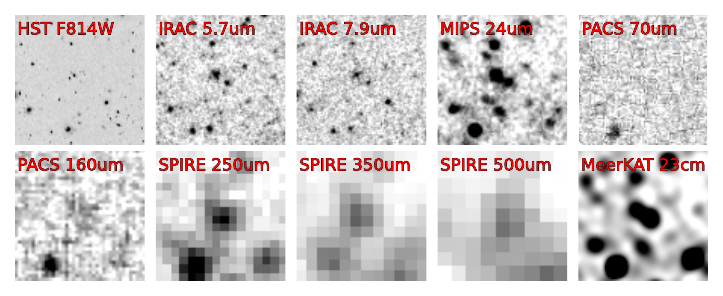}
    \caption{$90''\times90''$ size cutouts of COS-SBCX1.}
    \label{fig:SBCX1_cutouts}
\end{figure}
\begin{figure}[!htbp]
    \centering
    \includegraphics[width=\columnwidth]{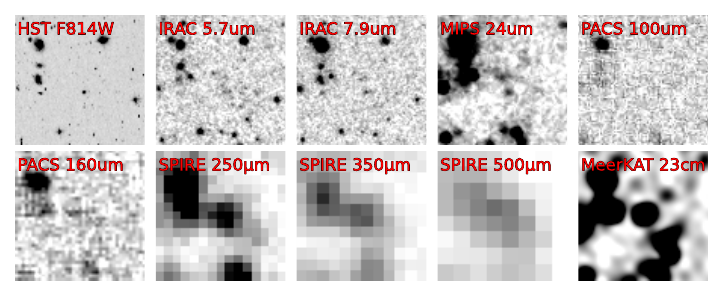}
    \caption{$90''\times90''$ size cutouts of COS-SBCX3.}
    \label{fig:SBCX3_cutouts}
\end{figure}
\begin{figure}[!htbp]
    \centering
    \includegraphics[width=\columnwidth]{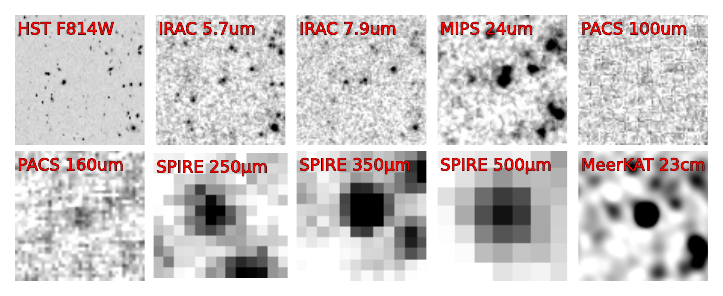}
    \caption{$90''\times90''$ size cutouts of COS-SBCX4.}
    \label{fig:SBCX4_cutouts}
\end{figure}
\begin{figure}[!htbp]
    \centering
    \includegraphics[width=\columnwidth]{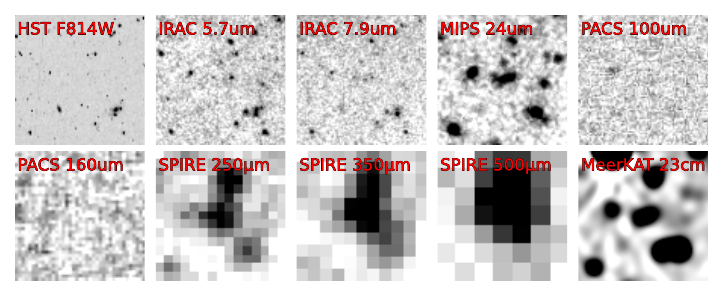}
    \caption{$90''\times90''$ size cutouts of COS-SBCX7.}
    \label{fig:SBCX7_cutouts}
\end{figure}
\begin{figure}[!htbp]
    \centering
    \includegraphics[width=\columnwidth]{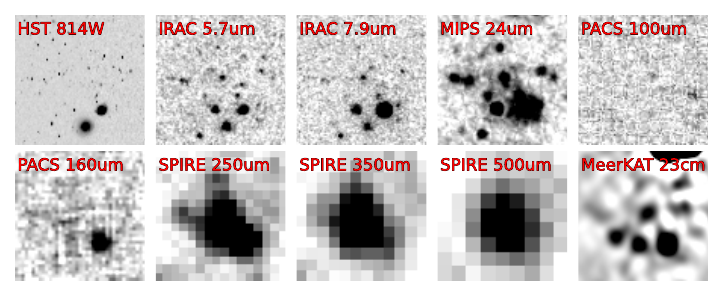}
    \caption{$90''\times90''$ size cutouts of COS-SBC3.}
    \label{fig:SBC3_cutouts}
\end{figure}
\begin{figure}[!htbp]
    \centering
    \includegraphics[width=\columnwidth]{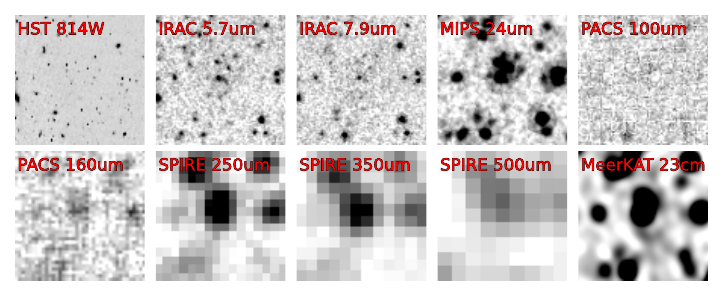}
    \caption{$90''\times90''$ size cutouts of COS-SBC4.}
    \label{fig:SBC4_cutouts}
\end{figure}
\begin{figure}[!htbp]
    \centering
    \includegraphics[width=\columnwidth]{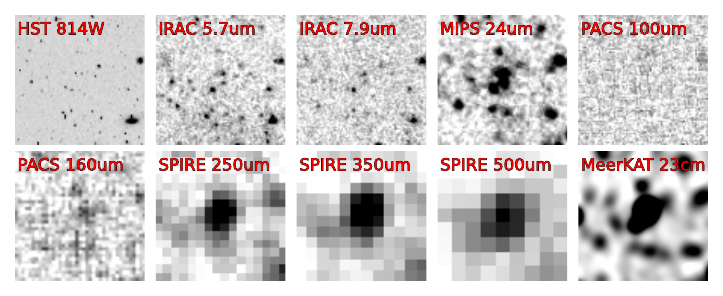}
    \caption{$90''\times90''$ size cutouts of COS-SBC6.}
    \label{fig:SBC6_cutouts}
\end{figure}


\onecolumn
\section{Physical properties of individual galaxies}
\label{sec:app-phys-prop}
\begin{table*}[!h]
    \centering
    \caption{Physical properties of HPC1001 members from COSMOS2020 Classic LePhare.}
    \renewcommand*{\arraystretch}{1.4}
    \label{tab:HPC1001_physpars}
    \begin{tabular}{c c c c c c c c c}
        \hline\hline
        Name & ID & RA & Dec. & $z_{\rm phot}$ & $z_{\rm spec}$ & $\log(M_\ast/{\rm M_{\odot}})$ & SFR & $d_{\rm core}$ \\
           & & [deg] & [deg] & & & & [${\rm M_\odot\,yr^{-1}}$] & [arcsec]\\
           \hline
        (Confirmed members)\\
        HPC1001.a & -- & 150.4659 & 2.6362 & -- & 3.613$^b$ & $10.3^{+0.2}_{-0.2}$ & $429^{+20}_{-20}$ & $1.5$ \\
        HPC1001.b & 1281317 & 150.4656 & 2.6360 & $3.48^{+0.10}_{-0.09}$ & 3.613$^b$ & $11.0^{+0.1}_{-0.1}$ & $152^{+30}_{-29}$ & $0.2$\\
        HPC1001.d & 1279885 & 150.4661 & 2.6361 & $3.66^{+0.14}_{-0.17}$ & 3.613$^b$ & $10.6^{+0.1}_{-0.1}$ & $81^{+15}_{-17}$ & $1.9$\\
        HPC1001.l$^a$ & 1274387 & 150.4647 & 2.6307 & $3.75^{+0.26}_{-0.16}$ & 3.604$\pm0.001$ & $10.3^{+0.1}_{-0.1}$ & $68^{+69}_{-20}$ & $19.1$ \\
        HPC1001.m$^a$ & 1272853 & 150.4618 & 2.6294 & $4.38^{+0.41}_{-0.35}$ & 3.610$\pm0.001$ & $11.3^{+0.1}_{-0.2}$ & $496^{+290}_{-241}$ & $27.3$ \\
        (Candidate members)\\
        HPC1001.c & 1279884 & 150.4652 & 2.6361 & $3.65^{+0.11}_{-0.15}$ & -- & $10.0^{+0.1}_{-0.1}$ & $18^{+15}_{-5}$ & $1.6$ \\
        HPC1001.e & 1280959 & 150.4652 & 2.6366 & $3.76^{+0.25}_{-0.18}$ & -- & $10.3^{+0.1}_{-0.1}$ & $57^{+37}_{-37}$ & $2.9$\\
        HPC1001.f & 1280265 & 150.4654 & 2.6370 & $3.82^{+0.29}_{-0.31}$ & -- & $10.0^{+0.1}_{-0.2}$ & $18^{+15}_{-6}$ & $3.9$\\
        HPC1001.g & 1278992 & 150.4650 & 2.6353 & $3.62^{+0.11}_{-0.13}$ & -- & $10.1^{+0.1}_{-0.1}$ & $18^{+4}_{-3}$ & 3.1 \\
        HPC1001.h$^c$ & -- & 150.4655 & 2.6356 & $3.79^{+0.37}_{-0.34}$ & -- & $10.0^{+0.2}_{-0.2}$ & $77^{+65}_{-42}$ & $1.1$ \\
        HPC1001.i & 1277620 & 150.4665 & 2.6344 & $3.26^{+0.53}_{-0.60}$ & -- & $9.9^{+0.1}_{-0.2}$ & $20^{+16}_{-14}$ & $6.2$ \\
        HPC1001.j & 1282867 & 150.4637 & 2.6384 & $3.76^{+0.15}_{-0.12}$ & -- & $10.7^{+0.1}_{-0.1}$ & $109^{+26}_{-79}$ & $11.2$ \\
        HPC1001.k$^a$ & 1280828 & 150.4673 & 2.6370 & $3.15^{+0.14}_{-0.11}$ & -- & $9.4^{+0.2}_{-0.1}$ & $14^{+4}_{-12}$ & 7.3\\
        &1271738 & 150.4664 & 2.6284 & $3.95^{+0.20}_{-0.26}$ & -- & $9.5^{+0.1}_{-0.2}$ & $11^{+6}_{-7}$ & $27.2$\\
        &1271880 & 150.4628 & 2.6283 & $3.64^{+0.11}_{-0.13}$ & -- & $9.6^{+0.1}_{-0.1}$ & $15^{+3}_{-2}$ & $29.2$\\
        & 1273801 & 150.4611 & 2.6305 & $3.69^{+0.23}_{-3.39}$ & -- & $8.4^{+0.2}_{-0.2}$ & $3^{+1}_{-2}$ & $25.2$\\
        & 1273919 & 150.4676 & 2.6300 & $3.84^{+0.24}_{-0.26}$ & -- & $10.0^{+0.1}_{-0.2}$ & $23^{+21}_{-6}$ & $22.5$ \\
        & 1274349 & 150.4646 & 2.6304 & $3.85^{+0.13}_{-0.13}$ & -- & $9.6^{+0.1}_{-0.1}$ & $15^{+2}_{-2}$ & $20.2$\\
        & 1275202 & 150.4629 & 2.6318 & $3.81^{+0.34}_{-0.17}$ & -- & $9.4^{+0.2}_{-0.1}$ & $16^{+4}_{-4}$ & $17.5$\\
        & 1278508 & 150.4686 & 2.6352 & $4.00^{+0.10}_{-0.11}$ & -- & $9.0^{+0.1}_{-0.2}$ & $4^{+4}_{-1}$ & $11.0$\\
        & 1278832 & 150.4713 & 2.6356 & $2.96^{+0.12}_{-0.12}$ & -- & $8.6^{+0.2}_{-0.2}$ & $4^{+1}_{-3}$ & $20.5$\\
        & 1282347 & 150.4695 & 2.6390 & $3.02^{+0.15}_{-0.15}$ & -- & $8.8^{+0.4}_{-0.4}$ & $1^{+1}_{-1}$ & $17.9$\\
        & 1282620 & 150.4688 & 2.6393 & $4.06^{+0.30}_{-0.37}$ & -- & $8.9^{+0.2}_{-0.2}$ & $4^{+1}_{-3}$ & $16.9$\\
        & 1286318 & 150.4700 & 2.6427 & $3.72^{+0.24}_{-0.18}$ & -- & $9.6^{+0.1}_{-0.2}$ & $15^{+15}_{-4}$ & $29.1$ \\
        & 1286772 & 150.4657 & 2.6426 & $3.04^{+0.06}_{-0.04}$ & -- & $9.6^{+0.1}_{-0.1}$ & $17^{+3}_{-3}$ & $24.0$\\
        & 1282703 & 150.4654 & 2.6393 & $0.45^{+3.23}_{-0.26}$ & -- & $7.2^{+0.2}_{-0.2}$ & $<1$ & 12.1 \\
        \hline
        Total & -- & 150.4656 & 2.6359 & $3.65\pm0.07$ & $3.613\pm0.001$ & $11.6\pm0.1$ & $1194^{+306}_{-260}$ & -- \\
        \hline
    \end{tabular}
{\\ Notes: $^a$Newly discovered members, not previously reported in \citet{Sillassen2022_HPC1001}. $^b$Sources are completely blended in the $4.5"$ NOEMA beam (\cref{fig:Sillassen-color-spectra}), with a strong (S/N=10.9) and broad (Width=1824${\rm km/s}$) emission line (\cref{tab:emission-lines}). $^c$Source is not in the COSMOS2020 catalogue, parameters adopted from COSMOS2015 \citep{Laigle2016,Sillassen2022_HPC1001}.}
\end{table*}

\clearpage

\onecolumn
    \renewcommand*{\arraystretch}{1.4}
    \begin{longtable}{c c c c c c c c c}
    \caption{\label{tab:SBCX1_physpars}Physical properties of COS-SBCX1 members from COSMOS2020 Classic LePhare.}\\
    
        \\\hline\hline
        ID & RA & Dec. & $z_{\rm phot}$ & $z_{\rm spec}$ & $\log(M_\ast/{\rm M_{\odot}})$ & SFR & $d_{\rm core}$ \\
           & [deg] & [deg] & & & & [${\rm M_\odot\,yr^{-1}}$] & [arcsec]\\
           \hline
           (Confirmed members)\\
           1408110 & 150.3480 & 2.7611 & $2.38^{+0.12}_{-0.17}$ & 2.422$\pm0.001$ & $11.4^{+0.1}_{-0.1}$ & $156^{+30}_{-31}$ & 4.9\\
           (Candidate members)\\
        --* & 150.3497 & 2.7619 & -- & -- & $10.8^{+0.2}_{-0.2}$ & -- & 1.9\\
        1398871 & 150.3534153 & 2.7532749 & $2.18^{+0.85}_{-0.64}$ & -- & $9.1^{+0.2}_{-0.2}$ & $3^{+2}_{-2}$ & 34.3 \\
        1399225 & 150.3483456 & 2.7530453 & $2.76^{+0.09}_{-0.11}$ & -- & $9.7^{+0.1}_{-0.2}$ & $17^{+6}_{-14}$ & 29.1 \\
        1399474 & 150.3479133 & 2.7530239 & $2.45^{+0.09}_{-0.08}$ & -- & $9.3^{+0.1}_{-0.1}$ & $15^{+2}_{-2}$ & 29.1 \\
        1399745 & 150.3487553 & 2.7523927 & $2.50^{+0.03}_{-0.06}$ & -- & $9.9^{+0.1}_{-0.1}$ & $62^{+10}_{-10}$ & 31.5 \\
        1399984 & 150.3541567 & 2.7542877 & $2.12^{+0.47}_{-0.91}$ & -- & $8.5^{+0.5}_{-0.4}$ & $1^{+0}_{-0}$ & 33.0 \\
        1400213 & 150.3415066 & 2.7543926 & $2.71^{+0.12}_{-0.17}$ & -- & $8.8^{+0.2}_{-0.2}$ & $2^{+2}_{-0}$ & 33.7 \\
        1400229 & 150.3470434 & 2.7545470 & $2.15^{+0.47}_{-1.62}$ & -- & $8.1^{+0.2}_{-0.2}$ & $1^{+0}_{-1}$ & 23.9 \\
        1400635 & 150.3417414 & 2.7547666 & $2.61^{+0.09}_{-0.11}$ & -- & $9.7^{+0.1}_{-0.1}$ & $22^{+5}_{-18}$ & 32.2 \\
        1401156 & 150.3540867 & 2.7554773 & $1.98^{+0.34}_{-0.25}$ & -- & $9.0^{+0.1}_{-0.2}$ & $1^{+1}_{-0}$ & 29.8 \\
        1401256 & 150.3502810 & 2.7545819 & $2.76^{+0.07}_{-0.07}$ & -- & $9.5^{+0.1}_{-0.2}$ & $34^{+6}_{-7}$ & 24.9 \\
        1401272 & 150.3395136 & 2.7556119 & $2.19^{+0.41}_{-0.44}$ & -- & $8.8^{+0.1}_{-0.2}$ & $1^{+1}_{-0}$ & 36.5 \\
        1401354 & 150.3515443 & 2.7550129 & $2.63^{+0.13}_{-0.16}$ & -- & $9.5^{+0.1}_{-0.2}$ & $10^{+9}_{-2}$ & 25.4 \\
        1401671 & 150.3472055 & 2.7559690 & $2.59^{+0.38}_{-0.52}$ & -- & $9.1^{+0.2}_{-0.2}$ & $3^{+1}_{-2}$ & 18.8 \\
        1401759 & 150.3431070 & 2.7559796 & $2.40^{+0.23}_{-0.43}$ & -- & $8.8^{+0.1}_{-0.1}$ & $3^{+0}_{-1}$ & 25.6 \\
        1401866 & 150.3403186 & 2.7561221 & $2.24^{+0.19}_{-0.18}$ & -- & $9.5^{+0.1}_{-0.1}$ & $1^{+0}_{-0}$ & 33.1 \\
        1401922 & 150.3439038 & 2.7551156 & $2.09^{+0.09}_{-0.06}$ & -- & $9.6^{+0.1}_{-0.2}$ & $33^{+7}_{-27}$ & 26.2 \\
        1402490 & 150.3503583 & 2.7558268 & $2.17^{+0.19}_{-0.16}$ & -- & $10.2^{+0.2}_{-0.2}$ & $62^{+51}_{-64}$ & 20.8 \\
        1402700 & 150.3389536 & 2.7568049 & $1.99^{+0.22}_{-0.22}$ & -- & $8.7^{+0.2}_{-0.2}$ & $2^{+0}_{-1}$ & 36.2 \\
        1402958 & 150.3452583 & 2.7569484 & $2.77^{+0.21}_{-0.28}$ & -- & $8.8^{+0.2}_{-0.1}$ & $2^{+0}_{-2}$ & 18.0 \\
        1403375 & 150.3393030 & 2.7569974 & $2.33^{+0.11}_{-0.11}$ & -- & $9.3^{+0.1}_{-0.1}$ & $7^{+1}_{-1}$ & 34.7 \\
        1404385 & 150.3465587 & 2.7584031 & $2.49^{+0.13}_{-0.17}$ & -- & $8.7^{+0.1}_{-0.1}$ & $1^{+0}_{-0}$ & 11.1 \\
        1404606 & 150.3524059 & 2.7583128 & $2.68^{+0.15}_{-0.22}$ & -- & $9.5^{+0.1}_{-0.1}$ & $18^{+4}_{-6}$ & 18.7 \\
        1404729 & 150.3519759 & 2.7575986 & $2.30^{+0.08}_{-0.15}$ & -- & $10.7^{+0.1}_{-0.1}$ & $217^{+39}_{-43}$ & 19.1 \\
        1404945 & 150.3447842 & 2.7582966 & $2.64^{+0.56}_{-1.78}$ & -- & $9.3^{+0.1}_{-0.2}$ & $12^{+10}_{-3}$ & 15.5 \\
        1405016 & 150.3467285 & 2.7591544 & $2.76^{+0.24}_{-0.30}$ & -- & $10.1^{+0.1}_{-0.1}$ & $9^{+7}_{-3}$ & 8.5 \\
        1405330 & 150.3412585 & 2.7584464 & $2.66^{+0.06}_{-0.14}$ & -- & $10.0^{+0.1}_{-0.1}$ & $58^{+11}_{-18}$ & 26.2 \\
        1405903 & 150.3473917 & 2.7600336 & $2.33^{+0.37}_{-0.38}$ & -- & $8.9^{+0.2}_{-0.2}$ & $3^{+1}_{-2}$ & 4.5 \\
        1406026 & 150.3419379 & 2.7597845 & $2.01^{+0.24}_{-0.17}$ & -- & $9.6^{+0.1}_{-0.2}$ & $13^{+4}_{-13}$ & 22.4 \\
        1406132 & 150.3487067 & 2.7592945 & $2.31^{+0.27}_{-0.30}$ & -- & $10.7^{+0.0}_{-0.0}$ & $0^{+0}_{-0}$ & 7.0 \\
        1406593 & 150.3444680 & 2.7606627 & $2.33^{+0.37}_{-0.46}$ & -- & $8.9^{+0.1}_{-0.2}$ & $1^{+1}_{-0}$ & 12.9 \\
        1406826 & 150.3496448 & 2.7606866 & $2.41^{+0.50}_{-0.34}$ & -- & $10.2^{+0.1}_{-0.2}$ & $40^{+39}_{-24}$ & 6.0 \\
        1406877 & 150.3437063 & 2.7608204 & $2.59^{+0.21}_{-0.31}$ & -- & $9.1^{+0.1}_{-0.1}$ & $2^{+0}_{-0}$ & 15.6 \\
        1407283 & 150.3481033 & 2.7607376 & $1.88^{+0.12}_{-0.10}$ & -- & $10.6^{+0.1}_{-0.1}$ & $14^{+9}_{-3}$ & 1.4 \\
        1407341 & 150.3534620 & 2.7613525 & $2.64^{+0.20}_{-0.25}$ & -- & $9.3^{+0.1}_{-0.2}$ & $6^{+3}_{-4}$ & 19.6 \\
        1407342 & 150.3397290 & 2.7614006 & $2.41^{+0.58}_{-1.79}$ & -- & $8.5^{+0.2}_{-0.2}$ & $1^{+1}_{-0}$ & 29.9 \\
        1407673 & 150.3477238 & 2.7616945 & $2.62^{+0.30}_{-0.35}$ & -- & $9.5^{+0.1}_{-0.2}$ & $6^{+5}_{-2}$ & 2.3 \\
        1407983 & 150.3500725 & 2.7614637 & $2.37^{+0.66}_{-0.75}$ & -- & $9.5^{+0.2}_{-0.2}$ & $11^{+6}_{-9}$ & 7.5 \\
        1408029 & 150.3475805 & 2.7619215 & $2.54^{+0.63}_{-0.44}$ & -- & $10.2^{+0.1}_{-0.2}$ & $40^{+32}_{-31}$ & 3.3 \\
        1408242 & 150.3542594 & 2.7620097 & $2.70^{+0.29}_{-0.35}$ & -- & $10.1^{+0.1}_{-0.2}$ & $29^{+23}_{-13}$ & 22.7 \\
        1408284 & 150.3480351 & 2.7617402 & $2.66^{+0.21}_{-0.29}$ & -- & $9.9^{+0.1}_{-0.2}$ & $15^{+12}_{-5}$ & 2.2 \\
        1408512 & 150.3457383 & 2.7623290 & $2.68^{+0.07}_{-0.09}$ & -- & $8.9^{+0.1}_{-0.1}$ & $7^{+1}_{-1}$ & 9.3 \\
        1408732 & 150.3522480 & 2.7618868 & $2.07^{+0.24}_{-0.17}$ & -- & $9.8^{+0.2}_{-0.2}$ & $28^{+7}_{-29}$ & 15.5 \\
        1409126 & 150.3457790 & 2.7631643 & $2.47^{+0.35}_{-0.46}$ & -- & $8.4^{+0.2}_{-0.2}$ & $1^{+0}_{-1}$ & 10.9 \\
        1409304 & 150.3524976 & 2.7633564 & $2.25^{+0.39}_{-0.40}$ & -- & $9.4^{+0.1}_{-0.2}$ & $3^{+2}_{-1}$ & 18.0 \\
        1409685 & 150.3470603 & 2.7637415 & $2.63^{+0.21}_{-0.32}$ & -- & $8.9^{+0.1}_{-0.1}$ & $3^{+0}_{-2}$ & 10.1 \\
        1409707 & 150.3496428 & 2.7630066 & $2.15^{+0.21}_{-0.23}$ & -- & $11.1^{+0.1}_{-0.1}$ & $125^{+20}_{-20}$ & 9.0 \\
        1409806 & 150.3532777 & 2.7638843 & $2.74^{+0.48}_{-2.28}$ & -- & $8.8^{+0.5}_{-0.5}$ & $1^{+1}_{-0}$ & 21.4 \\
        1410423 & 150.3554314 & 2.7638052 & $2.52^{+0.07}_{-0.05}$ & -- & $9.6^{+0.1}_{-0.1}$ & $14^{+11}_{-3}$ & 28.4 \\
        1410451 & 150.3502562 & 2.7637331 & $2.01^{+0.40}_{-0.35}$ & -- & $10.0^{+0.1}_{-0.1}$ & $19^{+11}_{-6}$ & 12.4 \\
        1410521 & 150.3419140 & 2.7645586 & $2.30^{+0.23}_{-0.38}$ & -- & $8.5^{+0.1}_{-0.2}$ & $2^{+0}_{-0}$ & 25.2 \\
        1412364 & 150.3508360 & 2.7661634 & $2.52^{+0.48}_{-0.45}$ & -- & $10.9^{+0.1}_{-0.1}$ & $108^{+72}_{-54}$ & 20.8 \\
        1412370 & 150.3461119 & 2.7664528 & $2.43^{+0.24}_{-1.19}$ & -- & $8.7^{+0.1}_{-0.1}$ & $2^{+0}_{-1}$ & 20.4 \\
        1412651 & 150.3538193 & 2.7667801 & $2.65^{+0.27}_{-0.41}$ & -- & $8.7^{+0.1}_{-0.1}$ & $2^{+0}_{-0}$ & 29.2 \\
        1412702 & 150.3517594 & 2.7665658 & $2.63^{+0.15}_{-0.19}$ & -- & $9.2^{+0.2}_{-0.1}$ & $11^{+3}_{-9}$ & 23.8 \\
        1412826 & 150.3524741 & 2.7663713 & $2.41^{+0.26}_{-0.24}$ & -- & $11.0^{+0.1}_{-0.1}$ & $144^{+116}_{-108}$ & 24.8 \\
        1413261 & 150.3450722 & 2.7673326 & $2.02^{+0.44}_{-0.29}$ & -- & $9.0^{+0.1}_{-0.2}$ & $1^{+1}_{-0}$ & 24.8 \\
        1413740 & 150.3536237 & 2.7679074 & $2.39^{+0.32}_{-1.21}$ & -- & $8.8^{+0.1}_{-0.1}$ & $1^{+0}_{-0}$ & 31.7 \\
        1415067 & 150.3523728 & 2.7691123 & $2.69^{+0.20}_{-0.34}$ & -- & $8.5^{+0.1}_{-0.1}$ & $3^{+0}_{-0}$ & 32.8 \\
        1415067 & 150.3523728 & 2.7691123 & $2.69^{+0.20}_{-0.34}$ & -- & $8.5^{+0.1}_{-0.1}$ & $3^{+0}_{-0}$ & 32.8 \\
        \hline
        Total & 150.3492 & 2.7619 & $2.48\pm0.03$ & $2.422\pm0.001$ & $12.0\pm0.1$ & $1438^{+170}_{-165}$ & --\\
\end{longtable}
Notes:*source not in the COSMOS2020 catalogue or in previous versions of COSMOS catalogues.

\clearpage
\twocolumn
\begin{table*}[!htbp]
    \centering
    \caption{Physical properties of COS-SBCX3 members from COSMOS2020 Classic LePhare.}
    \renewcommand*{\arraystretch}{1.4}
    \label{tab:SBCX3_physpars}
    \begin{tabular}{c c c c c c c c}
        \hline\hline
        ID & RA & Dec. & $z_{\rm phot}$ & $z_{\rm spec}$ & $\log(M_\ast/{\rm M_{\odot}})$ & SFR & $d_{\rm core}$ \\
           & [deg] & [deg] & & & & [${\rm M_\odot\,yr^{-1}}$] & [arcsec]\\
           \hline
           (Confirmed members)\\
           1088787 & 150.3105 & 2.4515 & $2.98^{+0.21}_{-0.20}$ & 3.030$\pm0.001$ & $10.7^{+0.1}_{-0.1}$ & $53^{+40}_{-12}$ & {2.7} \\
           1088927 & 150.3117 & 2.4510 & $3.15^{+0.28}_{-0.67}$ & 3.032$\pm0.001$ & $10.5^{+0.1}_{-0.2}$ & $100^{+45}_{-90}$ & {1.8} \\
           (Candidate members)\\
            1088602 & 150.3144 & 2.4515 & $2.71^{+0.13}_{-0.21}$ & -- & $8.5^{+0.2}_{-0.2}$ & $1^{+1}_{-0}$ & {11.5} \\
            1088952 & 150.3092 & 2.4516 & $2.77^{+0.12}_{-0.15}$ & -- & $8.6^{+0.1}_{-0.2}$ & $1^{+1}_{-0}$ & {7.3} \\
            1089223 & 150.3112 & 2.4511 & $1.94^{+0.48}_{-0.43}$ & -- & $9.9^{+0.1}_{-0.2}$ & $15^{+12}_{-10}$ & {0.5} \\
            1089223 & 150.3112 & 2.4511 & $1.94^{+0.48}_{-0.43}$ & -- & $9.9^{+0.1}_{-0.2}$ & $15^{+12}_{-10}$ & {0.5} \\
            1089511 & 150.3155 & 2.4522 & $3.09^{+0.14}_{-0.12}$ & -- & $8.4^{+0.1}_{-0.2}$ & $1^{+1}_{-0}$ & {15.8} \\
            1089759 & 150.3105 & 2.4518 & $3.12^{+0.11}_{-0.12}$ & -- & $10.1^{+0.1}_{-0.1}$ & $13^{+10}_{-4}$ & {3.3} \\
            1090209 & 150.3099 & 2.4529 & $3.01^{+0.27}_{-0.39}$ & -- & $9.3^{+0.1}_{-0.2}$ & $3^{+2}_{-0}$ & {8.0} \\
            1090976 & 150.3119 & 2.4537 & $2.74^{+0.11}_{-0.13}$ & -- & $9.1^{+0.1}_{-0.1}$ & $7^{+1}_{-1}$ & {9.4} \\
        \hline
        Total & 150.3113 & 2.4511 & $2.89\pm0.08$ & $3.031\pm0.001$ & $11.1\pm0.1$ & $213^{+63}_{-92}$ & -- \\
        \hline
    \end{tabular}
\end{table*}

\clearpage

\begin{table*}[!htbp]
    \centering
    \caption{Physical properties of COS-SBCX4 members from COSMOS2020 Classic LePhare.}
    \renewcommand*{\arraystretch}{1.4}
    \label{tab:SBCX4_physpars}
    \begin{tabular}{c c c c c c c c}
        \hline\hline
        ID & RA & Dec. & $z_{\rm phot}$ & $z_{\rm spec}$ & $\log(M_\ast/{\rm M_{\odot}})$ & SFR & $d_{\rm core}$ \\
           & [deg] & [deg] & & & & [${\rm M_\odot\,yr^{-1}}$] & [arcsec]\\
           \hline Core members \\
           \hline
           (Confirmed members)\\
           1049510 & 150.7512 & 2.4124 & $2.93^{+0.35}_{-0.38}$ & 2.642$\pm0.001$ & $11.1^{+0.1}_{-0.1}$ & $257^{+83}_{-186}$ & 2.2 \\
           1049929 & 150.7521 & 2.4140 & $3.85^{+0.66}_{-0.92}$ & 2.648$\pm0.001$ & $10.9^{+0.2}_{-0.2}$ & $195^{+120}_{-101}$ & 5.5 \\
           1050531 & 150.7504 & 2.4130 & $2.67^{+0.08}_{-0.12}$ & 2.647$\pm0.001$ & $11.2^{+0.1}_{-0.1}$ & $416^{+135}_{-265}$ & 1.8 \\
           (Candidate members)\\
        1042545 & 150.7528 & 2.4067 & $2.35^{+0.16}_{-1.13}$ & -- & $8.6^{+0.1}_{-0.1}$ & $7^{+1}_{-1}$ & 23.4 \\
        1042632 & 150.7517 & 2.4069 & $2.50^{+0.43}_{-0.80}$ & -- & $8.5^{+0.2}_{-0.2}$ & $1^{+0}_{-1}$ & 21.7 \\
        1044106 & 150.7463 & 2.4083 & $2.53^{+0.21}_{-0.26}$ & -- & $9.1^{+0.1}_{-0.1}$ & $3^{+0}_{-1}$ & 23.3 \\
        1044704 & 150.7447 & 2.4088 & $2.54^{+0.21}_{-0.31}$ & -- & $9.2^{+0.1}_{-0.1}$ & $3^{+0}_{-0}$ & 26.6 \\
        1045961 & 150.7463 & 2.4097 & $2.42^{+0.22}_{-0.63}$ & -- & $8.8^{+0.1}_{-0.1}$ & $6^{+1}_{-1}$ & 20.0 \\
        1046616 & 150.7468 & 2.4107 & $2.72^{+0.46}_{-2.53}$ & -- & $8.7^{+0.4}_{-0.4}$ & $2^{+1}_{-1}$ & 16.6 \\
        1047063 & 150.7557 & 2.4108 & $3.03^{+0.10}_{-0.07}$ & -- & $9.0^{+0.2}_{-0.2}$ & $6^{+6}_{-1}$ & 19.3 \\
        1048929 & 150.7519 & 2.4124 & $2.74^{+0.15}_{-0.14}$ & -- & $9.5^{+0.1}_{-0.1}$ & $8^{+7}_{-1}$ & 4.5 \\
        1049031 & 150.7508 & 2.4131 & $2.74^{+0.17}_{-0.13}$ & -- & $10.4^{+0.1}_{-0.1}$ & $24^{+4}_{-4}$ & 0.8 \\
        1049700 & 150.7550 & 2.4138 & $2.71^{+0.25}_{-0.41}$ & -- & $8.6^{+0.2}_{-0.2}$ & $2^{+0}_{-2}$ & 15.4 \\
        1049920 & 150.7499 & 2.4134 & $2.65^{+0.25}_{-0.19}$ & -- & $10.6^{+0.1}_{-0.1}$ & $41^{+17}_{-24}$ & 3.8 \\
        1049965 & 150.7511 & 2.4141 & $2.66^{+0.44}_{-1.34}$ & -- & $8.7^{+0.1}_{-0.2}$ & $2^{+0}_{-0}$ & 4.3 \\
        1050218 & 150.7547 & 2.4143 & $3.00^{+0.35}_{-0.40}$ & -- & $9.5^{+0.1}_{-0.2}$ & $7^{+6}_{-2}$ & 15.0 \\
        1050855 & 150.7533 & 2.4143 & $3.18^{+0.07}_{-0.12}$ & -- & $9.7^{+0.1}_{-0.1}$ & $15^{+2}_{-2}$ & 10.4 \\
        1050901 & 150.7528 & 2.4150 & $2.00^{+0.56}_{-0.47}$ & -- & $9.1^{+0.1}_{-0.2}$ & $2^{+1}_{-1}$ & 10.2 \\
        1051433 & 150.7561 & 2.4155 & $2.30^{+0.76}_{-0.81}$ & -- & $8.8^{+0.2}_{-0.2}$ & $1^{+1}_{-1}$ & 21.3 \\
        1053106 & 150.7552 & 2.4169 & $2.73^{+0.26}_{-0.49}$ & -- & $9.1^{+0.1}_{-0.1}$ & $4^{+3}_{-1}$ & 21.5 \\
        1053892 & 150.7502 & 2.4173 & $1.95^{+0.33}_{-0.19}$ & -- & $9.2^{+0.1}_{-0.2}$ & $4^{+4}_{-1}$ & 16.1 \\
        1055135 & 150.7515 & 2.4190 & $2.57^{+0.53}_{-2.20}$ & -- & $8.4^{+0.2}_{-0.2}$ & $1^{+0}_{-1}$ & 22.2 \\
        
    \hline
        Total & 150.7509 & 2.4132 & $2.85\pm0.09$ & $2.646\pm0.001$ & $11.7\pm0.1$ & $1017^{+200}_{-341}$ & --\\
    \hline
    \end{tabular}
\end{table*}

\clearpage

\begin{table*}[!htbp]
    \centering
    \caption{Physical properties of COS-SBCX7 members from COSMOS2020 Classic LePhare.}
    \renewcommand*{\arraystretch}{1.4}
    \label{tab:SBCX7_physpars}
    \begin{tabular}{c c c c c c c c}
        \hline\hline
        ID & RA & Dec. & $z_{\rm phot}$ & $z_{\rm spec}$ & $\log(M_\ast/{\rm M_{\odot}})$ & SFR & $d_{\rm core}$ \\
           & [deg] & [deg] & & & & [${\rm M_\odot\,yr^{-1}}$] & [arcsec]\\
           \hline
        (Confirmed memebers)\\
        392257 & 149.9910 & 1.7967 & $2.91^{+0.63}_{-0.58}$ & 2.415$\pm0.001$ & $10.7^{+0.1}_{-0.1}$ & $91^{+61}_{-49}$ & 6.0\\
        392639 & 149.9816 & 1.7960 & $2.70^{+0.04}_{-0.07}$ & 2.413$\pm0.001$ & $10.3^{+0.1}_{-0.1}$ & $114^{+25}_{-112}$ & 30.2\\
        394609 & 149.9897 & 1.7977 & $2.36^{+0.13}_{-0.14}$ & 2.416$\pm0.001$ & $10.8^{+0.1}_{-0.1}$ & $14^{+11}_{-4}$ & 0.6\\
        394944 & 149.9883 & 1.7980 & $2.38^{+0.15}_{-0.14}$ & 2.416$\pm0.001$ & $11.1^{+0.1}_{-0.1}$ & $92^{+70}_{-25}$ & 5.5\\
        (Candidate members)\\
        384203 & 149.9855 & 1.7889 & $1.83^{+0.17}_{-0.16}$ & -- & $8.8^{+0.1}_{-0.2}$ & $0^{+0}_{-0}$ & 33.8 \\
        385317 & 149.9929 & 1.7900 & $2.23^{+0.27}_{-0.29}$ & -- & $9.5^{+0.1}_{-0.1}$ & $2^{+0}_{-0}$ & 30.8 \\
        385439 & 149.9936 & 1.7900 & $2.13^{+0.18}_{-0.22}$ & -- & $8.9^{+0.1}_{-0.1}$ & $3^{+0}_{-0}$ & 32.1 \\
        385916 & 149.9957 & 1.7907 & $2.01^{+0.30}_{-0.26}$ & -- & $8.6^{+0.1}_{-0.2}$ & $1^{+0}_{-0}$ & 34.8 \\
        386678 & 149.9966 & 1.7911 & $1.99^{+0.13}_{-0.13}$ & -- & $9.2^{+0.1}_{-0.1}$ & $5^{+0}_{-0}$ & 35.9 \\
        389851 & 149.9848 & 1.7938 & $2.73^{+0.07}_{-0.07}$ & -- & $9.4^{+0.2}_{-0.1}$ & $23^{+5}_{-20}$ & 20.6 \\
        390151 & 149.9969 & 1.7944 & $2.75^{+0.04}_{-0.05}$ & -- & $8.6^{+0.2}_{-0.2}$ & $5^{+1}_{-1}$ & 30.7 \\
        390263 & 149.9957 & 1.7947 & $2.48^{+0.13}_{-0.17}$ & -- & $9.7^{+0.1}_{-0.1}$ & $13^{+2}_{-2}$ & 26.4 \\
        390843 & 149.9926 & 1.7944 & $2.30^{+0.18}_{-0.12}$ & -- & $9.9^{+0.1}_{-0.1}$ & $19^{+15}_{-4}$ & 17.4 \\
        391472 & 149.9889 & 1.7953 & $2.41^{+0.12}_{-0.10}$ & -- & $9.5^{+0.1}_{-0.2}$ & $9^{+8}_{-3}$ & 8.4 \\
        392030 & 149.9945 & 1.7965 & $2.31^{+0.28}_{-0.36}$ & -- & $8.6^{+0.2}_{-0.2}$ & $1^{+0}_{-1}$ & 20.1 \\
        392370 & 149.9939 & 1.7968 & $2.59^{+0.24}_{-0.31}$ & -- & $8.5^{+0.1}_{-0.1}$ & $2^{+0}_{-0}$ & 17.8 \\
        393623 & 149.9822 & 1.7968 & $2.22^{+0.05}_{-0.06}$ & -- & $10.8^{+0.1}_{-0.1}$ & $289^{+59}_{-73}$ & 24.6 \\
        393835 & 149.9879 & 1.7982 & $2.24^{+0.48}_{-2.15}$ & -- & $8.1^{+0.2}_{-0.2}$ & $1^{+0}_{-0}$ & 4.7 \\
        394391 & 149.9917 & 1.7981 & $2.85^{+0.19}_{-0.29}$ & -- & $11.1^{+0.1}_{-0.1}$ & $245^{+193}_{-76}$ & 9.8 \\
        396033 & 149.9907 & 1.8003 & $2.64^{+0.20}_{-0.23}$ & -- & $9.1^{+0.1}_{-0.2}$ & $3^{+2}_{-0}$ & 11.2 \\
        396633 & 149.9834 & 1.8002 & $2.08^{+0.04}_{-0.05}$ & -- & $8.8^{+0.1}_{-0.1}$ & $14^{+3}_{-3}$ & 22.2 \\
        397066 & 149.9847 & 1.8012 & $2.48^{+0.49}_{-0.61}$ & -- & $8.7^{+0.2}_{-0.2}$ & $2^{+0}_{-1}$ & 20.3 \\
        397363 & 149.9816 & 1.8016 & $2.71^{+0.18}_{-0.20}$ & -- & $9.7^{+0.1}_{-0.1}$ & $5^{+4}_{-1}$ & 30.2 \\
        397795 & 149.9903 & 1.8010 & $2.79^{+0.23}_{-0.20}$ & -- & $10.9^{+0.1}_{-0.1}$ & $132^{+93}_{-33}$ & 13.0 \\
        397953 & 149.9958 & 1.8021 & $2.51^{+0.15}_{-0.17}$ & -- & $9.1^{+0.1}_{-0.1}$ & $4^{+0}_{-0}$ & 29.2 \\
        398018 & 149.9815 & 1.8022 & $2.52^{+0.38}_{-0.56}$ & -- & $9.5^{+0.1}_{-0.2}$ & $4^{+4}_{-1}$ & 31.5 \\
        398599 & 149.9917 & 1.8027 & $2.83^{+0.27}_{-0.55}$ & -- & $9.1^{+0.2}_{-0.2}$ & $5^{+2}_{-3}$ & 20.7 \\
        398634 & 149.9882 & 1.8018 & $2.61^{+0.14}_{-0.14}$ & -- & $9.4^{+0.1}_{-0.2}$ & $8^{+7}_{-2}$ & 15.2 \\
        398787 & 149.9821 & 1.8019 & $2.37^{+0.12}_{-0.09}$ & -- & $10.8^{+0.1}_{-0.1}$ & $50^{+7}_{-7}$ & 29.3 \\
        399896 & 149.9934 & 1.8040 & $2.54^{+0.27}_{-0.32}$ & -- & $8.7^{+0.1}_{-0.2}$ & $1^{+0}_{-0}$ & 27.8 \\
        400170 & 149.9961 & 1.8043 & $2.84^{+0.38}_{-0.54}$ & -- & $9.3^{+0.2}_{-0.2}$ & $7^{+3}_{-5}$ & 34.8 \\
        400185 & 149.9890 & 1.8042 & $2.49^{+0.16}_{-0.22}$ & -- & $8.6^{+0.2}_{-0.2}$ & $1^{+1}_{-0}$ & 23.5 \\
        400413 & 149.9956 & 1.8042 & $1.84^{+0.36}_{-0.31}$ & -- & $8.7^{+0.2}_{-0.2}$ & $2^{+1}_{-2}$ & 33.5 \\
        400526 & 149.9865 & 1.8041 & $2.97^{+0.18}_{-0.18}$ & -- & $9.1^{+0.1}_{-0.2}$ & $4^{+3}_{-1}$ & 25.0 \\
        401372 & 149.9923 & 1.8054 & $1.83^{+0.56}_{-0.37}$ & -- & $8.3^{+0.2}_{-0.2}$ & $0^{+0}_{-0}$ & 30.3 \\
        401840 & 149.9904 & 1.8053 & $2.49^{+0.20}_{-0.16}$ & -- & $9.7^{+0.1}_{-0.2}$ & $13^{+12}_{-3}$ & 27.8 \\
        \hline
        Total & 149.9898 & 1.7978 & $2.57\pm0.03$ & $2.415\pm0.001$ & $11.8\pm0.1$ & $1097^{+244}_{-169}$ & -- \\
        \hline
    \end{tabular}
\end{table*}

\clearpage

\begin{table*}[!htbp]
    \centering
    \caption{Physical properties of COS-SBC3 members from COSMOS2020 Classic LePhare.}
    \renewcommand*{\arraystretch}{1.4}
    \label{tab:SBC3_physpars}
    \begin{tabular}{c c c c c c c c}
        \hline\hline
        ID & RA & Dec. & $z_{\rm phot}$ & $z_{\rm spec}$ & $\log(M_\ast/{\rm M_{\odot}})$ & SFR & $d_{\rm core}$\\
           & [deg] & [deg] & & & & [${\rm M_\odot\,yr^{-1}}$] & [arcsec]\\
           \hline
        (Confirmed members)\\
        1340799 & 150.7226 & 2.6963 & $3.48^{+0.80}_{-0.84}$ & 2.365$\pm0.001$ & $11.5^{+0.1}_{-0.1}$ & $190^{+90}_{-61}$ & 0.0\\
        (Candidate members)\\
        1334731 & 150.7227768 & 2.6905261 & $2.28^{+0.48}_{-0.99}$ & -- & $8.7^{+0.2}_{-0.2}$ & $2^{+2}_{-0}$ & 20.9 \\
        1335919 & 150.7306877 & 2.6917006 & $2.87^{+0.05}_{-0.05}$ & -- & $8.9^{+0.1}_{-0.1}$ & $4^{+0}_{-0}$ & 33.6 \\
        1337358 & 150.7297781 & 2.6932095 & $2.41^{+0.38}_{-0.66}$ & -- & $9.2^{+0.1}_{-0.2}$ & $4^{+1}_{-3}$ & 28.2 \\
        1337546 & 150.7273308 & 2.6934159 & $2.38^{+0.22}_{-0.25}$ & -- & $9.1^{+0.1}_{-0.1}$ & $2^{+0}_{-2}$ & 20.0 \\
        1337712 & 150.7234355 & 2.6935964 & $2.03^{+0.61}_{-0.50}$ & -- & $8.4^{+0.2}_{-0.2}$ & $0^{+0}_{-0}$ & 10.3 \\
        1337749 & 150.7271582 & 2.6936214 & $2.36^{+0.24}_{-0.54}$ & -- & $9.0^{+0.1}_{-0.2}$ & $2^{+0}_{-1}$ & 19.1 \\
        1338883 & 150.7310329 & 2.6941766 & $2.52^{+0.19}_{-0.16}$ & -- & $10.6^{+0.1}_{-0.0}$ & $0^{+0}_{-0}$ & 31.4 \\
        1338945 & 150.7233826 & 2.6943064 & $3.02^{+0.12}_{-0.11}$ & -- & $9.6^{+0.1}_{-0.2}$ & $9^{+8}_{-2}$ & 7.8 \\
        1338973 & 150.7240954 & 2.6945682 & $2.70^{+0.19}_{-0.15}$ & -- & $10.2^{+0.1}_{-0.1}$ & $23^{+15}_{-15}$ & 8.3 \\
        1339741 & 150.7217645 & 2.6943989 & $2.60^{+0.28}_{-0.25}$ & -- & $11.1^{+0.1}_{-0.1}$ & $100^{+36}_{-1044}$ & 7.5 \\
        1339910 & 150.7245508 & 2.6957013 & $2.92^{+0.12}_{-0.15}$ & -- & $9.0^{+0.2}_{-0.2}$ & $3^{+1}_{-2}$ & 7.4 \\
        1340398 & 150.7227948 & 2.6957280 & $2.07^{+0.31}_{-0.33}$ & -- & $10.3^{+0.1}_{-0.2}$ & $30^{+24}_{-22}$ & 2.3 \\
        1342336 & 150.7172687 & 2.6947672 & $3.01^{+0.17}_{-2.80}$ & -- & $9.7^{+0.1}_{-0.1}$ & $28^{+6}_{-6}$ & 19.9 \\
        1343209 & 150.7207002 & 2.6986974 & $3.04^{+0.18}_{-0.20}$ & -- & $9.4^{+0.1}_{-0.1}$ & $4^{+3}_{-1}$ & 10.9 \\
        1345593 & 150.7217632 & 2.7006506 & $2.79^{+0.31}_{-0.35}$ & -- & $10.0^{+0.1}_{-0.2}$ & $27^{+13}_{-20}$ & 15.9 \\
        1345985 & 150.7173665 & 2.7015039 & $2.70^{+0.06}_{-0.07}$ & -- & $9.8^{+0.1}_{-0.1}$ & $24^{+4}_{-3}$ & 26.5 \\
        1346174 & 150.7183122 & 2.7007088 & $3.05^{+0.29}_{-0.22}$ & -- & $10.1^{+0.1}_{-0.2}$ & $26^{+13}_{-19}$ & 22.1 \\
        1350310 & 150.7214095 & 2.7048155 & $2.85^{+0.06}_{-0.05}$ & -- & $9.7^{+0.1}_{-0.1}$ & $57^{+12}_{-14}$ & 30.9 \\
        1332724 & 150.7252518 & 2.6887611 & $2.32^{+0.31}_{-0.30}$ & -- & $9.1^{+0.1}_{-0.2}$ & $3^{+3}_{-0}$ & 28.9 \\
        1334717 & 150.7259156 & 2.6906676 & $2.15^{+0.30}_{-0.61}$ & -- & $9.1^{+0.1}_{-0.1}$ & $4^{+0}_{-1}$ & 23.6 \\
        1336266 & 150.7228321 & 2.6917998 & $2.07^{+0.29}_{-0.53}$ & -- & $8.5^{+0.2}_{-0.2}$ & $2^{+1}_{-0}$ & 16.3 \\
        1336604 & 150.7225462 & 2.6921015 & $1.82^{+0.18}_{-0.14}$ & -- & $9.2^{+0.1}_{-0.2}$ & $4^{+4}_{-1}$ & 15.2 \\
        1337924 & 150.7219464 & 2.6937675 & $2.27^{+0.50}_{-1.41}$ & -- & $8.4^{+0.2}_{-0.2}$ & $2^{+0}_{-0}$ & 9.5 \\
        1343661 & 150.7188833 & 2.6993108 & $1.95^{+0.30}_{-0.26}$ & -- & $8.9^{+0.1}_{-0.2}$ & $1^{+1}_{-0}$ & 17.1 \\
        1344343 & 150.7199699 & 2.6994621 & $2.29^{+0.78}_{-0.67}$ & -- & $10.1^{+0.2}_{-0.2}$ & $44^{+29}_{-29}$ & 14.7 \\
        1344398 & 150.7190970 & 2.6982136 & $2.10^{+0.10}_{-0.25}$ & -- & $10.5^{+0.1}_{-0.1}$ & $166^{+29}_{-39}$ & 14.3 \\
        1345139 & 150.7153071 & 2.7000313 & $2.04^{+0.15}_{-0.15}$ & -- & $9.5^{+0.1}_{-0.2}$ & $13^{+3}_{-11}$ & 29.4 \\
        1345894 & 150.7270190 & 2.7014442 & $2.20^{+0.61}_{-0.45}$ & -- & $9.3^{+0.1}_{-0.2}$ & $3^{+2}_{-2}$ & 24.4 \\
        1344398 & 150.7190970 & 2.6982136 & $2.10^{+0.10}_{-0.25}$ & -- & $10.5^{+0.1}_{-0.1}$ & $166^{+29}_{-39}$ & 14.3 \\
        \hline
        Total & 150.7196 & 2.6995 & $2.72\pm0.04$ & $2.365\pm0.001$ & $11.8\pm0.1$ & $955^{+116}_{-1048}$ & --\\
        \hline
        (Rejected members)\\
        1345246 & 150.7193 & 2.6998 & $3.18^{+0.55}_{-0.48}$ & 2.723$\pm0.001$ & $11.8^{+0.1}_{-0.1}$ & $243^{+158}_{-65}$ & 15.6\\
        \hline
    \end{tabular}
\end{table*}

\clearpage

\onecolumn
\renewcommand*{\arraystretch}{1.4}
\begin{longtable}{c c c c c c c c}
    \caption{\label{tab:SBC4_physpars}Physical properties of COS-SBC4 members from COSMOS2020 Classic LePhare.}\\
        \\\hline\hline
        ID & RA & Dec. & $z_{\rm phot}$ & $z_{\rm spec}$ & $\log(M_\ast/{\rm M_{\odot}})$ & SFR & $d_{\rm core}$ \\
           & [deg] & [deg] & & & & [${\rm M_\odot\,yr^{-1}}$] & [arcsec]\\
           \hline
           (Confirmed members)\\
           840072 & 150.0368 & 2.2178 & $1.82^{+0.04}_{-0.09}$ & $1.6543\pm0.0006^a$ & $11.4^{+0.1}_{-0.1}$ & $100^{+18}_{-18}$ & 1.4\\
           840456 & 150.0366 & 2.2192 & $1.64^{+0.05}_{-0.03}$ & $1.6381\pm0.0006^a$ & $10.2^{+0.1}_{-0.1}$ & $52^{+8}_{-9}$ & 5.4\\
           840660 & 150.0325 & 2.2195 & $1.65^{+0.04}_{-0.03}$ & $1.6388\pm0.0006^a$ & $10.4^{+0.1}_{-0.1}$ & $76^{+14}_{-14}$ & 15.6\\
           (Candidate members)\\
            824041 & 150.0366923 & 2.2061331 & $1.74^{+0.25}_{-0.18}$ & -- & $8.7^{+0.1}_{-0.1}$ & $3^{+0}_{-0}$ & 42.1 \\
            824902 & 150.0356076 & 2.2068429 & $1.35^{+0.08}_{-0.11}$ & -- & $8.8^{+0.1}_{-0.1}$ & $1^{+1}_{-0}$ & 39.8 \\
            825606 & 150.0349940 & 2.2077652 & $1.39^{+0.43}_{-0.19}$ & -- & $8.2^{+0.1}_{-0.2}$ & $0^{+0}_{-0}$ & 36.8 \\
            826524 & 150.0388680 & 2.2086292 & $1.81^{+0.05}_{-0.04}$ & -- & $9.1^{+0.1}_{-0.1}$ & $22^{+4}_{-4}$ & 34.0 \\
            826583 & 150.0294105 & 2.2087104 & $1.90^{+0.40}_{-0.37}$ & -- & $8.9^{+0.1}_{-0.2}$ & $0^{+0}_{-0}$ & 42.2 \\
            828463 & 150.0318645 & 2.2093383 & $1.88^{+0.11}_{-0.12}$ & -- & $9.1^{+0.1}_{-0.1}$ & $11^{+2}_{-2}$ & 35.3 \\
            830066 & 150.0394925 & 2.2117027 & $1.78^{+0.20}_{-0.28}$ & -- & $8.5^{+0.1}_{-0.1}$ & $1^{+0}_{-0}$ & 24.2 \\
            830220 & 150.0379440 & 2.2119681 & $1.69^{+1.66}_{-1.21}$ & -- & $7.9^{+0.2}_{-0.2}$ & $0^{+0}_{-0}$ & 21.6 \\
            830689 & 150.0275325 & 2.2092421 & $1.89^{+0.13}_{-0.16}$ & -- & $9.4^{+0.2}_{-0.2}$ & $19^{+5}_{-18}$ & 45.4 \\
            830722 & 150.0437771 & 2.2118636 & $1.39^{+0.04}_{-0.04}$ & -- & $9.1^{+0.1}_{-0.1}$ & $3^{+0}_{-0}$ & 33.2 \\
            831134 & 150.0363399 & 2.2124207 & $1.29^{+0.16}_{-0.21}$ & -- & $9.2^{+0.1}_{-0.2}$ & $2^{+1}_{-2}$ & 19.6 \\
            832008 & 150.0422681 & 2.2131902 & $1.62^{+0.10}_{-0.11}$ & -- & $8.7^{+0.1}_{-0.1}$ & $2^{+0}_{-0}$ & 25.9 \\
            832093 & 150.0353196 & 2.2132796 & $1.83^{+0.18}_{-0.11}$ & -- & $9.3^{+0.2}_{-0.2}$ & $8^{+2}_{-7}$ & 17.2 \\
            832129 & 150.0385824 & 2.2128016 & $1.91^{+0.20}_{-0.22}$ & -- & $9.5^{+0.1}_{-0.1}$ & $7^{+6}_{-1}$ & 19.3 \\
            833467 & 150.0429095 & 2.2139294 & $1.45^{+0.01}_{-0.02}$ & -- & $9.0^{+0.1}_{-0.1}$ & $5^{+0}_{-0}$ & 26.2 \\
            835280 & 150.0353156 & 2.2155435 & $1.37^{+0.08}_{-0.34}$ & -- & $9.5^{+0.1}_{-0.1}$ & $12^{+8}_{-5}$ & 9.8 \\
            835641 & 150.0353282 & 2.2149129 & $1.51^{+0.18}_{-0.13}$ & -- & $10.7^{+0.1}_{-0.1}$ & $22^{+7}_{-25}$ & 11.7 \\
            835706 & 150.0422240 & 2.2167433 & $1.77^{+0.12}_{-0.13}$ & -- & $9.0^{+0.1}_{-0.1}$ & $0^{+0}_{-0}$ & 20.1 \\
            835713 & 150.0446451 & 2.2168417 & $1.43^{+1.10}_{-0.40}$ & -- & $8.3^{+0.2}_{-0.2}$ & $0^{+0}_{-0}$ & 28.6 \\
            836069 & 150.0440213 & 2.2165815 & $1.81^{+0.18}_{-0.15}$ & -- & $9.2^{+0.1}_{-0.2}$ & $3^{+1}_{-2}$ & 26.5 \\
            836257 & 150.0366450 & 2.2172921 & $1.89^{+0.19}_{-0.19}$ & -- & $9.4^{+0.1}_{-0.2}$ & $6^{+5}_{-4}$ & 2.0 \\
            836352 & 150.0349150 & 2.2173707 & $1.87^{+2.54}_{-0.99}$ & -- & $8.3^{+0.2}_{-0.2}$ & $0^{+0}_{-0}$ & 6.8 \\
            836436 & 150.0339463 & 2.2167755 & $1.74^{+0.14}_{-0.31}$ & -- & $9.4^{+0.1}_{-0.1}$ & $8^{+3}_{-1}$ & 10.8 \\
            838011 & 150.0360963 & 2.2180155 & $1.91^{+0.07}_{-0.09}$ & -- & $10.1^{+0.1}_{-0.1}$ & $22^{+16}_{-5}$ & 2.5 \\
            838016 & 150.0430325 & 2.2189057 & $1.86^{+0.27}_{-0.24}$ & -- & $8.1^{+0.1}_{-0.2}$ & $0^{+0}_{-0}$ & 22.9 \\
            838184 & 150.0356083 & 2.2176223 & $1.87^{+0.16}_{-0.23}$ & -- & $11.0^{+0.0}_{-0.0}$ & $26^{+7}_{-20}$ & 4.2 \\
            838627 & 150.0353935 & 2.2170824 & $1.56^{+0.04}_{-0.08}$ & -- & $10.5^{+0.1}_{-0.1}$ & $33^{+16}_{-6}$ & 5.6 \\
            839179 & 150.0272763 & 2.2198195 & $1.66^{+0.23}_{-0.15}$ & -- & $8.9^{+0.1}_{-0.2}$ & $2^{+0}_{-1}$ & 34.9 \\
            839200 & 150.0346342 & 2.2186812 & $1.66^{+0.05}_{-0.04}$ & -- & $10.4^{+0.1}_{-0.1}$ & $77^{+14}_{-17}$ & 8.2 \\
            839308 & 150.0270367 & 2.2200170 & $1.86^{+0.23}_{-0.19}$ & -- & $9.0^{+0.1}_{-0.2}$ & $1^{+0}_{-1}$ & 35.9 \\
            839401 & 150.0304066 & 2.2200884 & $1.90^{+0.49}_{-0.38}$ & -- & $8.6^{+0.2}_{-0.2}$ & $1^{+0}_{-0}$ & 24.3 \\
            839911 & 150.0271665 & 2.2203543 & $1.84^{+0.24}_{-0.22}$ & -- & $8.6^{+0.1}_{-0.2}$ & $1^{+0}_{-0}$ & 35.7 \\
            840013 & 150.0450701 & 2.2178394 & $1.89^{+0.01}_{-0.02}$ & -- & $10.6^{+0.1}_{-0.1}$ & $251^{+39}_{-39}$ & 29.9 \\
            840139 & 150.0415085 & 2.2200229 & $1.84^{+0.22}_{-0.23}$ & -- & $9.2^{+0.1}_{-0.1}$ & $5^{+1}_{-1}$ & 18.8 \\
            840279 & 150.0256274 & 2.2186476 & $1.89^{+0.06}_{-0.06}$ & -- & $11.5^{+0.1}_{-0.1}$ & $168^{+45}_{-330}$ & 40.2 \\
            840962 & 150.0427285 & 2.2214865 & $1.65^{+0.54}_{-0.31}$ & -- & $8.5^{+0.2}_{-0.2}$ & $0^{+0}_{-0}$ & 25.2 \\
            841535 & 150.0411735 & 2.2214431 & $1.89^{+0.09}_{-0.09}$ & -- & $9.2^{+0.1}_{-0.1}$ & $5^{+1}_{-0}$ & 20.5 \\
            842081 & 150.0476079 & 2.2214240 & $1.89^{+0.28}_{-0.30}$ & -- & $8.5^{+0.2}_{-0.2}$ & $2^{+0}_{-2}$ & 41.1 \\
            842543 & 150.0370921 & 2.2209232 & $1.73^{+0.03}_{-0.03}$ & -- & $10.7^{+0.1}_{-0.1}$ & $191^{+34}_{-30}$ & 11.2 \\
            842559 & 150.0348747 & 2.2228335 & $1.68^{+0.21}_{-0.20}$ & -- & $8.9^{+0.1}_{-0.2}$ & $1^{+0}_{-0}$ & 19.2 \\
            842815 & 150.0378334 & 2.2231170 & $1.24^{+0.60}_{-0.19}$ & -- & $8.3^{+0.2}_{-0.2}$ & $1^{+0}_{-0}$ & 19.4 \\
            843056 & 150.0255753 & 2.2232906 & $1.87^{+0.30}_{-0.32}$ & -- & $8.6^{+0.1}_{-0.1}$ & $1^{+1}_{-0}$ & 44.8 \\
            843344 & 150.0403893 & 2.2224669 & $1.93^{+0.08}_{-0.11}$ & -- & $9.6^{+0.1}_{-0.1}$ & $17^{+4}_{-14}$ & 21.2 \\
            843499 & 150.0310788 & 2.2237657 & $1.98^{+0.38}_{-0.35}$ & -- & $8.6^{+0.1}_{-0.2}$ & $0^{+0}_{-0}$ & 29.6 \\
            843513 & 150.0358949 & 2.2221884 & $1.57^{+0.05}_{-0.05}$ & -- & $10.8^{+0.1}_{-0.1}$ & $36^{+9}_{-20}$ & 16.0 \\
            844408 & 150.0277642 & 2.2243840 & $1.49^{+0.14}_{-0.12}$ & -- & $9.3^{+0.1}_{-0.2}$ & $3^{+2}_{-2}$ & 40.1 \\
            844708 & 150.0363746 & 2.2241820 & $1.61^{+0.07}_{-0.08}$ & -- & $9.8^{+0.1}_{-0.1}$ & $9^{+2}_{-7}$ & 22.9 \\
            845635 & 150.0324820 & 2.2255824 & $1.56^{+0.34}_{-0.32}$ & -- & $7.9^{+0.1}_{-0.2}$ & $0^{+0}_{-0}$ & 31.9 \\
            845837 & 150.0438421 & 2.2257723 & $1.90^{+0.20}_{-0.19}$ & -- & $8.5^{+0.1}_{-0.2}$ & $0^{+0}_{-0}$ & 38.3 \\
            846014 & 150.0399675 & 2.2252185 & $1.44^{+0.09}_{-0.10}$ & -- & $9.0^{+0.1}_{-0.2}$ & $1^{+1}_{-0}$ & 29.0 \\
            846252 & 150.0289206 & 2.2260815 & $1.75^{+0.31}_{-0.24}$ & -- & $8.6^{+0.1}_{-0.2}$ & $0^{+0}_{-0}$ & 41.0 \\
            846409 & 150.0416911 & 2.2226615 & $1.86^{+0.05}_{-0.05}$ & -- & $10.8^{+0.0}_{-0.0}$ & $0^{+0}_{-0}$ & 24.8 \\
            846597 & 150.0425413 & 2.2241398 & $1.85^{+0.08}_{-0.08}$ & -- & $10.5^{+0.0}_{-0.0}$ & $39^{+6}_{-6}$ & 30.8 \\
            846766 & 150.0280304 & 2.2265595 & $1.65^{+0.57}_{-0.47}$ & -- & $7.9^{+0.2}_{-0.2}$ & $1^{+0}_{-0}$ & 44.4 \\
            846938 & 150.0344524 & 2.2267056 & $1.78^{+0.45}_{-0.35}$ & -- & $8.3^{+0.2}_{-0.2}$ & $0^{+0}_{-0}$ & 33.0 \\
            847198 & 150.0297779 & 2.2265661 & $1.82^{+0.13}_{-0.12}$ & -- & $9.1^{+0.1}_{-0.1}$ & $2^{+0}_{-2}$ & 40.2 \\
            847612 & 150.0388781 & 2.2273074 & $1.10^{+0.19}_{-0.16}$ & -- & $8.3^{+0.1}_{-0.2}$ & $0^{+0}_{-0}$ & 34.9 \\
            848783 & 150.0342761 & 2.2283128 & $1.72^{+0.40}_{-0.26}$ & -- & $8.4^{+0.1}_{-0.2}$ & $0^{+0}_{-0}$ & 38.8 \\
            838193 & 150.0358189 & 2.2182831 & $2.00^{+0.19}_{-0.20}$ & -- & $10.7^{+0.1}_{-0.1}$ & $25^{+10}_{-16}$ & 3.7 \\
            839977 & 150.0359993 & 2.2206550 & $3.32^{+2.87}_{-0.72}$ & -- & $10.5^{+0.1}_{-0.2}$ & $63^{+39}_{-37}$ & 10.5 \\
            \hline
            Total & 150.0364 & 2.2177 & $1.66\pm0.01$ & $1.65\pm0.01$ & $12.1\pm0.1$ & $1536^{+95}_{-342}$ & --\\
    \end{longtable}
Notes: $^a$spectroscopic redshifts from FMOS-COSMOS \citep{Kashino2019_FMOS_COSMOS}.

\clearpage

\begin{table*}[!htbp]
    \centering
    \caption{Physical properties of COS-SBC6 members from COSMOS2020 Classic LePhare.}
     \label{tab:SBC6_physpars}
    \renewcommand*{\arraystretch}{1.4}
    \begin{tabular}{c c c c c c c c}
        \hline\hline
        ID & RA & Dec. & $z_{\rm phot}$ & $z_{\rm spec}$ & $\log(M_\ast/{\rm M_{\odot}})$ & SFR & $d_{\rm core}$ \\
           & [deg] & [deg] & & & & [${\rm M_\odot\,yr^{-1}}$] & [arcsec] \\
           \hline
           (Confirmed members) \\
           835289 & 149.7053 & 2.2153 & $2.54^{+0.10}_{-0.15}$ & 2.323$\pm0.001$ & $11.1^{+0.1}_{-0.1}$ & $73^{+15}_{-19}$ & 2.9\\
           839791 & 149.7054 & 2.2171 & $2.28^{+0.04}_{-0.04}$ & 2.324$\pm0.001$ & $11.0^{+0.0}_{-0.1}$ & $39^{+6}_{-6}$ & 4.0\\
        (Candidate members) \\
        829302 & 149.7100619 & 2.2111172 & $1.98^{+0.96}_{-1.46}$ & -- & $8.2^{+0.5}_{-0.5}$ & $0^{+0}_{-0}$ & 24.4 \\
        829351 & 149.7063016 & 2.2111524 & $2.03^{+0.37}_{-0.34}$ & -- & $8.7^{+0.1}_{-0.2}$ & $0^{+0}_{-0}$ & 17.6 \\
        830266 & 149.6985180 & 2.2119558 & $2.52^{+0.25}_{-0.52}$ & -- & $8.9^{+0.1}_{-0.2}$ & $1^{+0}_{-0}$ & 28.5 \\
        832418 & 149.7008521 & 2.2138819 & $2.04^{+0.33}_{-0.28}$ & -- & $8.8^{+0.1}_{-0.1}$ & $0^{+0}_{-0}$ & 17.8 \\
        833030 & 149.7009141 & 2.2143979 & $2.10^{+0.44}_{-0.47}$ & -- & $8.2^{+0.2}_{-0.2}$ & $0^{+0}_{-0}$ & 16.9 \\
        833427 & 149.7109600 & 2.2140096 & $2.68^{+0.15}_{-0.12}$ & -- & $10.5^{+0.1}_{-0.1}$ & $56^{+15}_{-45}$ & 21.4 \\
        833554 & 149.7001036 & 2.2144192 & $2.00^{+0.16}_{-0.16}$ & -- & $9.3^{+0.1}_{-0.1}$ & $3^{+2}_{-0}$ & 19.6 \\
        835393 & 149.7022002 & 2.2155898 & $2.09^{+0.11}_{-0.12}$ & -- & $9.4^{+0.1}_{-0.2}$ & $4^{+3}_{-1}$ & 11.4 \\
        835756 & 149.7058336 & 2.2167914 & $2.59^{+0.13}_{-0.08}$ & -- & $10.4^{+0.1}_{-0.2}$ & $46^{+34}_{-12}$ & 3.5 \\
        836133 & 149.7118650 & 2.2156886 & $1.72^{+0.07}_{-0.04}$ & -- & $10.8^{+0.1}_{-0.1}$ & $12^{+12}_{-9}$ & 23.5 \\
        837254 & 149.7063679 & 2.2166370 & $2.65^{+0.09}_{-0.13}$ & -- & $9.8^{+0.0}_{-0.1}$ & $48^{+8}_{-7}$ & 4.4 \\
        837789 & 149.7140782 & 2.2165296 & $2.26^{+0.18}_{-0.24}$ & -- & $10.3^{+0.1}_{-0.1}$ & $8^{+6}_{-9}$ & 31.5 \\
        838317 & 149.7132336 & 2.2180115 & $2.15^{+0.11}_{-0.12}$ & -- & $9.7^{+0.1}_{-0.1}$ & $14^{+3}_{-9}$ & 29.4 \\
        839355 & 149.7096905 & 2.2200198 & $2.40^{+0.69}_{-0.71}$ & -- & $8.5^{+0.2}_{-0.2}$ & $1^{+0}_{-0}$ & 21.4 \\
        839473 & 149.7008990 & 2.2188627 & $2.13^{+0.07}_{-0.14}$ & -- & $10.1^{+0.1}_{-0.1}$ & $44^{+11}_{-42}$ & 19.1 \\
        839572 & 149.7007529 & 2.2195411 & $2.44^{+0.05}_{-0.08}$ & -- & $9.1^{+0.1}_{-0.2}$ & $7^{+7}_{-2}$ & 20.9 \\
        840635 & 149.7123006 & 2.2211582 & $2.05^{+0.45}_{-0.44}$ & -- & $8.3^{+0.2}_{-0.2}$ & $0^{+0}_{-0}$ & 31.3 \\
        841215 & 149.7103067 & 2.2213104 & $2.36^{+0.24}_{-0.48}$ & -- & $8.6^{+0.2}_{-0.1}$ & $2^{+0}_{-0}$ & 26.3 \\
        841575 & 149.7010166 & 2.2219712 & $2.60^{+0.21}_{-0.27}$ & -- & $9.0^{+0.1}_{-0.1}$ & $1^{+0}_{-0}$ & 26.7 \\
        841712 & 149.7063190 & 2.2212635 & $2.25^{+0.29}_{-0.17}$ & -- & $9.5^{+0.3}_{-0.1}$ & $33^{+8}_{-34}$ & 19.4 \\
        843027 & 149.7062909 & 2.2215255 & $2.17^{+0.05}_{-0.02}$ & -- & $10.4^{+0.1}_{-0.1}$ & $63^{+9}_{-9}$ & 20.3 \\
        843162 & 149.7037797 & 2.2230084 & $2.44^{+0.12}_{-0.14}$ & -- & $9.3^{+0.1}_{-0.2}$ & $4^{+3}_{-1}$ & 26.0 \\
        \hline
        Total & 149.7057 & 2.2160 & $2.22\pm0.03$ & $2.323\pm0.001$ & $11.6\pm0.1$ & $491^{+49}_{-78}$ & -- \\
        \hline
        (Rejected members)\\
        838104 & 149.7060 & 2.2177 & $2.41^{+0.15}_{-0.26}$ & 1.740$\pm0.001$ & $9.6^{+0.1}_{-0.1}$ & $21^{+6}_{-12}$ & 6.2\\
        \hline
    \end{tabular}
\end{table*}

\clearpage

\section{Redshift probability density functions}
\label{sec:pdfz}

\begin{figure*}[!h]
    \centering
    \includegraphics[width=\textwidth]{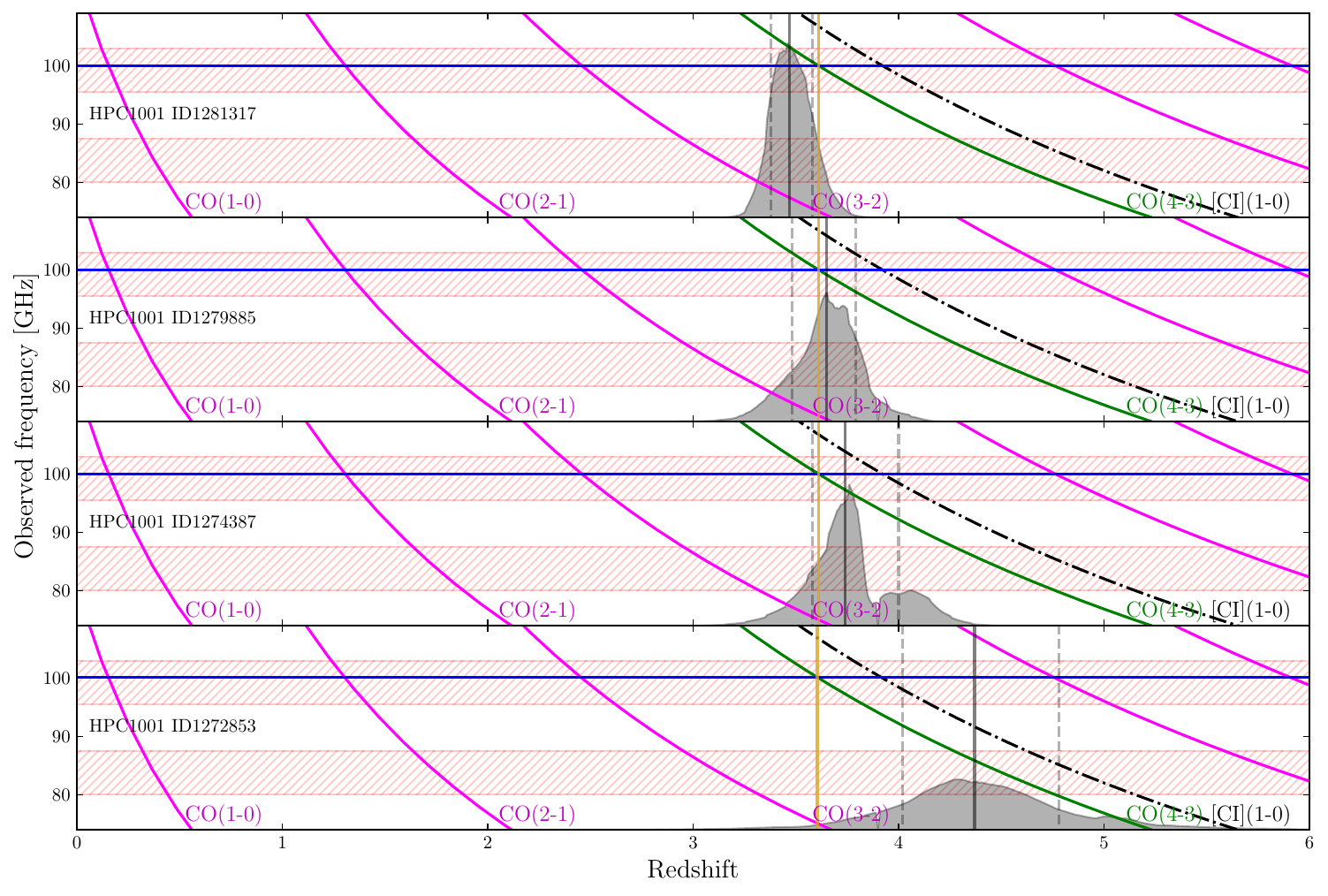}
    \caption{Line identification for sources in the HPC1001 pointing. For each source we show the PDF(z) from COSMOS2020 Classic \texttt{LePhare} as a grey shaded area, and mark its 16th, 50th, and 84th percentile with vertical grey lines. The red shaded areas show the spectral coverage of the observations. Dot-dashed magenta and black  curves show the observed frequencies of CO and CI lines as a function of redshift. The blue line marks the observed frequency of the detected line. We highlight the best redshift solution, $z_{\rm spec}$, with a vertical golden line. The identified emission lines are highlighted in green.}
    \label{fig:pdfz-hpc1001}
\end{figure*}

\begin{figure*}[!h]
    \centering
    \includegraphics[width=\textwidth]{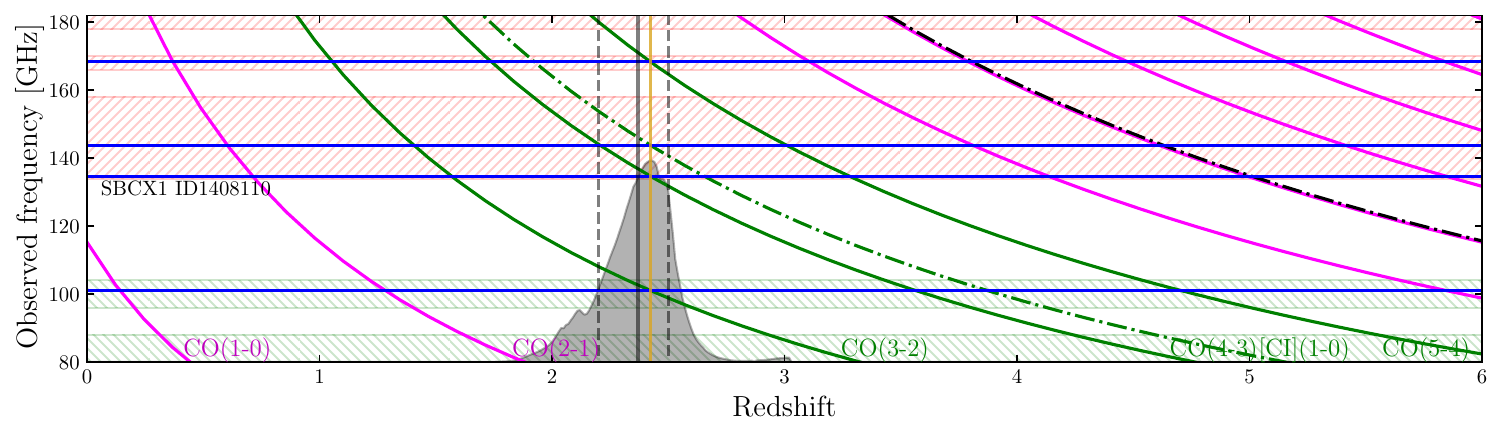}
    \caption{Same as \cref{fig:pdfz-hpc1001} but for COS-SBCX1.}
    \label{fig:pdfz-sbcx1}
\end{figure*}

\begin{figure*}[!h]
    \centering
    \includegraphics[width=\textwidth]{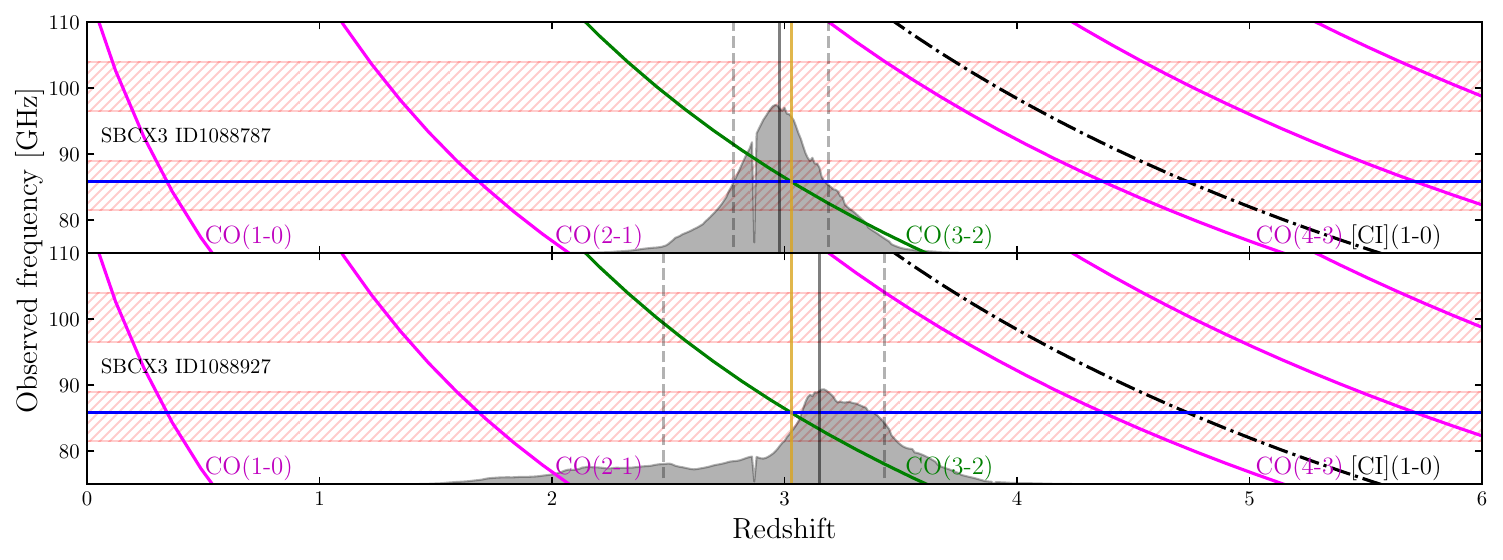}
    \caption{Same as \cref{fig:pdfz-hpc1001} but for COS-SBCX3.}
    \label{fig:pdfz-sbcx3}
\end{figure*}

\begin{figure*}[!h]
    \centering
    \includegraphics[width=\textwidth]{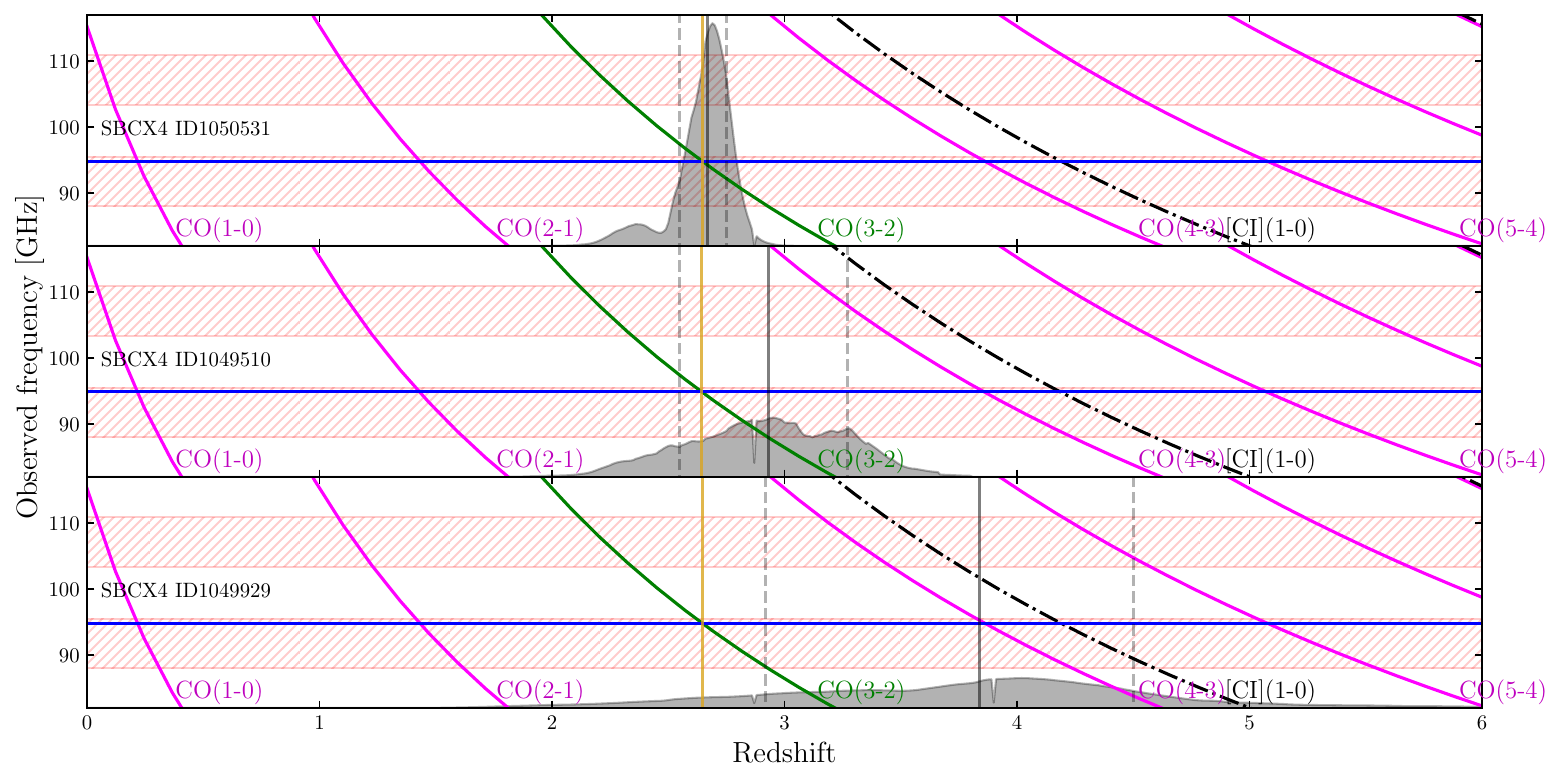}
    \caption{Same as \cref{fig:pdfz-hpc1001} but for COS-SBCX4.}
    \label{fig:pdfz-sbcx4}
\end{figure*}

\begin{figure*}[!h]
    \centering
    \includegraphics[width=\textwidth]{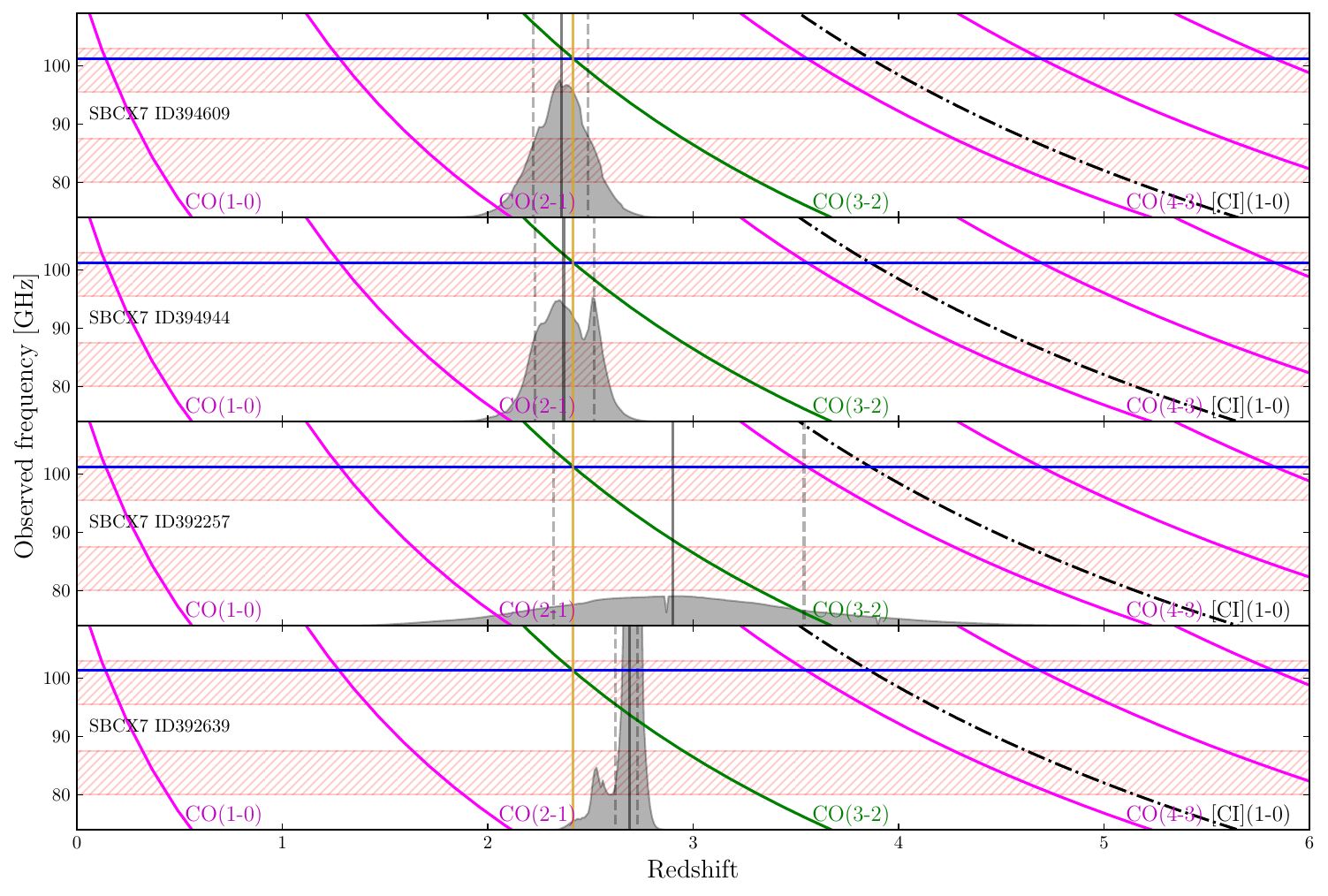}
    \caption{Same as \cref{fig:pdfz-hpc1001} but for COS-SBCX7.}
    \label{fig:pdfz-sbcx7}
\end{figure*}

\begin{figure*}[!h]
    \centering
    \includegraphics[width=\textwidth]{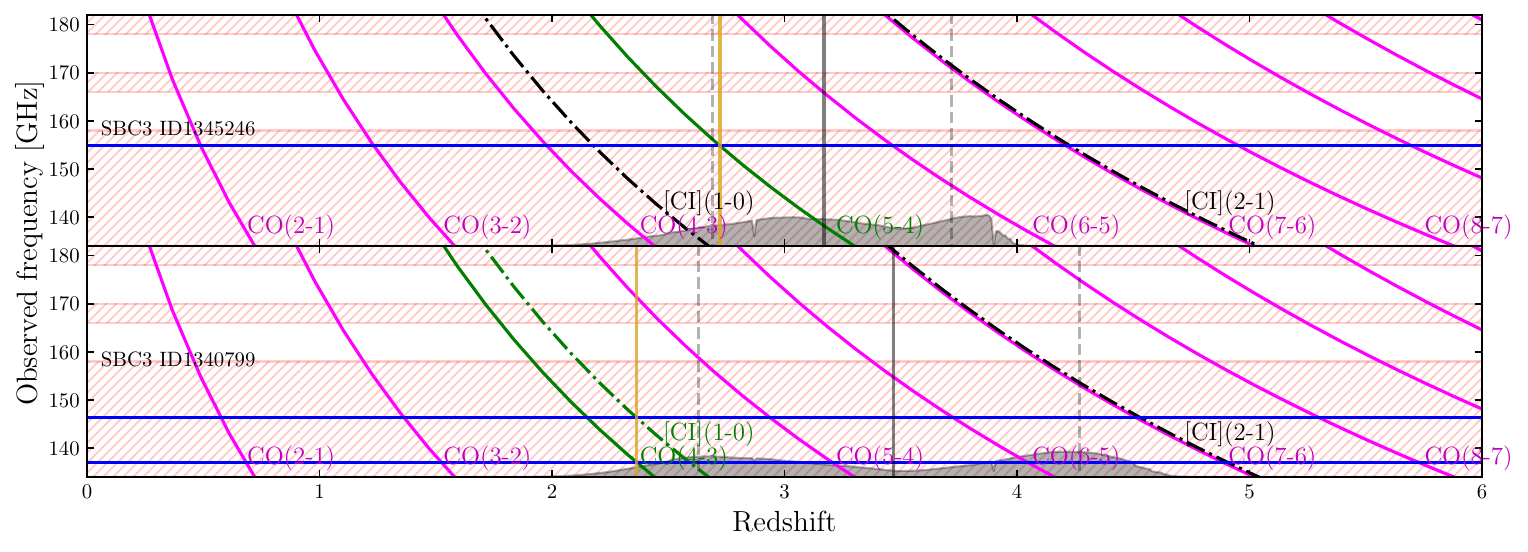}
    \caption{Same as \cref{fig:pdfz-hpc1001} but for COS-SBC3.}
    \label{fig:pdfz-sbc3}
\end{figure*}

\begin{figure*}[!h]
    \centering
    \includegraphics[width=\textwidth]{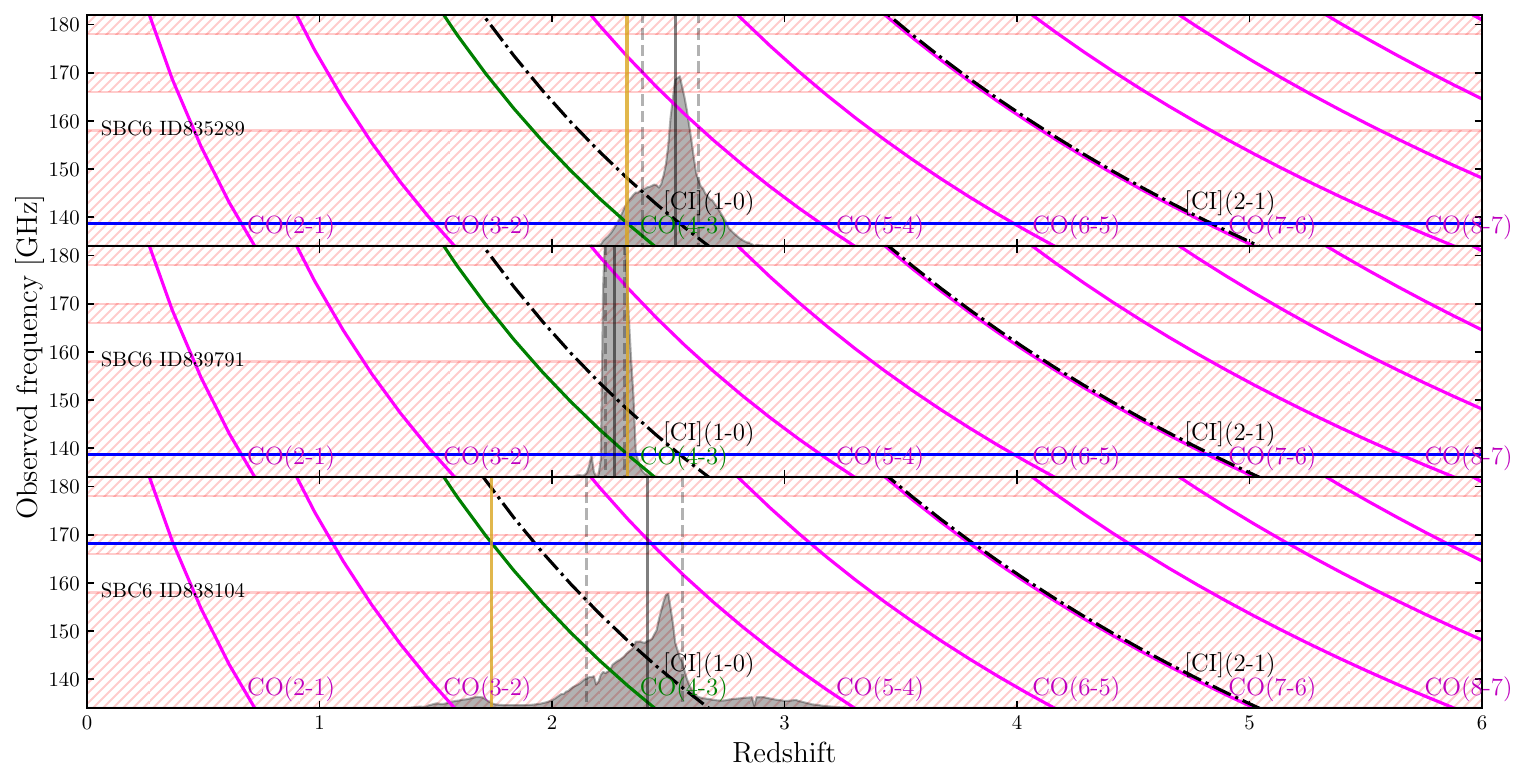}
    \caption{Same as \cref{fig:pdfz-hpc1001} but for COS-SBC6.}
    \label{fig:pdfz-sbc6}
\end{figure*}

\clearpage
\twocolumn
\section{CL-J1001 dark matter profile}
\begin{figure}[!htbp]
    \centering
    \includegraphics[width=1\columnwidth]{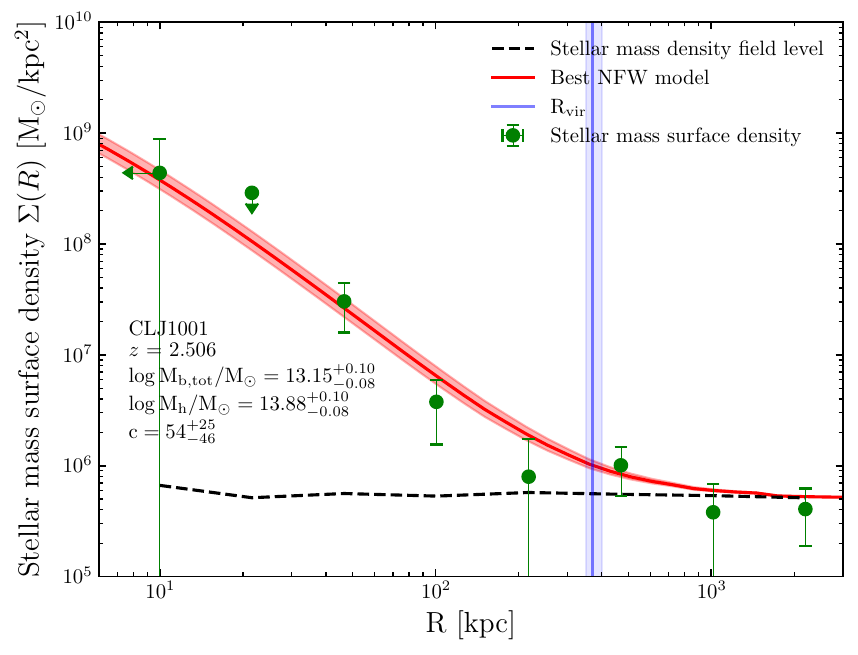}
    \caption{Projected density profile of CLJ1001 \citep{Wang_T2016cluster}. We fit a NFW profile to the radial stellar mass density. Green circles mark the projected
stellar mass density, with the field level shown as a dashed black line. The red curve shows the best-fit NFW profile with uncertainty as the red
shaded area, and the blue line indicates the derived virial radius \citep{Goerdt2010core} with uncertainty as the blue shaded area. The $z_{\rm spec}$, fitted halo mass, and fitted concentration parameters are indicated in text. Upper limits are at $3\sigma$ significances.}
    \label{fig:CLJ1001Mh5}
\end{figure}

\clearpage
\onecolumn
\section{FIR photometry}
\begin{table}[!htbp]
    \centering
    \setlength{\tabcolsep}{3pt}
    \renewcommand*{\arraystretch}{1.4}
    \rotatebox{90}{
    \begin{tabular}{c c c c c c c c c c c}
    Instrument & MIPS & PACS & PACS & SPIRE & SPIRE & SPIRE & SCUBA-2 & ALMA & ALMA/NOEMA & MeerKAT\\
    \hline\hline
    Structure & $24\,{\rm \mu m}$ & $100\,{\rm \mu m}$ & $160\,{\rm \mu m}$ & $250\,{\rm \mu m}$ & $350\,{\rm \mu m}$ & $500\,{\rm \mu m}$ & $850\,{\rm \mu m}$ & $1.3\,{\rm mm}$ & $3\,{\rm mm}$ & $23\,{\rm cm}$\\
    
    & ${\rm \mu Jy}$ & ${\rm mJy}$ & ${\rm mJy}$ & ${\rm mJy}$ & ${\rm mJy}$ & ${\rm mJy}$ & ${\rm mJy}$ & ${\rm mJy}$ & ${\rm \mu Jy}$ & ${\rm \mu Jy}$ \\ \hline
    HPC1001 & $<30$ & $<5.43$ & $<6.6$ & $12.99\pm2.57$ & $19.43\pm4.37$ & $17.38\pm2.41$ & $6.09\pm1.80$ & $2.46\pm0.07$ & $118.00\pm19.48$ & $46.34\pm1.30$\\
    COS-SBCX1 & $<57$ & $<5.97$ & $13.43\pm2.24$ & $17.21\pm2.78$ & $19.41\pm3.26$ & $14.19\pm2.76$ & $6.62\pm1.69$ & -- & $<46$ & $57.28\pm1.30$\\
    COS-SBCX3 & $<129$ & $<6.03$ & $16.59\pm2.61$ & $27.02\pm2.06$ & $36.51\pm2.38$ & $26.55\pm1.97$ & $5.86\pm1.03$ & $2.32\pm0.22$ & $97.00\pm26.72$ & $104.05\pm0.99$\\
    COS-SBCX4 & $<64.5$ & $7.03\pm2.05$ & $29.65\pm2.47$ & $48.49\pm2.35$ & $59.58\pm2.29$ & $49.21\pm1.37$ & $13.66\pm0.92$ & -- & $231.00\pm26.51$ & $170.73\pm1.63$\\
    COS-SBCX7 & $134.3\pm15.3$ & $<6.42$ & $18.19\pm2.70$ & $31.99\pm2.72$ & $35.93\pm3.07$ & $27.80\pm3.35$ & $4.86\pm1.37$ & -- & $169.00\pm33.67$ & $59.85\pm1.30$\\
    COS-SBC3 & $78.0\pm21.0$ & $<6.54$ & $7.66\pm2.41$ & $33.53\pm4.96$ & $45.99\pm4.85$ & $<41.64$ & $10.17\pm2.26$ & -- & $214.67\pm57.86$ & $73.45\pm1.36$\\
    COS-SBC4 & $<97.5$ & $<9.15$ & $19.23\pm2.49$ & $32.61\pm3.26$ & $26.86\pm5.09$ & $18.20\pm4.38$ & $6.68\pm1.37$ & -- & $<48$ & $95.93\pm1.30$\\ 
    COS-SBC6 & $151.3\pm23.5$ & $<6.57$ & $22.91\pm2.55$ & $41.32\pm2.75$ & $45.30\pm4.66$ & $27.29\pm2.02$ & $7.65\pm1.06$ & $1.85\pm0.31$ & -- & $383.08\pm2.81$\\
    \end{tabular}
    }
    \caption{Observed IR to radio photometry for all structures. Upper limits are at $3\sigma$ significances.}
    \label{tab:FIR-photo}
\end{table}

\clearpage
\twocolumn
\section{NFW Concentrations}
\begin{table}[!htbp]
    \centering
    \caption{Concentration parameters for all groups.}
    \renewcommand{\arraystretch}{1.25}
    \begin{tabular}{c c c}
        \hline\hline
        Structure & Concentration & Prediction$^a$\\
        \hline
        HPC1001 & $44^{+21}_{-42}$ & $21\pm4$ \\
        SBCX3 & $13^{+16}_{-10}$ & $20\pm4$\\
        SBCX4 & $55^{+23}_{-39}$ & $15\pm3$ \\
        SBCX1 & $55^{+23}_{-44}$ & $13\pm3$\\
        SBCX7 & $33^{+21}_{-15}$ & $13\pm3$\\
        SBC3 & $46^{+26}_{-46}$ & $13\pm3$\\
        SBC6 & $28^{+22}_{-21}$ & $13\pm3$\\
        SBC4 & $60^{+19}_{-35}$ & $10\pm2$\\
        \hline
    \end{tabular}
    \label{tab:concentration}
    {\\Notes:$^a$predicted concentration using concentration, halo mass, and redshift relation from \citet{Ludlow2016_concentrations}.}
\end{table}

\begin{figure}[!htbp]
    \centering
    \includegraphics[width=1\columnwidth]{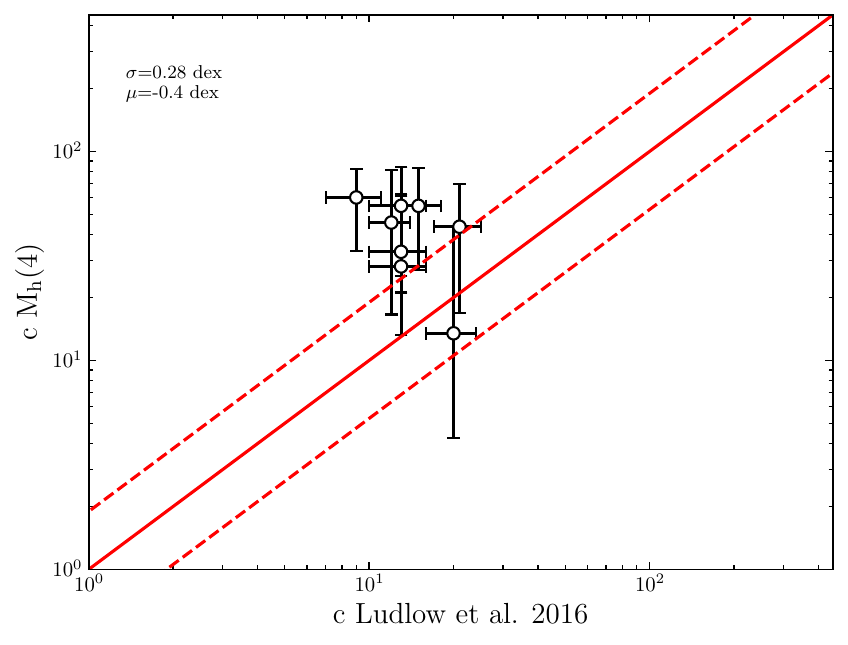}
    \caption{Comparison between the concentration parameter obtained from profile fitting and the predicted concentration parameters from \citet{Ludlow2016_concentrations}. }
    \label{fig:concentration}
\end{figure}

\clearpage
\onecolumn
\section{Interloper fractions}
\begin{figure}[!htbp]
    \centering
    \includegraphics[width=0.49\textwidth]{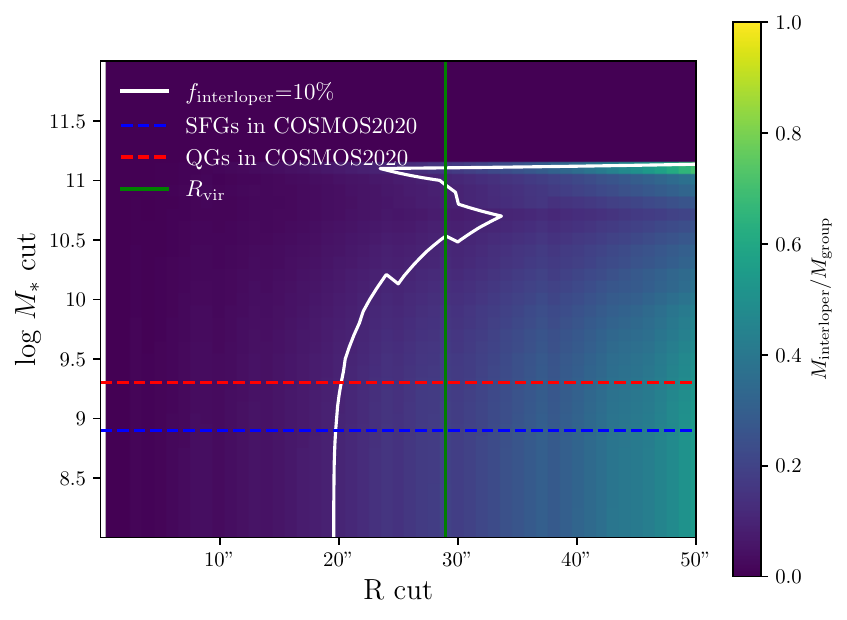}
    \includegraphics[width=0.49\textwidth]{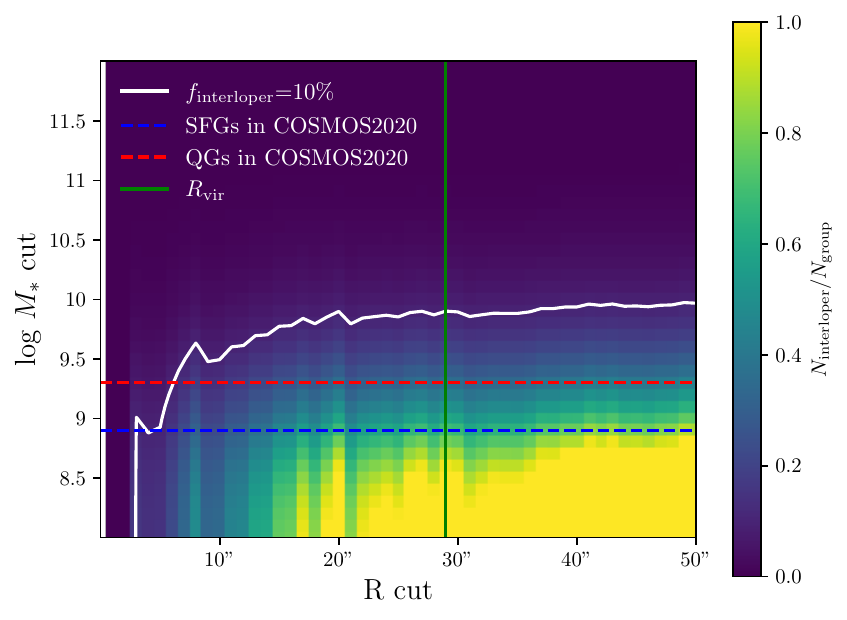}
    \caption{Interloper fractions as a function of the adopted mass cut and radius cut. We show one here as an example, for the structure COS-SBCX7. The white line shows $f_{\rm interloper}=10\%$, and the dashed red and blue lines show the completeness of quiescent galaxies and star-forming galaxies, respectively, in the COSMOS2020 catalogue. The green line shows the estimated virial radius. {\bf Left:} Stellar mass interloper fraction. {\bf Right:} Number interloper fraction.}
    \label{fig:2d-interlopers}
\end{figure}

\clearpage

\section{Projected NFW profiles}
\label{sec:nfw_profiles}
Formulae for projected NFW profiles in a flat Universe, adapted from \texttt{cluster\_toolkit} \citep{McClintock2019DES}\footnote{\url{https://github.com/tmcclintock/cluster_toolkit/}}.\\
First we define the critical density of a flat Universe at $z$:
\begin{align*}
    \rho_{\rm crit}(z) & = \frac{3}{8\pi G}\left(H_0\sqrt{\Omega_m(1+z)^3+\Omega_\Lambda}\right)^2, \\
\end{align*}
where $H_0$ is the Hubble constant, $G$ is the gravitational constant, $\Omega_m$ is the matter energy fraction, $\Omega_\Lambda$ is the vacuum energy fraction, and $\Omega_m+\Omega_\Lambda=1$.\\
Then use this to define the matter density of the universe at $z$:
\begin{align*}
    \rho_m(z) & = \Omega_m\rho_{\rm crit}(z).\\
\end{align*}
Then we define the overdensity of dark matter halo:
\begin{align*}
    \delta_c & =\frac{\Delta \frac{c^3}{3}}{\log(1+c)-\frac{c}{1+c}},\\
\end{align*}
where $c$ is the concentration parameter of the dark matter halo, and $\Delta$ is the overdensity constant.\\
Next, we define the radius where the dark matter halo is $\Delta$ times denser than the matter density of the Universe:
\begin{align*}
    R_\Delta & = \sqrt[3]{\frac{M_\Delta}{\frac{4}{3}\pi\rho_m\Delta}},\\
\end{align*}
where $M_\Delta$ is the mass contained within a sphere of $R_\Delta$ with density $\rho_m\Delta$.\\
Then we define the scale radius connecting $R_{\Delta}$ and $c$ of the halo:
\begin{align*}
    R_s & =\frac{R_\Delta}{c}.\\
\end{align*}
Next we define the shape of a line-of-sight-projected NFW profile:
\begin{align*}
    g(r) & = \left\{ 
    \begin{array}{c c}
      \frac{1-\frac{2}{\sqrt{1-\left(\frac{r}{R_s}\right)^2}}\tanh^{-1}\left(\sqrt{\frac{1-\frac{r}{R_s}}{1+\frac{r}{R_s}}}\right)}{\left(\frac{r}{R_s}\right)^2-1} & r < R_s \\
      \frac{1-\frac{2}{\sqrt{\left(\frac{r}{R_s}\right)^2-1}}\tan^{-1}\left(\sqrt{\frac{\frac{r}{R_s}-1}{1+\frac{r}{R_s}}}\right)}{\left(\frac{r}{R_s}\right)^2-1} & r \geq R_s. \\
    \end{array} \right.\\
\end{align*}
Finally we combine all these parameters to define the surface density profile at radius $r$:
\begin{align*}
    \Sigma(r) & = 2R_s\delta_c\rho_mg(r).\\
\end{align*}
And the cumulative surface density profile:
\begin{align*}
    \bar{\Sigma}(<R) & =\frac{2}{R^2}\int_0^RdR'R'\Sigma(R').
\end{align*}

\end{appendix}

\end{document}